\documentclass[%
 reprint,
 amsmath,amssymb,
prb,
]{revtex4-2}

\usepackage{lipsum}
\usepackage{bm}
\usepackage{amssymb}
\usepackage{amstext}
\usepackage{amsmath}
\usepackage{mathrsfs}
\usepackage{graphicx}
\usepackage{color}
\usepackage{amsthm}
\usepackage{floatrow}
\usepackage{array}

\usepackage{listings}
\usepackage{booktabs}
\usepackage{changepage}

\definecolor{codegreen}{rgb}{0,0.6,0.4}
\definecolor{codegray}{rgb}{0.5,0.5,0.5}
\definecolor{codepurple}{rgb}{0.58,0,0.82}
\definecolor{backcolour}{rgb}{0.97,0.97,0.97}
\definecolor{keywordcolor}{rgb}{0,0,0.8}

\lstdefinestyle{mystyle}{
    backgroundcolor=\color{backcolour},   
    commentstyle=\color{codegreen},
    keywordstyle=\color{keywordcolor},
    numberstyle=\tiny\color{codegray},
    stringstyle=\color{codepurple},
    basicstyle=\ttfamily\footnotesize,
    breakatwhitespace=false,         
    breaklines=true,                 
    captionpos=b,                    
    keepspaces=true,                 
    numbers=left,                    
    numbersep=5pt,                  
    showspaces=false,                
    showstringspaces=false,
    showtabs=false,                  
    tabsize=2
}
\usepackage{hyperref}

\begin{document}
\preprint{APS/123-QED}
\title{Simplex path integral and simplex renormalization group for high-order interactions}
\thanks{This work is a part of the Topophy program. Yang Tian is a lead and corresponding author, Aohua Cheng is a lead author, and Pei Sun is a corresponding author. }%

\author{Aohua Cheng}
\email{aohuacheng18@gmail.com}
\altaffiliation[]{Department of Psychology \& Tsinghua Laboratory of Brain and Intelligence, Tsinghua University, Beijing, 100084, China}
 \altaffiliation[Also at ]{Tsien Excellence in Engineering
Program, Tsinghua University, Beijing, 100084, China}

\author{Yunhui Xu}
\email{yhxu0916@gmail.com}
 \altaffiliation[]{Department of Physics, Tsinghua University, Beijing, 100084, China.}
 
\author{Pei Sun}%
\email{peisun@cityu.edu.mo}
\altaffiliation[]{Faculty of Health Sciences, City University of Macau, Macau, 999078, China.}

	\author{Yang Tian}
	\email{tyanyang04@gmail.com}
        \altaffiliation[]{Faculty of Health Sciences, City University of Macau, Macau, 999078, China}
        \altaffiliation[Also at ]{Faculty of Data Science, City University of Macau, Macau, 999078, China.}




\begin{abstract}
Modern theories of phase transitions and scale-invariance are rooted in path integral formulation and renormalization group (RG). Despite the applicability of these approaches on simple systems with only pairwise interactions, they are less effective on complex systems with un-decomposable high-order interactions (i.e., interactions among arbitrary sets of units). To precisely characterize the universality of high-order interacting systems, we propose simplex path integral and simplex renormalization group (SRG) as the generalizations of classic approaches to arbitrary high-order and heterogeneous interactions. We first formalize the trajectories of units governed by high-order interactions to define path integrals on corresponding simplices based on a high-order propagator. Then we develop a method to integrate out short-range high-order interactions in the momentum space, accompanied by a coarse graining procedure functioning on the simplex structure generated by high-order interactions. The proposed SRG, equipped with a divide-and-conquer framework, can deal with the absence of ergodicity arised from the sparse distribution of high-order interactions and renormalize a system with intertwined high-order interactions on the $p$-order according to its properties on the $q$-order ($p\leq q$). The associated scaling relation and its corollaries support to differentiate among scale-invariant, weakly scale-invariant, and scale-dependent systems across different orders. We have validated our theory in multi-order scale-invariance verification, topological invariance discovery, organizational structure identification, and information bottleneck analysis. These experiments demonstrate the capacity of our theory for identifying intrinsic statistical and topological properties of high-order interacting systems during system reduction.

\end{abstract}

\maketitle
\section{Introduction}
\subsection{Unknowns about high-order interactions}
Over the past decades, the studies on phase transition phenomena, especially the non-equilibrium ones, in different interacting systems have accomplished substantial progress \cite{henkel2008non,lubeck2004universal}. This progress should be credited to the development of path integral \cite{feynman2010quantum,kleinert2009path,chow2015path} and renormalization group \cite{pelissetto2002critical,goldenfeld2018lectures} theories, which significantly deepen our understanding of system dynamics and provide a precise formulation of scaling and criticality.

However, a theoretical vacancy can be found in the existing path integral and renormalization group approaches if we subdivide interactions into pairwise and high-order categories. As the name suggests, a pairwise interaction only involves a pair of units and does not require the participation of any other unit. This kind of interactions can be represented by edges in networks \cite{zhang2023higher,lucas2020multiorder} and are implicitly used in the derivations of classic path integral and renormalization group theories (e.g., see Ref. \cite{villegas2023laplacian}). A high-order interaction, on the contrary, is the mutual coupling among more than two units \cite{benson2016higher,lambiotte2019networks,majhi2022dynamics}. While some high-order interactions can be decomposed into a group of pairwise interactions, most high-order interactions are un-decomposable and in-equivalent to the direct sum of the pairwise ones \cite{lucas2020multiorder,battiston2021physics}. For instance, the triplet collaboration among three agents is not equivalent to the trivial sum of three pairs of individual collaborations. These un-decomposable high-order interactions have intricate effects on system dynamics \cite{majhi2022dynamics,battiston2021physics} and can not be trivially represented by ordinary networks \cite{lambiotte2019networks}. Although notable efforts have been devoted to studying the optimal characterization of high-order interactions (e.g., using simplicial complexes \cite{baccini2022weighted,torres2020simplicial,reitz2020higher,millan2020explosive} or hypergraphs \cite{lotito2022higher,carletti2020random,carletti2020dynamical}), there are fewer works focusing on developing specialized path integral and renormalization group theories for complex systems with high-order interactions \cite{bianconi2020spectral,reitz2020higher}. The difficulty of proposing these specialized frameworks arises from the intrinsic dependence on pairwise interactions when researchers constrain the discrete replicas of interacting systems as conventional networks. Once this constraint is relaxed, the intertwined effects of newly included high-order interactions would make the analysis on emerging dynamics highly non-trivial and even lead to novel results that can not be directly derived by classic theories (e.g., see instances in diffusion and random walks \cite{torres2020simplicial,schaub2020random,carletti2020random}, social dynamics \cite{alvarez2021evolutionary,iacopini2019simplicial,de2020impact}, and neural dynamics \cite{giusti2016two,petri2014homological}). 

\subsection{Related works and remaining challenges}
To suggest a possible direction for developing path integral and renormalization group theories for high-order interactions, we summarize the accomplishments and limitations of previous works. 

To date, most progress of the computational implementations of path integral and renormalization group is achieved on real systems with pairwise interactions \cite{meshulam2018coarse,garcia2018multiscale,villegas2023laplacian,bradde2017pca,lahoche2022generalized,zheng2020geometric}. Although box-covering methods \cite{song2005self,song2006origins,goh2006skeleton,kim2007fractality} may be the most straightforward ways for coarse graining while preserving the organizational properties of a system represented by a network, they depend on the assumption of internal fractal properties and convey no information about system dynamics. Different from box-covering, spectral method \cite{gfeller2007spectral} pays more attention to inherent dynamics properties (e.g., random walks) during coarse graining but may suffer from the obstacles of finding macro-units implied by small-world effects \cite{klemm2002growing,zheng2020geometric,villegas2023laplacian}. Aiming at resolving the coexisting and correlated scales, a geometric renormalization group is developed to identify the potential geometric scaling rather than path-distance scaling \cite{zheng2020geometric}. Nevertheless, this approach essentially relies on the \emph{a-priori}-knowledge of the hidden metric space of networks (e.g., the hyperbolic space \cite{krioukov2010hyperbolic}). With the pursuit of realizing the metric-free coarse graining whose iterability is not limited by small-world effects, a Laplacian renormalization group (LRG) \cite{villegas2023laplacian} has been derived from the diffusion on networks, a continuum counterpart of the block-spin transformation \cite{matsumoto2021renormalization}, to deal with the network topology, dynamics, and geometry encoded by the graph Laplacian operator \cite{de2016spectral}. This approach has a natural connection with path integrals as well \cite{villegas2022laplacian}. Although these properties make the LRG a promising foundation for in-depth explorations, this framework suffers from several non-negligible limitations. First, there is no apparent correspondence between the LRG and an appropriate approach for analyzing un-decomposable high-order interactions because the Laplacian operator can not characterize polyadic relations directly \cite{benson2016higher,lambiotte2019networks,majhi2022dynamics,lucas2020multiorder,battiston2021physics}. Second, the intrinsic dependence of the LRG on the ergodic property of system dynamics reflected by network connectivity limits the applicability of this approach in analyzing high-order interactions, which are more sparsely distributed in the system than pairwise interactions and do not necessarily ensure ergodicity.

There exist fewer pioneering works devoted to developing renormalization group theories for high-order interactions, among which, a notable framework is the real space renormalization group proposed by Refs. \cite{bianconi2020spectral,reitz2020higher}. This approach is initially defined on the Laplacian operators of the network skeletons of the Apollonian \cite{andrade2005apollonian} and the pseudo-fractal \cite{dorogovtsev2002pseudofractal} simplicial complexes to calculate spectral dimensions \cite{bianconi2020spectral}. Then, it is generalized to the normalized up-Laplacian operator \cite{schaub2020random,torres2020simplicial} of simplicial complexes to deal with high-order spectra. Despite the effectiveness of this approach, its dependence on the non-trivial manual derivations using the Gaussian model \cite{hwang2010spectral} and the specialized application scope about spectral dimensions \cite{bianconi2020spectral,reitz2020higher} make it less applicable to the real scenarios where a programmatic implementation is demanded for renormalizing the complex systems generated by empirical data automatically.

In sum, while representative works such as the LRG \cite{villegas2023laplacian} have suggested a promising way to renormalize complex systems, the appropriate path integral and renormalization group theories for high-order interactions remain clouded by numerous theoretical unknowns and technical difficulties.

\begin{figure*}[!t]
\includegraphics[width=1\columnwidth]{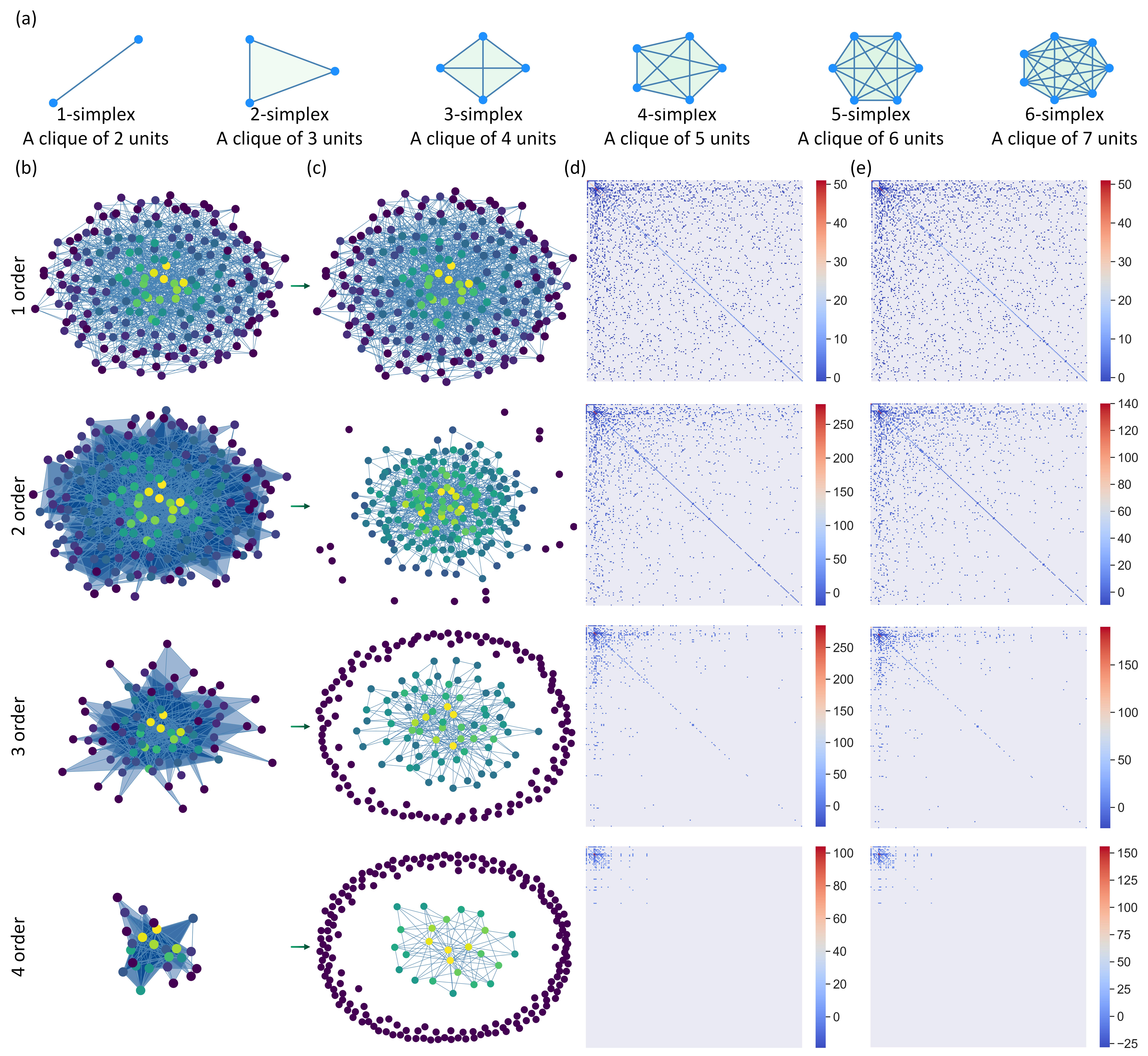}
\caption{\label{G1}System evolution governed by high-order interactions. (a) The simplicial complexes of different orders are presented. A $n$-order simplex manifests itself as a clique with $n+1$ units in the network and represents the $n$-order interactions among these $n+1$ units (i.e., each $n$-order interaction is formed by the simultaneous participation of $n+1$ units). As an instance, the $1$-order simplex is a pairwise interaction. (b) An instance of the interacting system with high-order interactions is presented, where there are $200$ units whose pairwise interactions follow a Barab{\'a}si-Albert network (the number of edges between a new unit and existing units is $c=6$, see definitions in Appendix \ref{ASec1}) \cite{albert2002statistical,barabasi2009scale}. The units engaged with the first four orders of interactions are extracted using simplicial complexes. (c) The associated network sketches of the system shown in (b) are presented on each order, where two units share an edge in the network sketch on $n$-order only if they are engaged with a $n$-order interaction in the system. (d-e) The corresponding multi-order Laplacian (d) and high-order path Laplacian (e) operators of the system shown in (b) are illustrated in every order.} 
\end{figure*}

\subsection{Our frameworks and contributions}

To fill these gaps, we generalize classic path integral and renormalization group theories from dyadic to polyadic relations. To realize such generalizations, we are required to derive the appropriate definitions of the canonical density operators (i.e., the functional over fields) \cite{pathria2016statistical,thompson2015mathematical,villegas2023laplacian} and the coarse graining processes in real and moment spaces \cite{pelissetto2002critical,goldenfeld2018lectures} for high-order interactions. 

Our work suggests a way to define these concepts on simplices, in which simplices can join any sub-set of units to characterize un-decomposable high-order interactions among them (see Fig. \ref{G1}(a) for instances and Appendix \ref{ASec1} for definitions). We propose two kinds of high-order propagators as the generalized path integral formulation of the abstract diffusion processes driven by multi-order interactions, which further creates a system mode description tantamount to the moment space. The coarse graining in moment space is implemented by progressively integrating out short-range system modes corresponding to some certain order of interactions. Parallel to this process, we propose a real space coarse graining procedure to reduce the simplex structures associated with short-range high-order interactions, where we suggest a way to overcome the dependence on ergodicity. The correspondence between the moment and real spaces enables our framework to the system coarse graining according to both organizational structure and internal dynamics. Taken together, these definitions give birth to our simplex path integral and simplex renormalization group (SRG) theories, which are applicable to arbitrary orders of interactions.

The SRG can be used to uncover the potential scale invariance of high-order interactions, whose non-trivial fixed point may suggest the existence of critical phenomena. Different from conventional renormalization groups, the SRG is designed to renormalize $p$-order interactions (each $p$-order interaction requires the simultaneous participation of $p+1$ units) in the system according to the structure and dynamics associated with $q$-order interactions ($p\leq q$). Therefore, we can compare the renormalization flow of $p$-order interactions guided by $p$-order interactions with that guided by arbitrary higher-order interactions. This feature allows us to investigate the impacts of higher-order interactions on lower-order ones and study the differences between lower-order and higher-order interactions in terms of scaling properties. In a special case with $p=q=1$ (i.e., pairwise interactions), the two kinds of high-order propagators developed in our work both reduce to the classic counterpart used in the LRG \cite{villegas2023laplacian} and our proposed SRG can reproduce the outputs of the LRG in the ergodic case. 

To comprehensively demonstrate the capacity of our approaches in identifying the latent characteristics of diverse real complex systems, we validate our framework from the perspectives of multi-order scale-invariance detection, topological invariance discovery, latent organizational structure identification, and information bottleneck optimization, respectively. An efficient code implementation of this framework is provided in \cite{aohua2023toolbox}. One can also see a detailed explanation about the programmatic implementation and usage of the SRG in Appendix \ref{ASec13}.

\section{Un-decomposable high-order interactions}\label{Sec2}

The Laplacian operator is a natural choice for characterizing interactions and their spreading among units \cite{newman2018networks,newman2006structure,newman2003random}. In a special case with pairwise interactions, the Laplacian has a classic expression, $\mathbf{L}_{ij}=\delta_{ij}\sum_{k}\mathbf{A}_{ik}-\mathbf{A}_{ij}$, where $\delta$ denotes the Kronecker delta function and $\mathbf{A}$ is a weighted adjacency matrix that describes the non-negative coupling strengths among all units in unit set $V$ (here the non-negativity is required by the Laplacian). Under the autonomous condition, the evolution of system dynamics given an initial state $\mathbf{x}_{0}$ is defined by $\mathbf{x}=\exp\left(-\tau\mathbf{L}\right)\mathbf{x}_{0}$, where $\tau\in\left(0,\infty\right)$ denotes a time scale \cite{bullo2020lectures}. However, in more general cases with un-decomposable high-order interactions, the classic Laplacian operator is no longer applicable. To fill the gap, previous studies have explored the generalization of the Laplacian operator on uniform hypergraphs \cite{hu2015laplacian,zhou2014some} and simplicial complexes \cite{horak2013spectra,chebbi2018discrete,schaub2020random,lucas2020multiorder}, where diverse variants of the Laplacian (e.g., the combinatorial Laplacian \cite{ribando2022graph}, the Hodge Laplacian \cite{lim2020hodge,schaub2020random}, and the multi-order Laplacian \cite{lucas2020multiorder}) have been proposed to characterize high-order interactions. 

In our work, we first derive a multi-order Laplacian following the spirit of Ref. \cite{lucas2020multiorder}. Then, we develop a new operator, referred to as the high-order path Laplacian, as an alternative description of high-order interactions from a different perspective. As shown in Fig. \ref{G1}, these two operators are defined on clique complexes, a kind of commonly used simplicial complexes in the persistent homology analysis \cite{wang2020persistent,ribando2022graph} (see Appendix \ref{ASec1} for clique complexes). The derivations of these two operators are elaborated in Appendix \ref{ASec2}, where we summarize the similarities and differences between them. Although these two operators have different physics meanings and derivation processes, they can be reformulated into similar mathematical forms as presented in Secs. \ref{SecII-A}-\ref{SecII-B}. Thus, the theoretical differences between these two operators will not impede their unified numerical calculations and programmatic implementations.

\subsection{High-order interactions through faces}\label{SecII-A}
Let us consider the $n$-order interactions that manifest as the simultaneous participation of $n+1$ related units. The un-decomposability of these interactions lies in that the couplings among $n+1$ units are required to occur simultaneously and none of them is dispensable. For instance, the triplet collaboration requires the participation of all three agents together. As discussed in Ref. \cite{lucas2020multiorder}, these interactions can be represented by the $n$-faces of the $n$-simplex where $n+1$ units are placed on. For the sake of clarity, we define $\{i\}_{n+1}\subset V$ as an arbitrary set of $n+1$ units in which unit $i$ is included and denote $S^{\{i\}}_{n+1}$ as the set of all permutations (i.e., rearrangements) on $\{i\}_{n+1}$. The multi-order Laplacian $\mathbf{L}_{M}^{\left(n\right)}$ associated with $n$-simplices is defined as
\begin{align}
\left[\mathbf{L}_{M}^{\left(n\right)}\right]_{ij}=n \delta_{ij}\mathbf{D}_{i}^{\left(n\right)}-\mathbf{A}_{ij}^{\left(n\right)}.\label{EQ1}
\end{align}
In Eq. (\ref{EQ1}), the term $\mathbf{D}_{i}^{\left(n\right)}$ counts the number of $n$-simplices that contains unit $i$ 
\begin{align}
\mathbf{D}_{i}^{\left(n\right)}=\sum_{\{i\}_{n+1}\subset V}\prod_{\omega\in 
S^{\{i\}}_{n+1}}\mathbf{M}_{\omega},\label{EQ2}
\end{align}
Meanwhile, the term $\mathbf{A}_{ij}^{\left(n\right)}$ counts the number of $n$-simplices that contains units $i$ and $j$
\begin{align}
\mathbf{A}_{ij}^{\left(n\right)}= \left(1-\delta_{ij}\right)\sum_{\{i,j\}_{n+1}\subset V}\prod_{\omega\in S^{\{i,j\}}_{n+1}}\mathbf{M}_{\omega},\label{EQ3}
\end{align}
where $\{i,j\}_{n+1}\subset V$ is an arbitrary set of $n+1$ units in which units $i$ and $j$ are included. The general definition of $\mathbf{M}_{\omega}$ used in Eq. (\ref{EQ2}-\ref{EQ3}) is given as a product of a series of matrices for any permutation $\omega=\left(k_{1}\ldots k_{n}\right)$
\begin{align}
\mathbf{M}_{\omega}=\mathbf{A}_{k_{1}k_{2}}\cdots\mathbf{A}_{k_{n-1}k_{n}},\label{EQ4}
\end{align}
where $\mathbf{A}=\mathbf{A}^{\left(1\right)}$ is the classic adjacency matrix of units. It can be seen that $\prod_{\omega\in S^{\{i,j\}}_{n+1}}\mathbf{M}_{\omega}$ serves as an indicator function, which equals $1$ if the units in $\{i,j\}_{n+1}$ form a $n$-simplex and equals $0$ otherwise.

Please see Appendix \ref{ASec2} for the full derivations of the multi-order Laplacian. In Eq. (\ref{EQ2}), vector $\mathbf{D}^{\left(n\right)}$ is similar to the degree vector while matrix $\mathbf{A}^{\left(n\right)}$ can be understood as the adjacency matrix. The coefficient $n$ in Eq. (\ref{EQ2}) arises from the fact that $\sum_{j}\mathbf{A}_{ij}^{\left(n\right)}=n\mathbf{D}^{\left(n\right)}_{i}$ (i.e., each simplex is repeatedly counted $n$ times). In a special case where $n=1$, Eq. (\ref{EQ2}) is equivalent to the classic Laplacian for pairwise interactions. In the more general cases where $n>1$, high-order interactions can be characterized by Eq. (\ref{EQ2}). Please see Fig. \ref{G1}(d) for the illustrations of the multi-order Laplacian. Although the definitions of Eqs. (\ref{EQ2}-\ref{EQ4}) are different from the approach proposed by Ref. \cite{lucas2020multiorder}, they are mathematically consistent with Ref. \cite{lucas2020multiorder}. 

\subsection{High-order interactions along paths}\label{SecII-B}
Then, we turn to another type of $n$-order interactions, which manifest as the sequential actions of $n+1$ related units after they all participate in (i.e., sequential participation). The un-decomposability of these interactions arises from that the mutual coupling effects emerge only after $n+1$ units have been engaged with the interplay progressively. For example, a pipeline-like collaboration may require all agents to communicate with each other and behave in order. Each previous action is globally known to all agents and affects all subsequent actions. This property can be characterized by the paths (i.e., a sequence of $1$-simplices) that successively pass through the $n+1$ units placed on the $n$-simplex. Consequently, to model the $n$-order interactions arisen from sequential participation, we need to analyze both the faces of simplices (i.e., to ensure that all related units are indispensable) and the paths defined on these faces (i.e., to describe the sequential participation).

To describe the interactions propagating along the paths on $n$-simplices, we develop an operator named as the high-order path Laplacian $\mathbf{L}_{H}^{\left(n\right)}$
\begin{align}
\left[\mathbf{L}_{H}^{\left(n\right)}\right]_{ij}=\frac{1}{n}\left(\delta_{ij}\mathbf{P}_{i}^{\left(n\right)}-\mathbf{B}_{ij}^{\left(n\right)}\right).\label{EQ5}
\end{align}
In Eq. (\ref{EQ5}), the coefficient $\frac{1}{n}$ denotes the normalization of the propagation speed of interactions along paths. The term $\mathbf{P}_{i}^{\left(n\right)}$ counts the number of paths that traverse all $n+1$ units on the $n$-simplex without repetition and terminate at unit $i$
\begin{align}
\mathbf{P}_{i}^{\left(n\right)}=n!\sum_{\{i\}_{n+1}\subset V}\prod_{\omega\in 
S^{\{i\}}_{n+1}}\mathbf{M}_{\omega},\label{EQ6}
\end{align}
where $n!$ counts the possibilities for the rest part of units in $\{i\}_{n+1}$ to form unique sequences after we fix unit $v_{i}$ as the endpoint of these sequences. The term $\mathbf{B}_{ij}^{\left(n\right)}$ counts the number of paths that traverse all $n+1$ units on the $n$-simplex without repetition, initiate at unit $j$, and terminate at unit $i$
\begin{align}
\mathbf{B}_{ij}^{\left(n\right)}= \left(1-\delta_{ij}\right)\left(n-1\right)!\sum_{\{i,j\}_{n+1}\subset V}\prod_{\omega\in 
S^{\{i,j\}}_{n+1}}\mathbf{M}_{\omega},\label{EQ7}
\end{align}
where the coefficient $\left(n-1\right)!$ is derived from the fact that the rest part of units in $\{i,j\}_{n+1}$ can form $\left(n-1\right)!$ unique sequences if we fix units $v_{i}$ and $v_{j}$ as the endpoints.

The construction process of the high-order path Laplacian $\mathbf{L}_{H}^{\left(n\right)}$ is elaborated in Appendix \ref{ASec2}. Please see Fig. \ref{G1}(e) for several instances of the high-order path Laplacian. In a special case with $n=1$, Eqs. (\ref{EQ6}-\ref{EQ7}) reduce to the classic degree vector and adjacency matrix, respectively (i.e., we have $\mathbf{P}^{\left(1\right)}=\mathbf{D}^{\left(1\right)}$ and $\mathbf{B}^{\left(1\right)}=\mathbf{A}$). Therefore, Eq. (\ref{EQ5}) denotes the classic Laplacian when interactions are pairwise. In most general cases with $n>1$, the $n$-order interactions formed by sequential participation are not equivalent to those caused by simultaneous participation (see Appendix \ref{ASec2} for details).

In general, one can choose the multi-order Laplacian $\mathbf{L}_{M}^{\left(n\right)}$ and the high-order path Laplacian $\mathbf{L}_{H}^{\left(n\right)}$ according to application demands. As we have suggested above, while $\mathbf{L}_{M}^{\left(n\right)}$ conveys information about the high-order interactions that units are simultaneously engaged with, the definition of $\mathbf{L}_{H}^{\left(n\right)}$ is more applicable to the cases where a high-order interaction is formed by an irreducible sequence of the participation of related units. In Appendix \ref{ASec2}, we explain how these two operators differ from each other in governing the dynamics of system evolution (e.g., with different decay rates on different orders). Specifically, the decay rate of system evolution maintains the same no matter units follow the simultaneous (i.e., $\mathbf{L}_{M}^{\left(n\right)}$) or the sequential (i.e., $\mathbf{L}_{H}^{\left(n\right)}$) participation when interactions are pairwise (i.e., $n=1$). When units exhibit $2$-order interactions, the sequential participation implies slower decays of system evolution compared with the simultaneous participation. When interactions happen on higher orders (i.e., $n>2$), the sequential participation leads to a larger decay rate. These differences highlight the dissimilar physics meanings and characteristics of $\mathbf{L}_{M}^{\left(n\right)}$ and $\mathbf{L}_{H}^{\left(n\right)}$.

Based on the unified mathematical forms presented in Secs. \ref{SecII-A}-\ref{SecII-B}, we have developed efficient code implementations of these two kinds of operators, which can be seen in \cite{aohua2023toolbox}. Meanwhile, in Appendix \ref{ASec3}, we explain the similarities and differences between our considered Laplacian operators and other Laplacian operators used in topology theories (e.g., the combinatorial Laplacian \cite{ribando2022graph} and the Hodge Laplacian \cite{lim2020hodge,schaub2020random} in persistent homology analysis \cite{xia2015multiscale,bramer2018multiscale,wang2020persistent,davies2023persistent,memoli2022persistent,wang2023persistent}), which helps us understand the theoretical significance of designing a renormalization group on $\mathbf{L}_{M}^{\left(n\right)}$ and $\mathbf{L}_{H}^{\left(n\right)}$. For convenience, we uniformly denote them as $\mathbf{L}^{\left(n\right)}$ in our subsequent derivations. One can specify the actual definition of $\mathbf{L}^{\left(n\right)}$ in the application.

\begin{figure*}[!t]
\includegraphics[width=1\columnwidth]{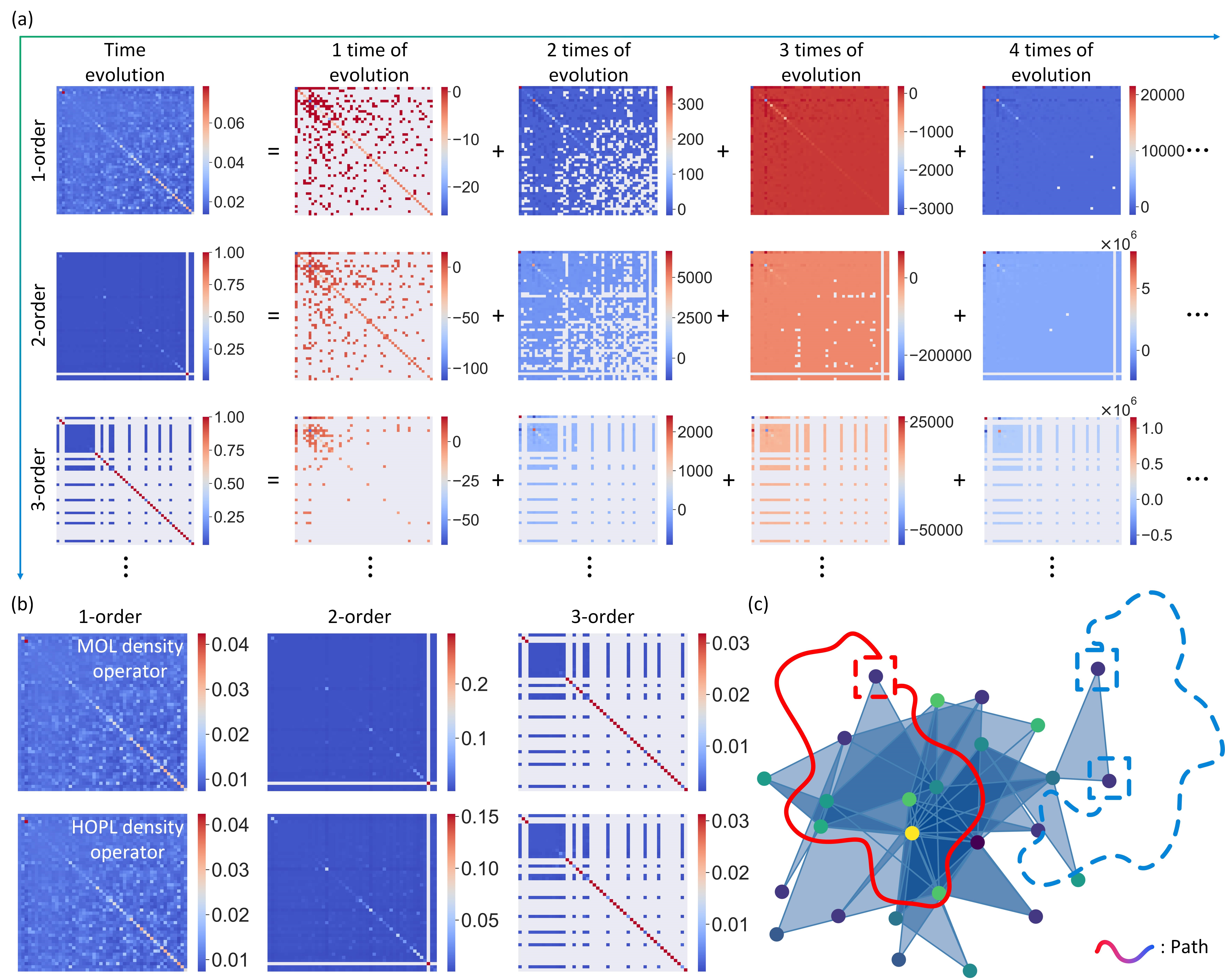}
\caption{\label{G2}Illustrations of the simplex path integral. (a) An interacting system of $50$ units is generated, where the pairwise interactions of units follow a Barab{\'a}si-Albert network ($c=4$). This system is used to show the derivation of the time evolution operators of the first three orders, where we use the multi-order Laplacian (MOL) for calculation and set $\tau=1$. For each order of interactions, we illustrate the first four terms of the series expansion of time evolution operator in Eq. (\ref{EQ9}). (b) For the system defined in (a), we show the density operators calculated based on the MOL or the high-order path Laplacian (HOPL) on each order. (c) We show that values in the density operator can be related to the evolution paths of system states. The blue path leads from one state to another while the red path starts from and ends at the same state.} 
\end{figure*}

\section{Simplex path integral}\label{Sec3}
Given the high-order interactions described by $\mathbf{L}^{\left(n\right)}$, we turn to explore the path integral formulation of the trajectories of units governed by high-order interactions. As first suggested by Ref. \cite{de2016spectral} and then discussed in Refs. \cite{ghavasieh2021unraveling,ghavasieh2021multiscale,villegas2023laplacian,moretti2019network,villegas2022laplacian}, the Laplacian can be understood as a Hamiltonian if we consider the system dynamics governed by it. In the case of high-order interactions, we can consider the eigendecomposition, $\mathbf{L}^{\left(n\right)}=\sum_{\lambda^{\left(n\right)}}\lambda^{\left(n\right)}\big\vert\lambda^{\left(n\right)}\big\rangle\big\langle\lambda^{\left(n\right)}\big\vert$, using the bra-ket notation and define a time evolution operator 
\begin{align}
\mathbf{U}^{\left(n\right)}_{\tau}=&\exp\left(-\tau\mathbf{L}^{\left(n\right)}\right)\label{EQ8}\\=&\sum_{k=0}^{\infty}\frac{1}{k!}\left(-\tau\mathbf{L}^{\left(n\right)}\right)^{k}\label{EQ9},\\=&\sum_{\lambda^{\left(n\right)}}\exp\left(-\tau\lambda^{\left(n\right)}\right)\big\vert\lambda^{\left(n\right)}\big\rangle\big\langle\lambda^{\left(n\right)}\big\vert.\label{EQ10}
\end{align}

As suggested by Eq. (\ref{EQ9}), this operator characterizes the probability for each unit to evolve from one state to another state within a given time scale $\tau$, i.e., $\big\vert\mathbf{x}\big\rangle=\mathbf{U}^{\left(n\right)}_{\tau}\big\vert\mathbf{x}_{0}\big\rangle$. The characterization is realized by considering the distributions of units, $\frac{\left(-\tau\mathbf{L}^{\left(n\right)}\right)^{k}}{k!}$, generated by all the possible ways of evolution, $k\in\left(0,\infty\right)$, during a time period of $\tau$. See Fig. \ref{G2}(a) for illustrations. As indicated by Eq. (\ref{EQ10}), the probability amplitude is intrinsically determined by the exponential decay rates of diffusion modes in a moment space. Short-range interactions associated with fast decays (i.e., large eigenvalues) in the moment space have small impacts on units.

At equilibrium, the Gibbs state of the system associated with $n$-order interactions is 
\begin{align}
\mathbf{\rho}^{\left(n\right)}=\frac{\mathbf{U}^{\left(n\right)}_{\tau}}{\operatorname{tr}\left(\mathbf{U}^{\left(n\right)}_{\tau}\right)}=\frac{\sum_{\lambda^{\left(n\right)}}\exp\left(-\tau\lambda^{\left(n\right)}\right)\big\vert\lambda^{\left(n\right)}\big\rangle\big\langle\lambda^{\left(n\right)}\big\vert}{\sum_{\lambda^{\left(n\right)}}\exp\left(-\tau\lambda^{\left(n\right)}\right)},\label{EQ11}
\end{align}
where $\operatorname{tr}\left(\cdot\right)$ denotes the trace and $\tau$ serves as the inverse of a finite temperature. Please see Fig. \ref{G2}(b) for instances. The Boltzmann constant is assumed as unitary for simplicity. The partition function, $\sum_{\lambda^{\left(n\right)}}\exp\left(-\tau\lambda^{\left(n\right)}\right)$, is proportional to the average return probability of random walks within a time scale $\tau$ \cite{de2016spectral}. See Figs. \ref{G2}(b-c) for illustrations.

Eq. (\ref{EQ11}) defines a density operator of system evolution characterized by high-order interactions
\begin{align}
\big\langle\mathbf{x}\big\vert\mathbf{\rho}^{\left(n\right)}\big\vert\mathbf{x}_{0}\big\rangle=\frac{1}{\operatorname{tr}\left(\mathbf{U}^{\left(n\right)}_{\tau}\right)}\big\langle\mathbf{x}\big\vert\mathbf{U}^{\left(n\right)}_{\tau}\big\vert\mathbf{x}_{0}\big\rangle,\label{EQ12}
\end{align}
which naturally relates to the path integral formulation. Let us consider a minimum time step $\varepsilon$ such that $N=\frac{\tau}{\varepsilon}$. At the limit $N\rightarrow\infty$ (or $\varepsilon\rightarrow 0$), the numerator of the density operator (i.e., the time evolution operator) implies a power form of the corresponding high-order propagator $\mathbf{K}^{\left(n\right)}\left(\mathbf{x},\mathbf{x}_{0},\tau\right)=\big\langle\mathbf{x}\big\vert\mathbf{U}^{\left(n\right)}_{\tau}\big\vert\mathbf{x}_{0}\big\rangle=\lim_{N\rightarrow\infty}\big\langle\mathbf{x}\big\vert\left(\mathbf{U}^{\left(n\right)}_{\varepsilon}\right)^{N}\big\vert\mathbf{x}_{0}\big\rangle$. On the one hand, this power form expression directly leads to a path integral in conventional quantum field theory
\begin{align}
&\mathbf{K}^{\left(n\right)}\left(\mathbf{x},\mathbf{x}_{0},\tau\right)\notag\\
=&\lim_{N\rightarrow\infty}\int\prod_{k=0}^{N-1}\big\langle\mathbf{x}_{\left(k+1\right)\varepsilon}\big\vert\mathbf{U}^{\left(n\right)}_{\varepsilon}\big\vert\mathbf{x}_{k\varepsilon}\big\rangle\mathsf{d}\mathbf{x}_{\varepsilon}\cdots\mathsf{d}\mathbf{x}_{\left(N-1\right)\varepsilon},\label{EQ13}
\end{align}
where we mark $\mathbf{x}=\mathbf{x}_{N\varepsilon}$. On the other hand, this power form expression can also be reformulated using the imaginary time and Eqs. (\ref{EQ8}-\ref{EQ9})
\begin{align}
\mathbf{K}^{\left(n\right)}\left(\mathbf{x},\mathbf{x}_{0},\tau\right)=&\sum_{k=0}^{\infty}\frac{1}{k!}\Bigg\langle\mathbf{x}\Bigg\vert\left(-\frac{i}{\hbar}\int_{0}^{\tau}-i\hbar\mathbf{L}^{\left(n\right)}\mathsf{d}\varepsilon\right)^{k}\Bigg\vert\mathbf{x}_{0}\Bigg\rangle,\label{EQ14}
\end{align}
where $\hbar$ denotes the reduced Planck constant \cite{feynman2010quantum,kleinert2009path}. Eq. (\ref{EQ14}) is closely related to the imaginary time evolution governed by operator $\mathbf{L}^{\left(n\right)}$ in the infinite-dimensional path space. Specifically, the probability amplitude for the system to evolve from $\big\vert\mathbf{x}_{0}\big\rangle$ to $\big\vert\mathbf{x}\big\rangle$ during a period of $\tau$ conveyed by the propagator in Eq. (\ref{EQ13}) can be measured by summing over all possible evolution paths connecting between these two states (i.e., the discrete counterpart of path integrals) in Eq. (\ref{EQ14}). This is the reason why the $n$-order network topology conveyed by $\mathbf{L}^{\left(n\right)}$ intrinsically shapes path integrals. See Fig. \ref{G2}(c) for illustrations.

Moreover, the denominator of the density operator (i.e., the partition function) can be interpreted by considering the expectation $\big\langle\mathbf{U}^{\left(n\right)}_{\tau}\big\rangle=\operatorname{tr}\left(\rho^{\left(n\right)}\mathbf{U}^{\left(n\right)}_{\tau}\right)$
\begin{align}
&\operatorname{tr}\left(\rho^{\left(n\right)}\mathbf{U}^{\left(n\right)}_{\tau}\right)\notag\\=&\int \big\langle\mathbf{x}_{0}\big\vert\mathbf{U}^{\left(n\right)}_{\tau}\big\vert\mathbf{x}_{0}\big\rangle\mathsf{d}\mathbf{x}_{0},\label{EQ15}\\
=&\int\Bigg\langle\mathbf{x}_{0}\Bigg\vert\left(\oint_{\Gamma} \exp\left(-\frac{i}{\hbar}\int_{0}^{\tau}-i\hbar\mathbf{L}^{\left(n\right)}\mathsf{d}\varepsilon\right)\mathsf{d}\mathbf{x}_{\varepsilon}\right)\Bigg\vert\mathbf{x}_{0}\Bigg\rangle\mathsf{d}\mathbf{x}_{0},\label{EQ16}
\end{align}
where $\Gamma$ is an arbitrary closed curve that starts from and ends at $\mathbf{x}_{0}$. Eq. (\ref{EQ16}) suggests why the partition function is proportional to the average return probability of random walks within a time scale $\tau$ from the perspective of path integral formulation. See Fig. \ref{G2}(c) for illustrations.

The path integrals in Eqs. (\ref{EQ13}-\ref{EQ16}) are fully characterized by $\mathbf{L}^{\left(n\right)}$, which describes latent simplex structures of the system. They are integrals over all the possible evolution paths of a system governed by high-order interactions. Consequently, they are referred to as the simplex path integrals in our work. Although our framework is presented using the terminologies of quantum field theory \cite{feynman2010quantum,kleinert2009path} for convenience, it is generally applicable to diverse classical systems (e.g., the brain in neuroscience) in the world. In fact, we can naturally find the Maxwell-Boltzmann statistics, a common description of classical systems, in Eqs. (\ref{EQ11}-\ref{EQ12}). 

\section{Simplex renormalization group}\label{Sec4}
The connection between Eqs. (\ref{EQ8}-\ref{EQ16}) and the moment space naturally inspires us to consider the generalization of renormalization group theories. The freedom degrees corresponding to short-range interactions (e.g., fast diffusion within the clusters of strongly correlated units) can be safely coarse grained without significant information loss, which is consistent with our interpretation of Eq. (\ref{EQ10}). Here we develop a framework to generalize renormalization group to high-order interactions, where we also propose possible ways to overcome the limitations of the Laplacian renormalization group (LRG) \cite{villegas2023laplacian}.

\subsection{Renormalization procedure}\label{Sec-IVA}
To offer a clear vision, we first introduce our approach, referred to as the simplex renormalization group (SRG), in an ergodic case as assumed by Ref. \cite{villegas2023laplacian}. The SRG renormalizes a system on the $p$-order according to the properties of $q$-order interactions ($p\leq q$). Specifically, the renormalization is realized as the following:

\begin{figure*}[!t]
\includegraphics[width=1\columnwidth]{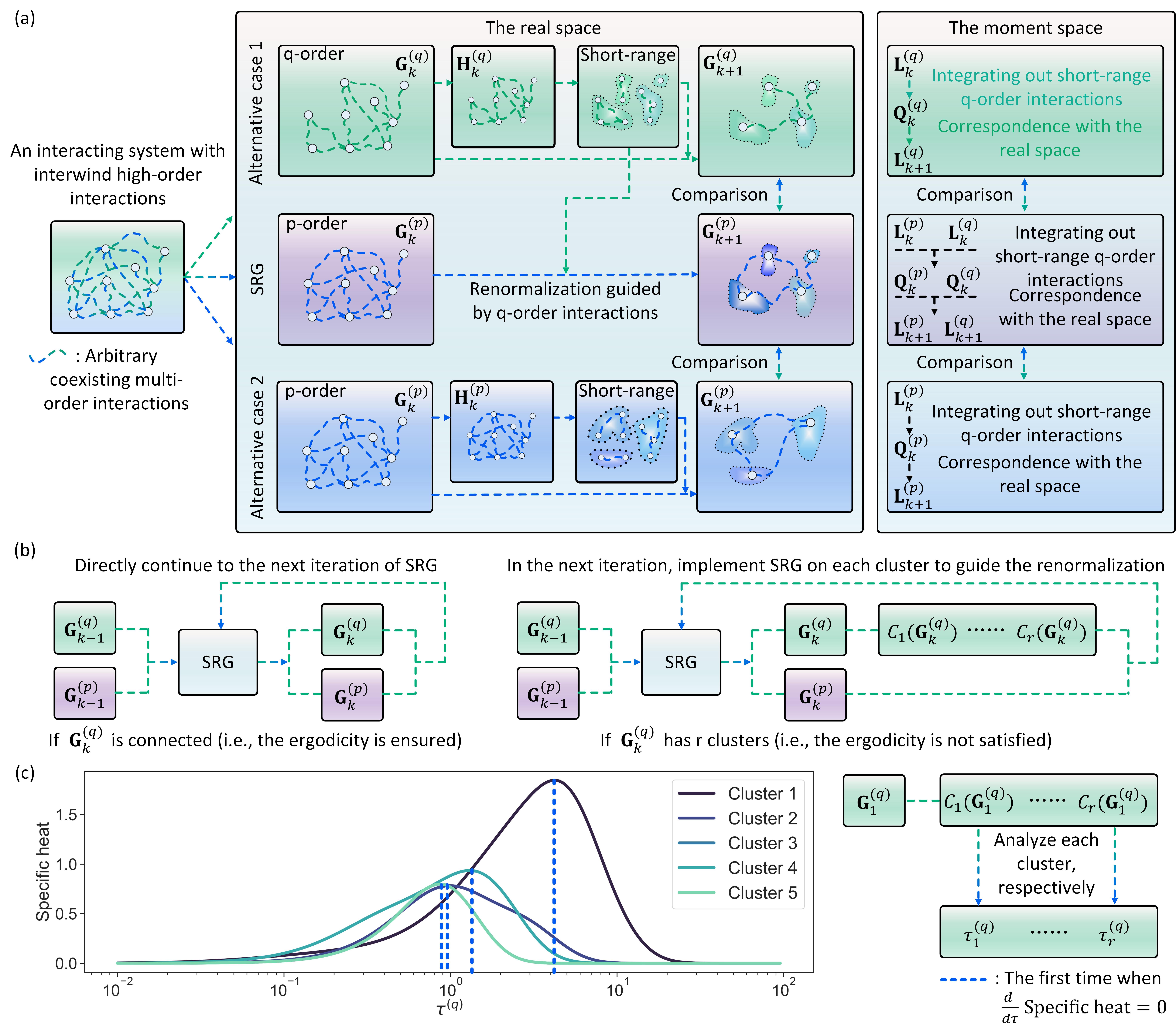}
\caption{\label{G3}Illustrations of the simplex renormalization group (SRG). (a) The conceptual illustration of the SRG is presented. For the sake of legibility, the simplicial complexes of different orders are not elaborated and are abstractly represented by dashed lines with different colors. Given a system where multi-order interactions coexist, the SRG can be applied to renormalize the system on the $p$-order based on the structure and dynamics associated with $q$-order interactions ($p\leq q$). (b) Conceptual illustrations of our divide-and-conquer procedure for dealing with the dependence of renormalization flows on the ergodicity. (c) A system with $500$ units is defined, where pairwise interactions follow a Barab{\'a}si-Albert network ($c=2$). We use this system to illustrate the selection of time scale based on the specific heat ($q=2$). } 
\end{figure*}

\begin{itemize}
    \item[(1) ] In each $k$-th iteration, we have two iterative Laplacian operators, $\mathbf{L}^{\left(p\right)}_{k}$ and $\mathbf{L}^{\left(q\right)}_{k}$, and their associated high-order network sketches, $\mathbf{G}^{\left(p\right)}_{k}$ and $\mathbf{G}^{\left(q\right)}_{k}$, with $N^{\left(p\right)}_{k}=N^{\left(q\right)}_{k}$ units (there is an edge connecting between two units in $\mathbf{G}^{\left(p\right)}_{k}$ or $\mathbf{G}^{\left(q\right)}_{k}$ only if they share a $p$- or $q$-order interaction). 
    \item[(2) ] In the real space, we first search targets to coarse grain. Specifically, we search for every cluster of units sharing short-range high-order interactions within the time scale $\widehat{\tau}^{\left(q\right)}$, such that we can coarse grain these units into a macro-unit. To this end, we generate a reference network, $\mathbf{H}^{\left(q\right)}_{k}$, in the following way. We first initialize $\mathbf{H}^{\left(q\right)}_{k}$ as a null network (i.e., with no edge). We follow Eq. (\ref{EQ11}) to calculate the density operator associated with time scale $\widehat{\tau}^{\left(q\right)}$ as $\widehat{\rho}^{\left(q\right)}_{k}=\exp\left(-\widehat{\tau}^{\left(q\right)}\mathbf{L}^{\left(q\right)}_{k}\right)/\operatorname{tr}\left(\exp\left(-\widehat{\tau}^{\left(q\right)}\mathbf{L}^{\left(q\right)}_{k}\right)\right)$. In operator $\widehat{\rho}^{\left(q\right)}_{k}$, its main diagonal elements, related to self-interactions of units, are frequently greater than most of its off-diagonal elements (e.g., see Fig. \ref{G2}(b)). If the $\left(i,j\right)$-th element is greater than the $\left(i,i\right)$-th or the $\left(j,j\right)$-th element, then the $q$-order interactions between units $v_{i}$ and $v_{j}$ are sufficiently quick such that their accumulations within time scale $\widehat{\tau}^{\left(q\right)}$ are faster than those of self-interactions. In this case, we add an edge between units $v_{i}$ and $v_{j}$ in network $\mathbf{H}^{\left(q\right)}_{k}$. By progressively implement the edge adding procedure for every pair of units, we can form $n_{k}^{\left(q\right)}$ clusters in $\mathbf{H}^{\left(q\right)}_{k}$. Each cluster contains a set of strongly correlated units with short-range high-order interactions. In other words, these clusters can be understood as the Kadanoff blocks (i.e., the quasi-independent blocks whose mean size serves as the correlation length of the system) in the real space renormalization procedure \cite{wilson1971renormalization,burkhardt2012real,efrati2014real}.
    \item[(3) ] After we find the targets for coarse graining in the real space, we implement a reormalization procedure. Specifically, in $\mathbf{G}^{\left(q\right)}_{k}$, each set of units that belong to the same cluster in $\mathbf{H}^{\left(q\right)}_{k}$ are aggregated into a macro-unit to generate a new network $\mathbf{G}^{\left(q\right)}_{k+1}$. Consequently, there remain $n^{\left(q\right)}_{k}$ macro-units in $\mathbf{G}^{\left(q\right)}_{k+1}$ after coarse graining. In $\mathbf{G}^{\left(q\right)}_{k+1}$, two macro-units, $v_{i}$ and $v_{j}$, are connected if at least one unit aggregated into $v_{i}$ is connected with one unit aggregated into $v_{j}$ in $\mathbf{G}^{\left(q\right)}_{k}$. If a unit or an interaction in $\mathbf{G}^{\left(q\right)}_{k}$ does not participate in any coarse graining according to $\mathbf{H}^{\left(q\right)}_{k}$, it is adopted into $\mathbf{G}^{\left(q\right)}_{k+1}$ without modification. Parallel to this process, we also follow the same rules to aggregate the units in $\mathbf{G}^{\left(p\right)}_{k}$ that belong to the same cluster of $\mathbf{H}^{\left(q\right)}_{k}$ into a macro-unit in $\mathbf{G}^{\left(p\right)}_{k+1}$. The generated $\mathbf{G}^{\left(p\right)}_{k+1}$ contains $n^{\left(q\right)}_{k}$ macro-units as well. Therefore, we define $N^{\left(p\right)}_{k+1}=N^{\left(q\right)}_{k+1}=n^{\left(q\right)}_{k}$. 
\item[(4) ] In the moment space, we first search modes to reduce. Specifically, we find the eigenvalues of $\mathbf{L}^{\left(q\right)}_{k}$ that are smaller than $\frac{1}{\widehat{\tau}^{\left(q\right)}}$, where $\widehat{\tau}^{\left(q\right)}$ is the time scale for $q$-order simplicial complexes (see Sec. \ref{Sec-IVB} for the selection of $\widehat{\tau}^{\left(q\right)}$). These eigenvalues correspond to long-range high-order interactions because the contributions of their eigenvectors have slow decays in Eq. (\ref{EQ10}). We denote the number of these eigenvalues as $s_{k}^{\left(q\right)}$. Other eigenvalues correspond to short-range high-order interactions and can be coarse grained. Moreover, according to Eq. (\ref{EQ11}), the totality of long-range high-order interactions (i.e., corresponding to smaller eigenvalues) is generally greater than that of short-range ones (i.e., associated with larger eigenvalues). Therefore, integrating out short-range high-order interactions does not damage the effective information contained in the system significantly. The only problem is that the eigenvalue spectrum cutoff defined above may overestimate the number of short-range high-order interactions and make $s_{k}^{\left(q\right)}$ too small. This is because the eigenvalue spectrum of $\mathbf{L}^{\left(q\right)}_{k}$ does not reflect a possibility that the short-range $q$-order interactions between units $v_{i}$ and $v_{j}$ can be slower than their self-interactions. For instance, although the interactions between units $v_{i}$ and $v_{j}$ are classified as short-range, the $\left(i,j\right)$-th element of $\widehat{\rho}^{\left(q\right)}_{k}$ can be smaller than both the $\left(i,i\right)$-th and the $\left(j,j\right)$-th elements. In this case, the time scale $\widehat{\tau}^{\left(q\right)}$ may be too short for the correlations between $v_{i}$ and $v_{j}$ to accumulate to a non-negligible level. Compared with self-interactions, the interactions between units $v_{i}$ and $v_{j}$ are less important in determining the behaviours of $v_{i}$ and $v_{j}$ since these interactions are not strong enough. To avoid the possible overestimation, we use $n_{k}^{\left(q\right)}$, the number of Kadanoff blocks in $\mathbf{H}^{\left(q\right)}_{k}$, as a bound. In the case with overestimation, the number of long-range interactions is underestimated (i.e., $s_{k}^{\left(q\right)}\leq n_{k}^{\left(q\right)}$) and we correct it by defining $m_{k}^{\left(q\right)}=\max\{s_{k}^{\left(q\right)},n_{k}^{\left(q\right)}\}$. The value of $m_{k}^{\left(q\right)}$ defines the number of modes to keep during renormalization.
\item[(5) ] After we find the modes to reduce in the moment space, we exclude these modes with fast decays from the system such that there remain $m_{k}^{\left(q\right)}$ modes. Specifically, the current $q$-order Laplacian, $\mathbf{L}^{\left(q\right)}_{k}$, is reduced to the re-scaled contributions of the eigenvectors associated with long-range high-order interactions 
\begin{align}
\mathbf{Q}^{\left(q\right)}_{k}=\sum_{\lambda^{\left(q\right)}_{k}}I_{ \Lambda^{\left(q\right)}_{k}}\left(\lambda^{\left(q\right)}_{k}\right)\lambda^{\left(q\right)}_{k}\big\vert\lambda^{\left(q\right)}_{k}\big\rangle\big\langle\lambda^{\left(q\right)}_{k}\big\vert,\label{EQ17}
\end{align}
where $\Lambda^{\left(q\right)}_{k}$ is the set of the $m_{k}^{\left(q\right)}$ smallest eigenvalues of $\mathbf{L}^{\left(q\right)}_{k}$ (i.e., the first $m_{k}^{\left(q\right)}$ eigenvalues after we sort the eigenvalue spectrum in a decreasing order) and $I_{A}\left(\cdot\right)$ is the indicator function defined on an arbitrary set $A$. Meanwhile, the current $p$-order Laplacian, $\mathbf{L}^{\left(p\right)}_{k}$, is reduced to the contributions of the $m_{k}^{\left(q\right)}$ smallest eigenvalues (i.e., long-range $p$-order interactions)
\begin{align}
\mathbf{Q}^{\left(p\right)}_{k}=\sum_{\omega^{\left(p\right)}_{k}}I_{ \Omega^{\left(p\right)}_{k}}\left(\omega^{\left(p\right)}_{k}\right)\omega^{\left(p\right)}_{k}\big\vert\omega^{\left(p\right)}_{k}\big\rangle\big\langle\omega^{\left(p\right)}_{k}\big\vert,\label{EQ18}
\end{align}
where $\Omega^{\left(p\right)}_{k}$ is the set of the $m_{k}^{\left(q\right)}$ smallest eigenvalues of $\mathbf{L}^{\left(p\right)}_{k}$. Note that the eigenvalues in $\Omega^{\left(p\right)}_{k}$ are not necessarily smaller than $\frac{1}{\widehat{\tau}^{\left(q\right)}}$.

\item[(6) ] The correspondence between moment space and real space for $q$-order interactions is ensured by a Laplacian $\mathbf{L}^{\left(q\right)}_{k+1}$. For $v_{i}$ and $v_{j}$, two macro-units in $\mathbf{G}^{\left(q\right)}_{k+1}$, we define  
\begin{align}
\left[\mathbf{L}^{\left(q\right)}_{k+1}\right]_{v_{i}v_{j}}=\big\langle v_{i}\big\vert\mathbf{Q}^{\left(q\right)}_{k}\big\vert v_{j}\big\rangle,\label{EQ19}
\end{align}
if $v_{i}$ and $v_{j}$ are connected in $\mathbf{G}^{\left(q\right)}_{k+1}$. We set $\left[\mathbf{L}^{\left(q\right)}_{k+1}\right]_{v_{i}v_{j}}=0$ if $v_{i}$ and $v_{j}$ are disconnected in $\mathbf{G}^{\left(q\right)}_{k+1}$. Here $\big\vert v_{j}\big\rangle$ is a $N^{\left(q\right)}_{k}$-dimensional ket where unitary components correspond to all the units in $\mathbf{G}^{\left(q\right)}_{k}$ that are aggregated into $v_{i}$ and zero components correspond to all other units in $\mathbf{G}^{\left(q\right)}_{k+1}$ that are not covered by $\nu_{i}$ (i.e., the value of $\left[\mathbf{L}^{\left(q\right)}_{k+1}\right]_{v_{i}v_{j}}$ is the sum of all $\left[\mathbf{Q}^{\left(q\right)}_{k}\right]_{xy}$ if $v_{i}$ and $v_{j}$ are connected, where $x$ goes through all the units in $\mathbf{G}^{\left(q\right)}_{k}$ that are aggregated into $v_{i}$ and $y$ goes through all units in $\mathbf{G}^{\left(q\right)}_{k}$ that are aggregated into $v_{j}$). After defining all off-diagonal elements, the diagonal elements of $\mathbf{L}^{\left(q\right)}_{k+1}$ are defined following 
\begin{align}
\left[\mathbf{L}^{\left(q\right)}_{k+1}\right]_{v_{i}v_{i}}=-\sum_{v_{j}}\left[\mathbf{L}^{\left(q\right)}_{k+1}\right]_{v_{i}v_{j}}.\label{EQ20}
\end{align}
Mathematically, this procedure actually defines a similarity transformation $T$ between $\mathbf{Q}^{\left(q\right)}_{k}$ and $\mathbf{L}^{\left(q\right)}_{k+1}$
\begin{align}
T^{-1}\mathbf{Q}^{\left(q\right)}_{k}T=\operatorname{diag}\left(\left[\mathbf{L}^{\left(q\right)}_{k+1},\mathbf{0}\right]\right),\label{EQ21}
\end{align}
where $\mathbf{0}$ is a $\left(N^{\left(q\right)}_{k}-n^{\left(q\right)}_{k}\right)$-dimensional all-zero square matrix. It is clear that the similarity transformation is given as $T=\left(\big\vert v_{1}\big\rangle,\ldots, \big\vert v_{n^{\left(q\right)}_{k}}\big\rangle, \big\vert \eta_{n^{\left(q\right)}_{k}+1}\big\rangle,\ldots, \big\vert \eta_{N^{\left(q\right)}_{k}}\big\rangle\right)$, where each ket $\eta_{j}$ is selected such that the columns of $T$ define a group of orthonormal bases. Similarly, we can derive the Laplacian $\mathbf{L}^{\left(p\right)}_{k+1}$ for $p$-order interactions based on $\mathbf{Q}^{\left(p\right)}_{k}$. The derived $\mathbf{L}^{\left(p\right)}_{k+1}$ and $\mathbf{L}^{\left(q\right)}_{k+1}$ are used in the $\left(k+1\right)$-th iteration.
\end{itemize} 

The renormalization of $p$-order interactions in both the real space (i.e., see steps (2-3)) and the moment space (i.e., see steps (4-5)) is realized under the guidance of $q$-order interactions. After renormalizing $\mathbf{L}^{\left(p\right)}_{k}$ by repeating steps (1-6), the SRG progressively drives the system towards an intrinsic scale of $p$-order interactions that exceeds the microscopic scales. More specifically, the SRG creates a $q$-order-guided iterative scale transformation that, irrespective of starting from which local scale, always leads to the latent critical point associated with $p$-order interactions (if there is any). It provides opportunities to verify whether the concerned thermodynamic functions are scale-invariant or not. See Fig. \ref{G3}(a) for a summary of key steps in our approach. Note that the using of a reference network in step (2) is not necessary for theoretical derivations, yet it is favorable for computer programming.

In a non-ergodic case, the presented procedure can not be directly adopted. This is because our proposed renormalization approach is rooted in the relations among the time evolution operators defined by Eqs. (\ref{EQ8}-\ref{EQ10}), the exponential decay of diffusion modes in the moment space, and the evolution paths in the path space, which implicitly depends on the assumption of the ergodicity of system states. In the non-ergodic case where a set of states can never evolve to or be transformed from another set of states (i.e., the network connectivity is absent), path integrals between these state sets are ill-posed and the diffusion over the network can be subdivided into several irrelevant sub-processes (i.e., the diffusion processes defined on two disconnected clusters are independent from each other). Dealing with these properties is crucial for our work since the SRG is oriented toward analyzing high-order interactions, which are usually sparsely distributed and may lack ergodicity.

\begin{figure*}[!t]
\includegraphics[width=1\columnwidth]{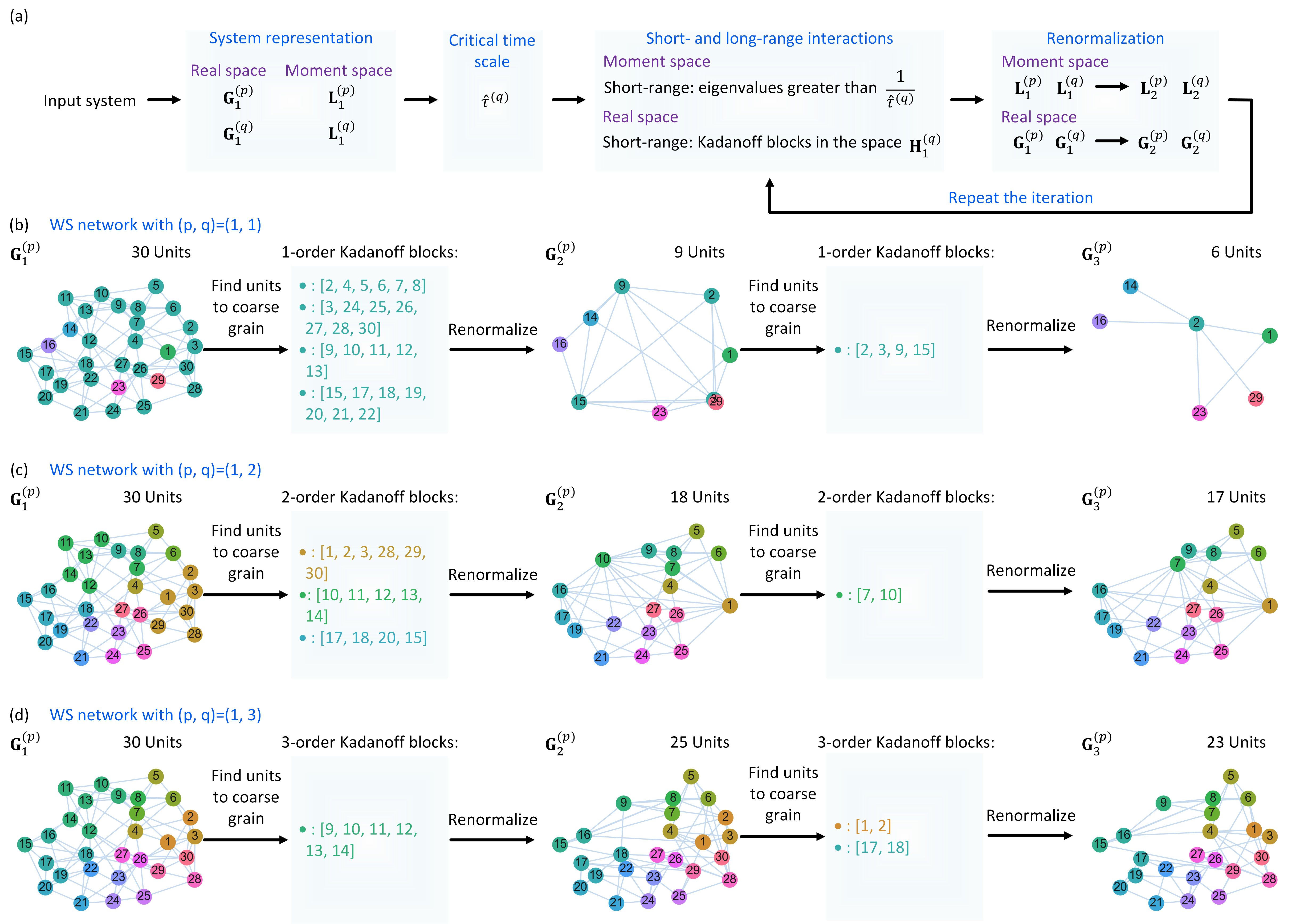}
\caption{\label{G4} Illustrations of the iteration process and coarse graining procedure in the simplex renormalization group (SRG). (a) summarizes key processing pipelines of the SRG proposed in Secs. \ref{Sec-IVA}-\ref{Sec-IVB}. Specifically, given an input system and a selected combination of $\left(p,q\right)$, we follow step (1) to characterize the system on the $p$- and $q$-orders to generate its representations in the real (i.e., $\mathbf{G}^{\left(p\right)}_{1}$ and $\mathbf{G}^{\left(q\right)}_{1}$) and the moment (i.e., $\mathbf{L}^{\left(p\right)}_{1}$ and $\mathbf{L}^{\left(q\right)}_{1}$) spaces. Then, we follow Sec. \ref{Sec-IVB} to calculate the time scale $\widehat{\tau}^{\left(q\right)}$. If the ergodicity does not holds in $\mathbf{G}^{\left(q\right)}_{1}$, we repeat the time scale calculation on each cluster of $\mathbf{G}^{\left(q\right)}_{1}$. After obtaining $\widehat{\tau}^{\left(q\right)}$, we follow steps (2-3) to find the Kadanoff blocks and implement coarse graining in the real space for generating $\mathbf{G}^{\left(p\right)}_{2}$ and $\mathbf{G}^{\left(q\right)}_{2}$. We also follow steps (4-5) to find the modes corresponding to short-range high-order interactions and reduce the system in the moment space into the contributions of long-range high-order interactions. Finally, we calculate $\mathbf{L}^{\left(p\right)}_{2}$ and $\mathbf{L}^{\left(q\right)}_{2}$ according to the correspondence between moment space and real space. The generated $\mathbf{G}^{\left(p\right)}_{2}$, $\mathbf{G}^{\left(q\right)}_{2}$, $\mathbf{L}^{\left(p\right)}_{2}$, and $\mathbf{L}^{\left(q\right)}_{2}$ are used for the next iteration of the SRG. (b-d) use a system whose pairwise interactions follow the Watts-Strogatz network to show the real space renormalization flows on $1$-order when the SRG is guided by $q\in\{1,2,3\}$, respectively. In these instances, the SRG runs two iterations and we assign each macro-unit in $G^{\left(q\right)}_{3}$ with a unique color. If a unit is aggregated into a certain macro-unit, it adopts the color from this macro-unit. By recursively repeat this procedure, we can assign units in each iteration with appropriate colors to indicate which macro-unit in $G^{\left(q\right)}_{3}$ contains them. Meanwhile, to highlight the effects of different values of $q$ on renormalization flows, we assign each unit in $G^{\left(q\right)}_{1}$ with an index to trace the units grouped into the Kadanoff blocks in the step (2) of the SRG framework. For instance (see (c)), the units $1$, $2$, $3$, $28$, $29$, and $30$ in $G^{\left(1\right)}_{1}$ are aggregated into a macro-unit when $q=2$, whose index is assigned as $1$ in $G^{\left(1\right)}_{2}$. Meanwhile, the units $10$, $11$, $12$, $13$, and $14$ in $G^{\left(1\right)}_{1}$ are coarse grained into a macro-unit in $G^{\left(1\right)}_{2}$, which is indexed as $10$. Moreover, the units $17$, $18$, $20$, and $15$ in $G^{\left(1\right)}_{1}$ are grouped into the macro-unit $17$ in $G^{\left(1\right)}_{2}$. Therefore, the system size is reduced from $30$ to $18$ after this time of renormalization.} 
\end{figure*}

Here we develop a divide-and-conquer approach. Our idea arises from the fact that we can treat the diffusion from one cluster (i.e., a connected component) to another cluster on any order as being decelerated to a condition with zero velocity (i.e., being infinitely slow). This infinitely long-term diffusion process contains no short-range high-order interactions and, therefore, should not be integrated out during renormalization. Consequently, the SRG does not need to deal with the diffusion across any pair of clusters and can focus on the diffusion within each cluster severally. In the $k$-iteration, we consider the following two cases:
\begin{itemize}
    \item[(A) ] If $\mathbf{G}^{\left(q\right)}_{k}$ is connected (i.e., the ergodicity holds), we directly apply steps (1-6) on $\mathbf{G}^{\left(q\right)}_{k}$ to guide the renormalization of $p$-order interactions and derive the inputs of the $\left(k+1\right)$-iteration. 
     \item[(B) ] If $\mathbf{G}^{\left(q\right)}_{k}$ is disconnected (i.e., the ergodicity does not hold) and has $r$ clusters denoted by $C_{1},\;\ldots,C_{r}$, we respectively apply steps (1-6) on each cluster of $\mathbf{G}^{\left(q\right)}_{k}$ to guide the renormalization of $p$-order interactions formed among the units that belong to this cluster. Specifically, we treat each cluster $C_{i}$ as a network and input it into steps (1-6) with $\mathbf{G}^{\left(p\right)}_{k}$. Please note that the above procedure does not affect $\mathbf{G}^{\left(p\right)}_{k}$ if the selected cluster $C_{i}$ is trivial (i.e., contains only one unit). After dealing with all $r$ clusters, the obtained results are used for the $\left(k+1\right)$-iteration. 
\end{itemize}
In Fig. \ref{G3}(b), we conceptually illustrate our approach, which enables us to relax the constraint on system ergodicity while implementing the SRG.

In Appendix \ref{ASec3}, we explain how the SRG is related to the persistent homology theory \cite{wang2020persistent,edelsbrunner2008persistent,aktas2019persistence,horak2009persistent} and suggest its potential applicability in analyzing the lifetime of topological properties. In Appendix \ref{ASec4}, we discuss the SRG in a form of conventional renormalization group, suggesting the benefits of including high-order interactions into renormalization group analysis and indicate the relations between the SRG and other frameworks \cite{bradde2017pca,kopietz2010introduction,polonyi2003lectures,dupuis2021nonperturbative}. Specifically, we find that the ideas underlying the SRG can be clearly interpreted by relating the SRG with a PCA-like renormalization group theory \cite{bradde2017pca}.

\subsection{Critical time scale}\label{Sec-IVB}
After developing the renormalization procedure, we attempt to present a detailed analysis of the time scale $\tau^{\left(q\right)}$ and its associated scaling relation.

In the ergodic case, the setting of $\tau^{\left(q\right)}$ in the SRG (as we have mentioned in step (1)) is finished in $1$-st iteration and required to make the specific heat, an indicator
of the transitions of diffusion processes, maximize (see similar ideas in Ref. \cite{villegas2023laplacian}). In Appendix \ref{ASec4}, we explain why it is valid to define renormalization flows based on the specific heat. Specifically, we consider the $q$-order spectral entropy in the $1$-st iteration of the SRG (see Ref. \cite{de2016spectral} for the definition of spectral entropy)
\begin{align}
&S^{\left(q\right)}_{1}\left(\tau^{\left(q\right)}\right)\notag\\=&-\operatorname{tr}\left[\rho^{\left(q\right)}_{1}\log \left(\rho^{\left(q\right)}_{1}\right)\right],\label{EQ22}\\
=&\log \left(\sum_{\lambda^{\left(q\right)}_{1}}\exp\left(-\tau^{\left(q\right)}\lambda^{\left(q\right)}_{1}\right)\right)+\tau^{\left(q\right)}\big\langle\lambda^{\left(q\right)}_{1}\big\rangle_{\rho},\label{EQ23}
\end{align}
where each $\lambda^{\left(q\right)}_{1}$ is an eigenvalue of $\mathbf{L}^{\left(q\right)}_{1}$ and $\rho^{\left(q\right)}_{1}$ is the density operator derived from $\mathbf{L}^{\left(q\right)}_{1}$ following Eq. (\ref{EQ11}). Note that we have $\big\langle\lambda^{\left(q\right)}_{1}\big\rangle_{\rho}=\operatorname{tr}\left(\mathbf{L}^{\left(q\right)}_{1}\rho^{\left(q\right)}_{1}\right)$. The $q$-order specific heat is defined according to the first derivative of the $q$-order spectral entropy
\begin{align}
&X^{\left(q\right)}_{1}\left(\tau^{\left(q\right)}\right)\notag\\=&-\frac{\mathsf{d}}{\mathsf{d}\log\left(\tau^{\left(q\right)}\right)}S^{\left(q\right)}_{1}\left(\tau^{\left(q\right)}\right),\label{EQ24}\\
=&-\left(\tau^{\left(q\right)}\right)^{2}\frac{\mathsf{d}}{\mathsf{d}\tau^{\left(q\right)}}\big\langle\lambda^{\left(q\right)}_{1}\big\rangle_{\rho}.\label{EQ25}
\end{align}
Please see Appendix \ref{ASec5} for detailed derivations. The value of $\tau^{\left(q\right)}$ is selected to ensure $\frac{\mathsf{d}}{\mathsf{d}\tau^{\left(q\right)}}X^{\left(q\right)}_{1}\left(\tau^{\left(q\right)}\right)=0$, which indicates an infinite deceleration condition of diffusion processes. See Fig. \ref{G3}(c) for illustrations. Apart from the infinite deceleration condition, determining the time scale value $\widehat{\tau}^{\left(q\right)}$ used in the step (2) of the SRG also requires us to consider the high-order properties of diffusion. For convenience, we denote $\widetilde{\tau}^{\left(q\right)}$ as the value of $\tau^{\left(q\right)}$ under the infinite deceleration condition (i.e., there is $\frac{\mathsf{d}}{\mathsf{d}\tau^{\left(q\right)}}X^{\left(q\right)}_{1}\left(\tau^{\left(q\right)}\right)\big\vert_{\widetilde{\tau}^{\left(q\right)}}=0$). The optimal time scale for the SRG is set as
\begin{align}
\widehat{\tau}^{\left(q\right)}=\frac{2\varepsilon}{q+1}\widetilde{\tau}^{\left(q\right)},\label{EQ26}
\end{align}
which is the average subdivided time scale for each $1$-simplex in a $n$-simplex (i.e., the average number of adjacent $1$-simplices between arbitrary units $i$ and $j$ is $\frac{n+1}{2}$). Note that $\widehat{\tau}^{\left(q\right)}=\widetilde{\tau}^{\left(q\right)}$ holds for pairwise interactions. The value of $\varepsilon$ is used to correct the errors of time scale calculation when system units share ultra-dense interactions, which arise from the numerical bias of diffusion rate estimation in ultra-dense systems. The definition of $\varepsilon$ can be seen in Appendix \ref{ASec6}.

The above procedure can be directly generalized to the non-ergodic case, where we treat each cluster $C_{i}$ as a network and derive a $\widehat{\tau}^{\left(q\right)}$ for it if it has not been assigned with a time scale yet. The time scale of a cluster does not change after it is assigned. See Fig. \ref{G3}(c) for illustrations.

\subsection{Renormalization flow}\label{Sec-IVC}

After determining the time scale, the SRG proposed in Sec. \ref{Sec-IVA} can be applied to renormalize interacting systems in a non-parametric manner. Please see Appendix \ref{ASec13} for the programmatic implementation of the SRG.

In Fig. \ref{G4}(a), we explain how the SRG iterates. For an input system, we define $p$ and $q$ to realize the renormalization guided by different orders of interactions. Given each pair of $\left(p,q\right)$, we first search $p$- and $q$-simplices to represent the system on the $p$- and $q$-order (i.e., generate the high-order network sketches, $\mathbf{G}^{\left(p\right)}_{1}$ and $\mathbf{G}^{\left(q\right)}_{1}$, as shown in Fig. \ref{G1}). Meanwhile, we derive the critical time scale, $\widehat{\tau}^{\left(q\right)}$, on $\mathbf{G}^{\left(q\right)}_{1}$ following Sec. \ref{Sec-IVB}. After the critical time scale is calculated, we iteratively realize the steps (1-6) introduced in Sec. \ref{Sec-IVA} to renormalize the system once, twice, and so on. For instance, in the first iteration of the SRG, we follow steps (2-3) to find the Kadanoff blocks on the $q$-order (i.e., clusters in $\mathbf{H}^{\left(q\right)}_{1}$) and coarse grain the units in each Kadanoff block into a macro-unit on the $p$ and $q$-order to derive $\mathbf{G}^{\left(p\right)}_{2}$ and $\mathbf{G}^{\left(q\right)}_{2}$. Then, we follow steps (4-6) to calculate the associated high-order Laplacian operators $\mathbf{L}^{\left(p\right)}_{2}$ and $\mathbf{L}^{\left(q\right)}_{2}$. The obtained $\mathbf{G}^{\left(p\right)}_{2}$, $\mathbf{G}^{\left(q\right)}_{2}$, $\mathbf{L}^{\left(p\right)}_{2}$, and $\mathbf{L}^{\left(q\right)}_{2}$ serve as inputs for the SRG to run the second iteration to generate $\mathbf{G}^{\left(p\right)}_{3}$, $\mathbf{G}^{\left(q\right)}_{3}$, $\mathbf{L}^{\left(p\right)}_{3}$, and $\mathbf{L}^{\left(q\right)}_{3}$. In this manner, the iteration continues and a renormalization flow is gradually generated.

\begin{figure*}[!t]
\includegraphics[width=1\columnwidth]{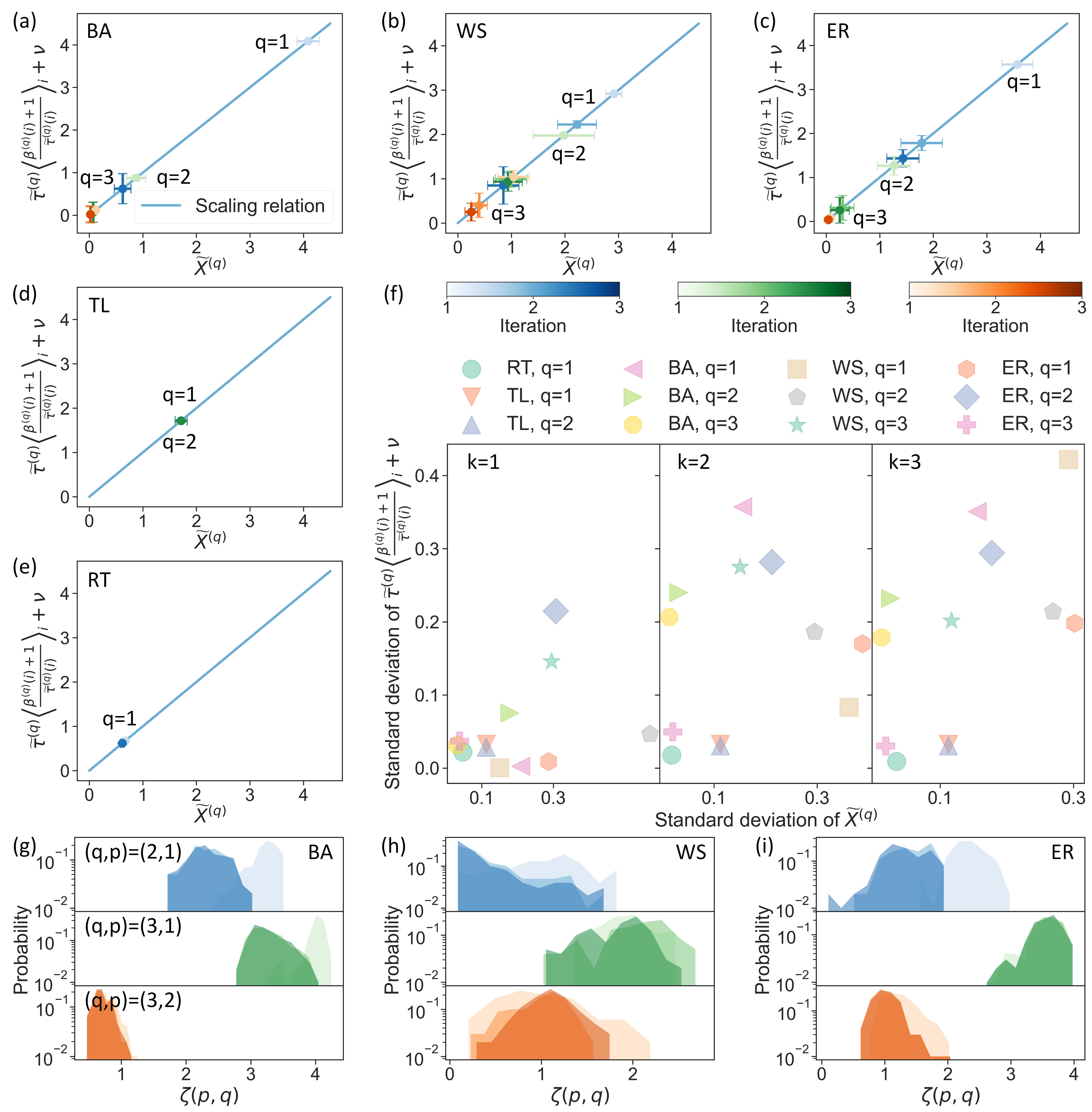}
\caption{\label{G5}Large-scale instances of the renormalization flows of the simplex renormalization group (SRG). The first three iterations of renormalization on the systems whose pairwise interactions follow the random tree (a), the square lattice (b), the square lattice with a periodic boundary condition (c), the triangular lattice (d), the triangular lattice with a periodic boundary condition (e), the Watts-Strogatz networks (f-h), and the Erdos-Renyi networks (i-j) are presented, where we set $\left(p,q\right)=\left(1,1\right)$, $\left(p,q\right)=\left(1,2\right)$ and $\left(p,q\right)=\left(1,3\right)$ for (a-c, f, i), (d-e, g, j) and (h), respectively. Please see Appendix \ref{ASec1} for the definitions of these random network models. For convenience, the periodic boundary condition is abbreviated to the PBC. The absence of the PBC implies that the definitions of different lattices (i.e., square and triangular) only hold within system boundaries. } 
\end{figure*}

In Figs. \ref{G4}(b-d), we mainly illustrate how the real space coarse graining in the SRG happens on different orders. We define $p=1$ and $q\in\{1,2,3\}$ to generate renormalization flows and visualize the changes caused by coarse graining the units within each Kadanoff block. Two important messages are conveyed by Fig. \ref{G4}. First, whenever a Kadanoff block of $m$ units is find in $\mathbf{H}^{\left(q\right)}_{k}$, the size of $\mathbf{G}^{\left(p\right)}_{k+1}$ is reduced by $m-1$ (this property also holds for $\mathbf{G}^{\left(q\right)}_{k+1}$). If there is no Kadanoff block in $\mathbf{H}^{\left(q\right)}_{k}$ (i.e., no unit to coarse grain), the SRG arrives at its fixed point and the system remains unchanged during subsequent renormalization. Second, as shown in the comparison across different values of $q$, the system reduction during renormalization is always slighter given a larger value of $q$, which mainly arises from the sparser distribution of high-order interactions. 

To demonstrate the meaning of distinguishing between $p$ and $q$ during renormalization, we compare between the cases with $p\neq q$ (i.e., renormalize the system on an order according to its properties on another order) and $p=q$ (i.e., equivalent to the situation where we do not distinguish between $p$ and $q$) in Appendix \ref{ASec7}. In Fig. \ref{AG2}, we observe significant differences in both real space (i.e., $\mathbf{G}^{\left(p\right)}_{k}$) and moment space (i.e., $\mathbf{L}^{\left(p\right)}_{k}$) between the cases of $p=q$ and $p\neq q$, demonstrating the effectiveness of supporting $p\neq q$ during renormalization. Therefore, by distinguishing between $p$ and $q$, the SRG is enabled to analyze more phenomena than classic frameworks with $p\equiv q$. Meanwhile, these results qualitatively suggest the effects of higher-order interactions on lower-order interactions during renormalization (i.e., in the case with $p\neq q$), whose quantification is explored in our subsequent analysis.

\begin{figure*}[!t]
\includegraphics[width=1\columnwidth]{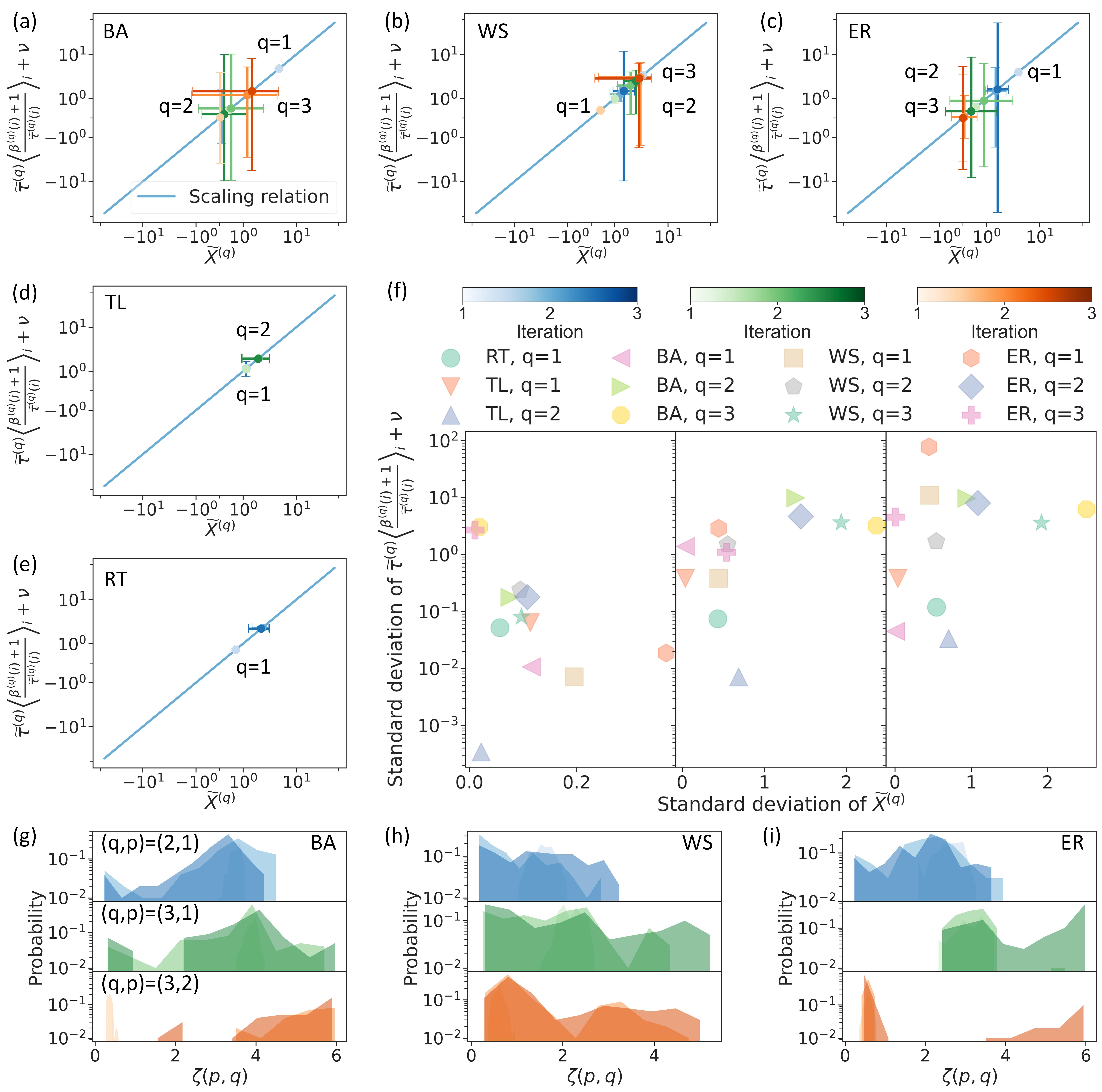}
\caption{\label{G6}Scaling relation of the simplex renormalization group. (a-e) We verify the scaling relation on the systems whose pairwise interactions follow a Barab{\'a}si-Albert network (BA, $c=3$), a Watts-Strogatz network (WS, each unit initially has $8$ neighbors and edges are rewired according to a probability of $0.3$), and an Erdos-Renyi network (ER, two units share an edge with a probability of $0.08$), the triangular lattice (TL, with the interactions of the first two orders), and the random tree (RT, with only pairwise interactions), respectively. Scatters denote the mean values of $\left(\widetilde{X}^{\left(q\right)}
,\widetilde{\tau}^{\left(q\right)}\Big\langle\frac{\beta^{\left(q\right)}\left(i\right)+1}{\widetilde{\tau}^{\left(q\right)}\left(i\right)}\Big\rangle_{i}+\nu\right)$ averaged across all replicas on each order and error bars denote standard deviations. See the definitions of random networks in Appendix \ref{ASec1}. (f) The standard deviations compared across different systems. (g-i) The high-order effects measured on the replicas in (a-c), where each color area denotes the distribution of high-order effects in the $k$-iteration of the SRG ($k\in\{1,2,3\}$).} 
\end{figure*}

In Fig. \ref{G5}, we present more instances of the renormalization flow generated by the SRG. We apply the SRG on synthetic interacting systems whose pairwise interactions follow the random tree (RT), the square lattice (SL), the triangular lattice (TL), the Watts-Strogatz networks (WS), and the Erdos-Renyi networks (ER). Among these random networks, the square and triangular lattices can be defined with or without the periodic boundary condition. Please see Appendix \ref{ASec1} for the definitions of these random networks. As shown in Fig. \ref{G5}, the key structural properties of RT, SL, and TL are principally preserved during renormalization (i.e., we can see qualitative similarities in $\mathbf{G}^{\left(p\right)}_{k}$ across different iterations) while the SRG implemented on WS and ER gradually reveals specific latent structures that are dissimilar to the original ones. Among these instances, the lattice systems with periodic boundary conditions are observed to make the renormalization ineffective. This is a phenomenon caused by the identical properties of all units under the periodic boundary condition (i.e., all units are exactly the same as each other on any $q$-order). Because the renormalizaton procedure in the SRG begins with distinguishing between short- and long-range interactions, the SRG can effectively reduce a system only when units and interactions are not identical. Otherwise, it is impossible to tell which interactions are long-range and which are short-range. In Fig. \ref{G5}(c) and Fig. \ref{G5}(e), the SRG does not coarse grain systems because it discover no short-range interaction. This phenomenon does not cause conflicts in our work because a lattice system with identical units and interactions can be treated as an extreme case of self-similarity to certain extent (i.e., each sub-system is identical to the whole system) and can be invariant for coarse graining. To avoid this trivial phenomenon in our analysis, all the lattice systems implemented in our subsequent experiments (e.g., Figs. \ref{G6}-\ref{G7}) are defined without the periodic boundary condition.

\subsection{Scaling relation and high-order effects}\label{Sec-IVD}

Given the SRG, a crucial task is to analyze the potential scaling relation and quantify the effects of high-order interactions on it.

In the ergodic case, we denote $\widetilde{X}^{\left(q\right)}$ and $\widetilde{\rho}^{\left(q\right)}$ as the specific heat and density operator associated with $\widetilde{\tau}^{\left(q\right)}$. We can insert these variables into Eq. (\ref{EQ25}) to derive the following relation
\begin{align}
\frac{\widetilde{X}^{\left(q\right)}}{\widetilde{\tau}^{\left(q\right)}}=\big\langle\lambda^{\left(q\right)}_{1}\big\rangle_{\widetilde{\rho}}+\mu,\;\forall \mu\in\mathbb{R}, \label{EQ27}
\end{align}
where $\big\langle\lambda^{\left(q\right)}_{1}\big\rangle_{\widetilde{\rho}}=\operatorname{tr}\left(\mathbf{L}^{\left(q\right)}_{1}\widetilde{\rho}^{\left(q\right)}\right)$ and the constant $\mu$ is obtained when we solve Eq. (\ref{EQ25}) as a differential equation. If the Laplacian eigenvalue spectrum (i.e., the density of the eigenvalues of $\mathbf{L}^{\left(q\right)}_{1}$) follows a general power-law form $\operatorname{Prob}\left(\lambda^{\left(q\right)}_{1}\right)\sim \left(\lambda^{\left(q\right)}_{1}\right)^{\beta^{\left(q\right)}}$, we can obtain a scaling relation
\begin{align}
\beta^{\left(q\right)}=\widetilde{X}^{\left(q\right)}-1-\nu,\;\exists \nu\in\left[0,\infty\right),\label{EQ28}
\end{align}
whose derivations are presented in Appendix \ref{ASec8}. The parameter $\nu$ is a specific constant. For a system with scale-invariance property on the $q$-order, the power-law form of the Laplacian eigenvalue spectrum holds and should be invariant under the transformation of the SRG. Therefore, we expect to see an approximately constant $\widetilde{X}^{\left(q\right)}$ in a certain interval of time scale (i.e., the infinite deceleration condition), accompanied by a fixed $\beta^{\left(q\right)}$ during renormalization. For a system without scale-invariance, the scaling relation does not hold true.

To generalize our analysis to the non-ergodic case, we respectively derive $\widetilde{\tau}^{\left(q\right)}\left(i\right)$ and $\beta^{\left(q\right)}\left(i\right)$ for each cluster $C_{i}$ of $\mathbf{G}^{\left(q\right)}_{1}$ following Eqs. (\ref{EQ27}-\ref{EQ28}). These definitions enable us to obtain 
\begin{align}
\big\langle\lambda^{\left(q\right)}_{1}\big\rangle_{\widetilde{\rho}}=\Big\langle\frac{\beta^{\left(q\right)}\left(i\right)+1}{\widetilde{\tau}^{\left(q\right)}\left(i\right)}\Big\rangle_{i}+\nu,\;\exists \nu\in\left[0,\infty\right),
\label{EQ29}
\end{align}
and
\begin{align}
\widetilde{X}^{\left(q\right)}
=\widetilde{\tau}^{\left(q\right)}\Big\langle\frac{\beta^{\left(q\right)}\left(i\right)+1}{\widetilde{\tau}^{\left(q\right)}\left(i\right)}\Big\rangle_{i}+\nu,\;\exists \nu\in\left[0,\infty\right),
\label{EQ30}
\end{align}
where $\langle\cdot\rangle_{i}$ denotes weighted averaging across all clusters of $\mathbf{G}^{\left(q\right)}_{1}$ (i.e., weighted according to cluster size). In the non-ergodic case, variables $\widetilde{X}^{\left(q\right)}$ and $\widetilde{\tau}^{\left(q\right)}$ refer to the global specific heat and the time scale measured on $\mathbf{L}^{\left(q\right)}_{1}$ under the infinite deceleration condition. Eqs. (\ref{EQ27}-\ref{EQ28}) actually serve as a special case of Eqs. (\ref{EQ29}-\ref{EQ30}) where there exists only one cluster. Please see Appendix \ref{ASec9} for derivations of these equations. Moreover, although the scaling relations in Eqs. (\ref{EQ27}-\ref{EQ30}) are derived in a case with $k=1$, we can also explore these scaling relations on $\mathbf{G}^{\left(q\right)}_{k}$ for $k>1$ (i.e., we directly replace $\mathbf{G}^{\left(q\right)}_{1}$ with $\mathbf{G}^{\left(q\right)}_{k}$ during calculating Eqs. (\ref{EQ27}-\ref{EQ30})). In summary, we can verify if a system obeys scaling relations and further explore whether the SRG preserves scaling relations.

Eqs. (\ref{EQ27}-\ref{EQ30}) are derived for $q$-order interactions during renormalization and may divide from the original scaling relation of $p$-order interactions. These $p$-order interactions, initially corresponding to $\widetilde{X}^{\left(p\right)}$ and $\beta^{\left(p\right)}$, are renormalized according to $\widetilde{X}^{\left(q\right)}$ and $\beta^{\left(q\right)}$ in the SRG. In the case with $p<q$, we can measure the effects of $q$-order interactions on the scaling relation of $p$-order ones as
\begin{align}
\zeta\left(p,q\right)
=\vert\widetilde{X}^{\left(p\right)}-\widetilde{X}^{\left(q\right)}\vert,
\label{EQ31}
\end{align}
which is applicable to both ergodic and non-ergodic cases. Meanwhile, similar to the scaling relation analysis, the measurement of high-order effects can be implemented in any iteration of the SRG.

Overall, the proposed SRG offers an opportunity to verify the potential existence of scale-invariance on different orders and classify interacting systems according to their high-order scaling properties. 

\section{Application of the simplex renormalization group}\label{Sec-V}
To this point, we have elaborated the SRG framework and its properties. Below, we validate its applicability in analyzing the systems with high-order interactions.

\subsection{Verification of the multi-order scale-invariance}\label{Sec-VA}
We aim at verifying the multi-order scale-invariance (i.e., being invariant under the SRG transformation on different orders). Specifically, being scale-invariant on an order requires the system to be located at a certain fixed point of the renormalization flow (i.e., the system behaves similarly across all scales), leading to scale-independent characteristics of the system described by the power-law scaling. In other words, if we describe system states by a set of observables, their measurements are expected to maintain invariant across all the iterations of the SRG. Due to the internal complexity of the system on different orders, we may see a rich variety of fixed points in empirical data. In an opposite where the system lacks a fixed point, its behaviours significantly change across scales. 

In Appendix \ref{ASec10}, we implement our verification according to the behaviours of the Laplacian eigenvalue spectrum (the behaviours of degree distributions are also presented as auxiliary information). Based on the existence of invariant power-law forms of Laplacian eigenvalue spectra under the SRG transformation, we can classify systems into scale-invariant, weakly scale-invariant, and scale-dependent types on different orders. 

Apart from the above approach, there exists a more direct way to realize the verification. Because a scale-invariant system on the $q$-order should invariably follow the scaling relation, being bound to satisfy Eq. (\ref{EQ29}) is one of the necessary conditions of being scale-invariant. Consequently, we can first measure the deviations from the scaling relation of concerned systems before renormalization. Given significantly large deviations, the system can not be scale-invariant no matter how it behaves during renormalization. In Fig. \ref{G6}, the measurement is conducted on multiple types of systems in each $k$-th iteration ($k\in\{1,2,3\}$) of the SRG (see Appendix \ref{ASec1} for random network definitions and see Appendix \ref{ASec11} for experiment details). As shown in Figs. \ref{G6}(a-e), although the mean observations of  $\left(\widetilde{X}^{\left(q\right)}
,\widetilde{\tau}^{\left(q\right)}\Big\langle\frac{\beta^{\left(q\right)}\left(i\right)+1}{\widetilde{\tau}^{\left(q\right)}\left(i\right)}\Big\rangle_{i}+\nu\right)$ averaged across all replicas collapse onto the scaling relation on every order and in each $k$-th iteration, the departures from the scaling relation measured by the standard deviations of $\widetilde{X}^{\left(q\right)}$ and $\widetilde{\tau}^{\left(q\right)}\Big\langle\frac{\beta^{\left(q\right)}\left(i\right)+1}{\widetilde{\tau}^{\left(q\right)}\left(i\right)}\Big\rangle_{i}+\nu$ in some systems can be non-negligible. Specifically, as suggested in Fig. \ref{G6}(f), the systems whose pairwise interactions follow known scale-invariant structures, such as the triangular lattice or the random tree, do ensure that all their replicas follow the scaling relation with vanishing deviations in each $k$-th iteration. The replicas of the systems whose pairwise interactions follow the Barab{\'a}si-Albert network, a structure with weak scale-invariant property (i.e., the Barab{\'a}si-Albert network can be renormalized within a specific set of scales but is not strictly scale-invariant \cite{villegas2023laplacian}), have relatively small deviations on the first three orders as well. The replicas of other kinds of systems whose pairwise interactions follow the Watts-Strogatz or the Erdos-Renyi networks \cite{erdHos1960evolution} exhibit larger deviations from the scaling relation on most orders, suggesting the absence of mutli-order scale-invariance.

In sum, the method reported in Fig. \ref{G6} can offer a convenient detection of scale-dependent systems in practice because scale-invariant systems are expect to satisfy Eq. (\ref{EQ29}) before and during renormalization. To seek a more precise verification, one should apply the approach in Appendix \ref{ASec10}.

Finally, we measure the high-order effects on the scaling relation applying Eq. (\ref{EQ31}). As shown in Figs. \ref{G6}(g-i), these high-order effects are non-negligible and generally increase with the difference between $p$ and $q$. These results explain how the SRG differs from conventional renormalization groups that focus on a single order.

\begin{figure*}[!t]
\includegraphics[width=1\columnwidth]{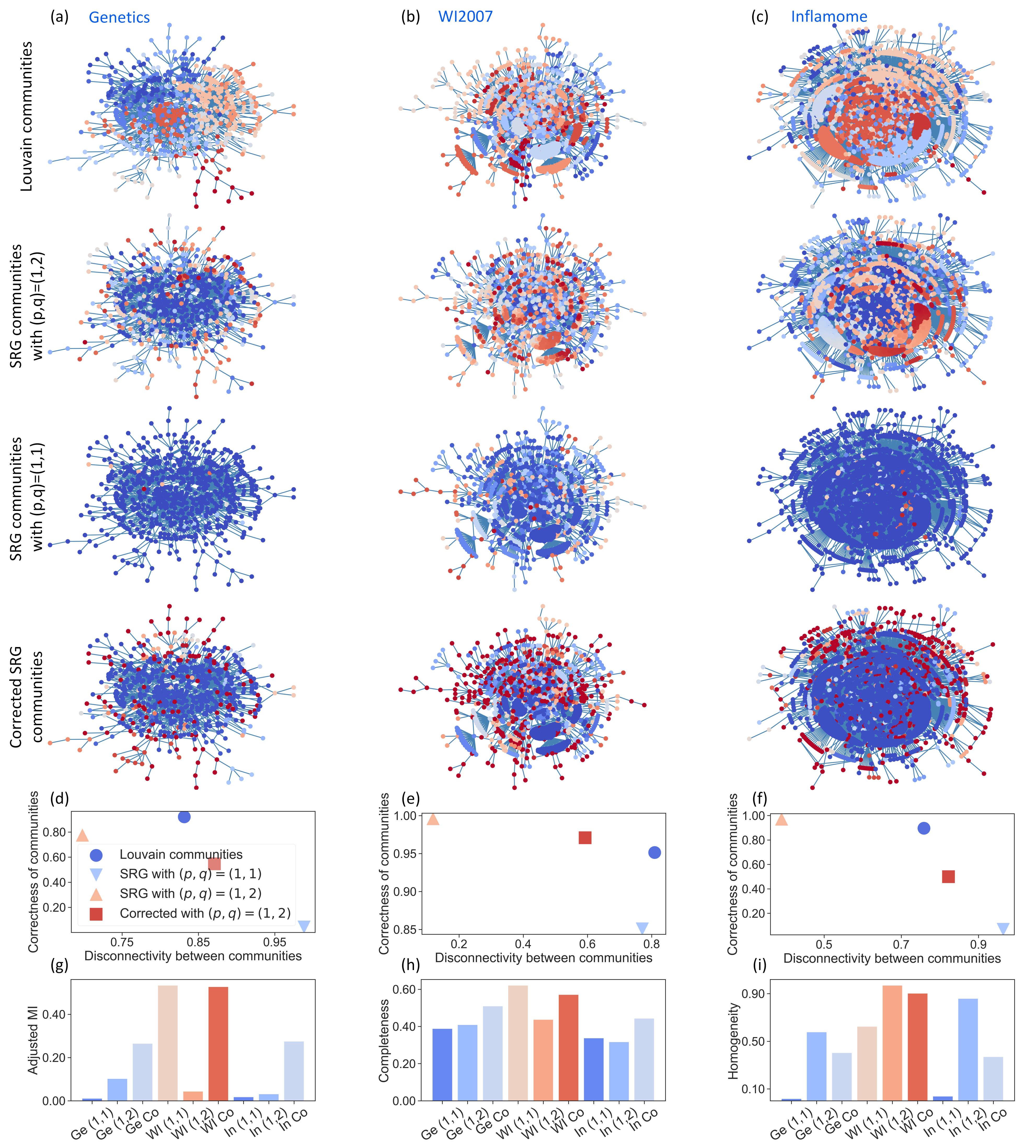}
\caption{\label{G7}The topological invariance explored by the SRG. (a-d) The $m$-order Betti number $\mathcal{B}_{m}$ ($m\in\{0,\ldots,3\}$) is calculated on $\mathbf{G}^{\left(p\right)}_{k}$ in each $k$-th iteration of the SRG ($k\in\{1,\ldots,5\}$), where the multi-order Laplacian (MOL) is used. (e-h) The same analysis is repeated using a SRG defined with the high-order path Laplacian (HOPL). Each line in (a-h) corresponds to an instance of the evolution of the $m$-order Betti number during renormalization. Please note that we define the $y$-axis as $\mathcal{B}_{m}+1$ to meet the demands of a log-scale axis.} 
\end{figure*}

\subsection{Exploration of the topological invariance}\label{Sec-VB}
In Appendix \ref{ASec3}, we have explained how the SRG is theoretically related to the persistent homology theory and other concepts, where we suggest the potential applicability of the SRG in analyzing persistent homology. Below, we validate this applicability by using the SRG to explore the topological invariance.

In brief, the persistent homology theory can be used to study the lifetimes of different topological properties in a filtration process, during which we progressively add or remove simplices from the system \cite{wang2020persistent,edelsbrunner2008persistent,aktas2019persistence,horak2009persistent} (i.e., this is similar to the network evolution process in complex network studies \cite{dorogovtsev2002evolution}). Although the SRG is not exactly same as the classic filtration approach because it modifies the system in a more complicated manner, it does offer an opportunity to study the evolution of topological properties during renormalization (i.e., we can measure different topological properties on $\mathbf{G}^{\left(p\right)}_{k}$ for each $k\geq 1$). Similar to the analysis of scale-invariance, a topological property can be treated as invariant if it is persistent across all the iterations of the SRG (i.e., referred to as the topological invariance). In our analysis, we mainly focus on the Betti number on different orders, a family of important topological properties used in biology \cite{chung2019exact,teramoto2018betti,topaz2015topological} and physics \cite{giri2021measuring,goel2019strong,robins2006betti} studies. In general, a $m$-order Betti number, $\mathcal{B}_{m}$, counts the number of $m$-dimensional holes in the system, which is a key topological characteristic of high-order structures and defines the dimension of $m$-th homology \cite{edelsbrunner2008persistent}. In the terminology of graph theory, a $0$-order Betti number counts the number of connected components and a $1$-order Betti number is the number of cycles. Computationally, the $m$-order Betti number can be calculated using the toolbox offered by Ref. \cite{mogutda2023toolbox}.

In Fig. \ref{G7}, we explore the potential existence of topological invariance in different random network models, including the Barab{\'a}si-Albert network ($c=4$), the Watts-Strogatz network (each unit initially has 10 neighbors and edges are rewired according to a probability of 0.3), the Erdos-Renyi network (two units share an edge with a probability of 0.1), the triangular lattice, and the random tree. One can see Appendix \ref{ASec1} for their definitions. We generate $100$ replicas for each random network model, where every replica consists of $200$ units. In the experiment, we design a SRG with a certain combination of $\left(p,q\right)$ and let it function on each random network to generate a sequence of $\mathbf{G}^{\left(p\right)}_{k}$ for $k\in\{1,\ldots,5\}$. Given each $\mathbf{G}^{\left(p\right)}_{k}$, we measure the $m$-order Betti number, $\mathcal{B}_{m}$, on it ($m\in\{0,\ldots,3\}$). If a $m$-order Betti number is generally constant across all iterations, its corresponding topological property can be treated as invariant during renormalization. 

As shown in Fig. \ref{G7}, the Betti numbers of scale-dependent systems (e.g., the Watts-Strogatz and the Erdos-Renyi networks) exhibit non-negligible changes on most orders when the SRG transforms systems across scales. Compared with these systems, weakly scale-invariant systems (e.g., the Barab{\'a}si-Albert network) have less variable Betti numbers during renormalization. When systems become scale-invariant (e.g., the triangular lattice and the random tree), the Betti numbers on all orders are generally invariant except that some fluctuations may occur in certain replicas due to random effects. This finding is interesting because, to our best knowledge, there is no theoretical guarantee that the multi-order scale-invariance is correlated with the topological invariance yet. According to our observation, we have the following speculation: there may exist specific topological invariance properties near the critical point that make the SRG, a framework dealing with high-order topological structures, have less effects on the information contained in the system. As a result, the behaviours of the system seem to be approximately invariant under the scale transformation of the SRG, leading to the inference that the system is born at a fixed point of the renormalization flow. Exploring this speculation may deepen our understanding of the relation between the persistent homology \cite{wang2020persistent,edelsbrunner2008persistent,aktas2019persistence,horak2009persistent} and the criticality studied by the renormalization group theory \cite{pelissetto2002critical,goldenfeld2018lectures}, which remains as a valuable direction of future works.

\subsection{Identification of organizational structure}\label{Sec-VC}
Next, we use the SRG to explore the roles of different order of interactions in preserving or defining the organizational structures (e.g., latent communities) of complex systems, where we also validate the applicability of the SRG in organizational structure identification.

When organizational structures are given, we explore the conditions under which these structures can be maintained by the SRG. In Appendix \ref{ASec12}, we implement the verification using protein-protein interactions (e.g., orthologous and genetic interactions) in Caenorhabditis elegans \cite{simonis2009empirically}. We first extract the initial community structures of protein-protein interactions by applying the Louvain community detection algorithm \cite{blondel2008fast} (see Appendix \ref{ASec1} for introduction). Then, we use the SRG to renormalize protein-protein interactions on a high order (i.e., with $\left(p,q\right)=\left(1,2\right)$) or a low order (i.e., with $\left(p,q\right)=\left(1,1\right)$). We define latent community structures according to unit aggregation during renormalization (see Appendix \ref{ASec12} for details). Consistent with Ref. \cite{gfeller2007spectral}, our results in Appendix \ref{ASec12} highlight that renormalization can never be treated as a trivial counterpart of clustering. Meanwhile, the SRG guided by high-order interactions (e.g., $q=2$) can generally preserve the properties of initial community structures while the SRG guided by pairwise interactions does not. Based on these results, we speculate that high-order interactions are essential in forming and characterizing organizational structures. Communities can not be treated as the trivial collections of pairwise interactions and it may be inappropriate to partition a system into sub-systems only based on pairwise relations. Given this observation, we naturally wonder if high-order interactions alone are sufficient to define ideal community structures (i.e., do not consider pairwise relations while determining system partition).

\begin{figure*}[!t]
\includegraphics[width=1\columnwidth]{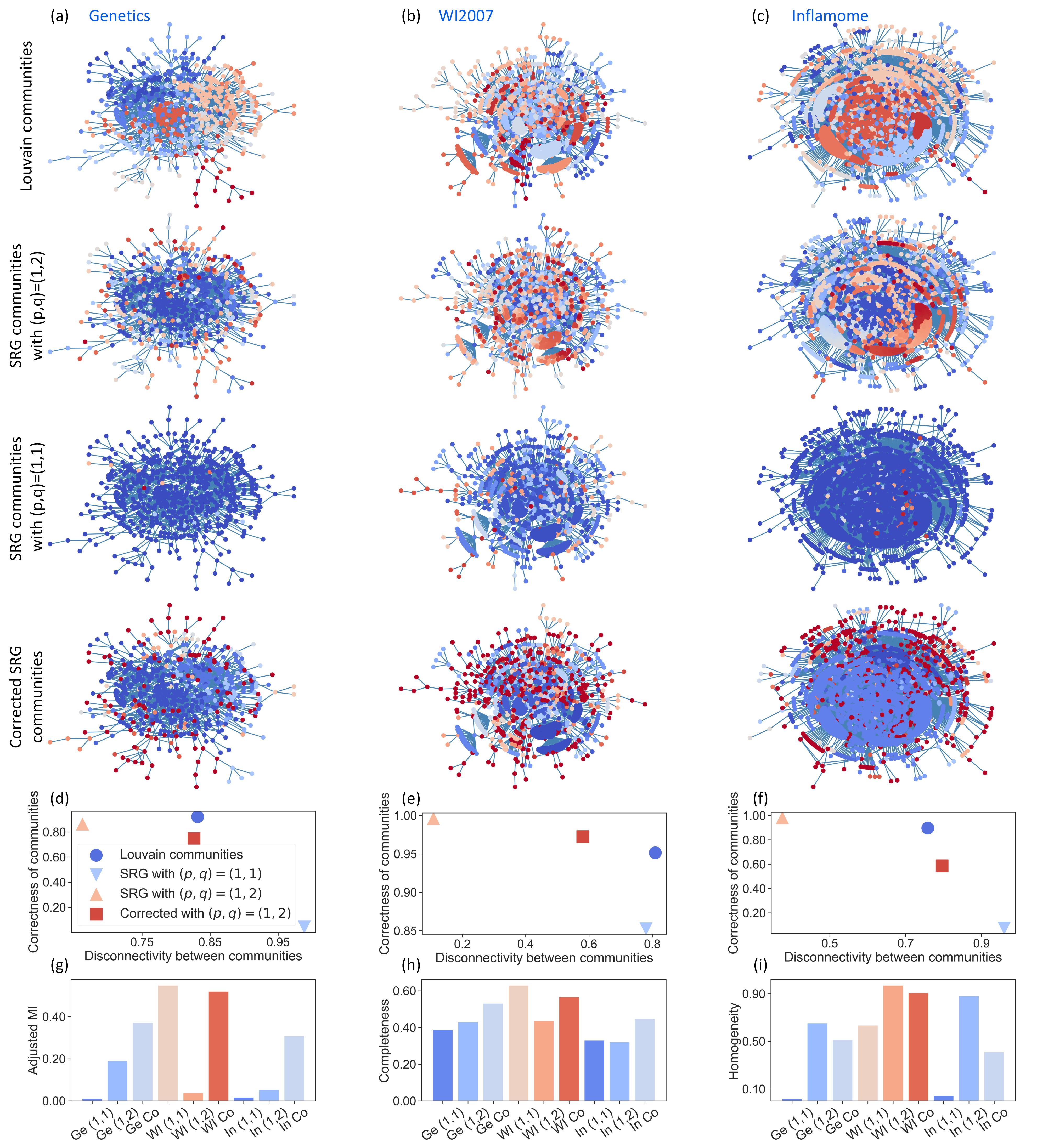}
\caption{\label{G8}The organizational structure discovered by renormalization flows. (a-c) The Louvain communities underlying the Genetics (Ge), WI2007 (WI), and Inflamome (In) data sets \cite{simonis2009empirically} are extracted using the Louvain community detection algorithm \cite{blondel2008fast}. The units of different communities are distinguished by colors. For comparison, the SRG is applied to renormalize these systems with $\left(p,q\right)=\left(1,2\right)$ and $\left(p,q\right)=\left(1,1\right)$. Units are determined as belonging to the same SRG community if they are aggregated into the same macro-unit. Meanwhile, the corrected SRG communities are presented. (d-f) The correctness of communities and disconnectivity between communities are shown. (g-i) The adjusted mutual information, completeness, and homogeneity are measured on SRG communities. ``Ge (1,1)" denotes the SRG communities with $\left(p,q\right)=\left(1,1\right)$ on the Genetics data set. ``Ge Co" denotes the corrected SRG communities on the Genetics data set. Other notions can be understood in a similar way. Please see Appendix \ref{ASec1} for the definitions of all measures used in (d-i). } 
\end{figure*}

To tackle this question, we explore whether a SRG defined with different combination of $\left(p,q\right)$ can be applied to identify latent communities when system partition is unknown. In Fig. \ref{G8}(a-c), we define communities according to the renormalization flows of the SRG and compare them with the Louvain communities (i.e., the communities detected by the Louvain algorithm \cite{blondel2008fast}, see Appendix \ref{ASec1}). Specifically, we classify the units into the same SRG community if they are aggregated into the same macro-unit during renormalization (e.g., we generate SRG communities after the first iteration of renormalization flows in Fig. \ref{G8}(a-c)).

To quantitatively compare the SRG communities with the Louvain communities, we analyze the correctness of communities (i.e., or referred to as the performance) \cite{fortunato2010community} and the disconnectivity between communities (i.e., or referred to as the coverage) \cite{fortunato2010community}. Note that both these concepts reflect the properties of communities on $1$-order because they are defined on pairwise interactions \cite{fortunato2010community} (see Appendix \ref{ASec1} for precise definitions). As shown in Fig. \ref{G8}(d-f), the behaviours of SRG are different from the Louvain community detection algorithm, which is designed to maximize the modularity of communities \cite{blondel2008fast,reichardt2006statistical,clauset2004finding}. When the SRG is guided by pairwise interactions, the generated SRG communities tend to be disconnected with each other (i.e., lack pairwise interactions across communities) and have low accuracy (i.e., units in the same community may lack pairwise interactions because lots of short-range pairwise interactions are reduced). When the SRG is guided by high-order interactions, the generated SRG communities tend to be precise (i.e., units in the same community always involve in high-order interactions) and non-isolated (i.e., there exist across-community pairwise interactions since the SRG mainly reduces high-order interactions). We are inspired to generate the SRG communities by combining the properties of these two cases. We suggest considering pairwise interactions to correct the SRG communities generated by the SRG guided by high-order interactions. Specifically, for two SRG communities generated with $\left(p,q\right)=\left(1,2\right)$, if at least half of the units in one community have pairwise interactions with the units in another community, then all units in these two SRG communities are merged into the same community. The corrected SRG communities are presented in Fig. \ref{G8}(d-f), exhibiting better capability in balancing between the correctness and the disconnectivity.

From another perspective, we quantify the consistency between the SRG communities and the Louvain communities in defining system partition in Fig. \ref{G8}(g-i). As measured by the adjusted mutual information, completeness, and homogeneity (see Appendix \ref{ASec1} for the definitions of these metrics, whose larger values suggest higher consistency extents), the partitions defined by the corrected SRG communities are generally consistent with, but not exactly equivalent to, those formed by the Louvain communities in all data sets. On the other hand, the SRG communities defined only based on high-order (i.e., with $\left(p,q\right)=\left(1,2\right)$) or low-order (i.e., with $\left(p,q\right)=\left(1,1\right)$) interactions do not necessarily keep consistency with the Louvain communities in these data sets. Because the Louvain communities are defined by maximizing the modularity \cite{blondel2008fast,reichardt2006statistical,clauset2004finding}, we suggest that the optimization of modular structures can not be realized only based on pairwise or high-order interactions. Instead, modular structures depend on multiple orders of interactions.

Taken together, high-order interactions are more effective than pairwise interactions in preserving the principal characteristics of organizational structures. However, merely high-order interactions alone are not sufficient to define ideal system partitions with optimal modular structures and clear separability among communities. For real systems, their organizational structures are defined by the intricate coexistence of various orders of interactions, exhibiting diverse behaviours across different orders. Although the SRG does not behave in a manner equivalent to the optimized algorithms for community detection (e.g., the Louvain algorithm \cite{blondel2008fast}), it offers a flexible way to analyze the roles of the interactions of every order in characterizing organizational structures since we can freely select different combinations of $\left(p,q\right)$ in analysis. Moreover, after considering the interactions of multiple orders, the corrected SRG communities have the potential to offer a system partition scheme competitive to optimized algorithms (see Fig. \ref{G8}(d-i)), which suggesting the applicability of the SRG in identifying latent community structures of real systems.

\subsection{Optimization of information bottleneck}\label{Sec-VD}
Finally, we analyze the SRG in terms of informational properties. In previous studies, the analysis using information theory has suggested that renormalization group frameworks essentially maximize the mutual information between relevant features and the environment \cite{koch2018mutual,lenggenhager2020optimal} or reduce the mutual information among irrelevant features \cite{hu2020machine}. In this work, we suggest exploring the SRG from the perspective of information bottleneck \cite{tishby2000information} because a renormalization group, similar to dimensionality reduction or representation learning approaches in machine learning \cite{alemi2016deep,saxe2019information,chechik2003information}, essentially deals with information encoding during data compression \cite{kline2022gaussian}. 

To realize this analysis, we follow one of our earlier works \cite{tian2023network} to represent the associated high-order network sketch in the $k$-th iteration of the renormalization flow, $\mathbf{G}^{\left(p\right)}_{k}$, by a Gaussian variable
\begin{align}
     \mathbf{X}^{\left(p\right)}_{k}\sim\mathcal{N}\left(\mathbf{0},\mathbf{L}^{\left(p\right)}_{k}+\frac{1}{N^{\left(p\right)}_{k}}\mathbf{J}\right),\label{EQ30}
\end{align}
where $\mathbf{J}$ is an all-one matrix and $N^{\left(p\right)}_{k}$ measures the number of units in $\mathbf{G}^{\left(p\right)}_{k}$. The derived Gaussian variable offers a optimal representation of $\mathbf{G}^{\left(p\right)}_{k}$ with network-topology-dependent smoothness and maximum entropy properties and enable us to compare $\mathbf{G}^{\left(p\right)}_{k}$ across different iterations (see Ref. \cite{tian2023network} for explanations). If necessary, one can further add a scaled unitary matrix, $a\mathbf{I}$ ($a>0$), to the covariance matrix of $\mathbf{X}^{\left(p\right)}_{k}$ to ensure its semi-positive property in the non-ergodic case (i.e., the case where $\mathbf{G}^{\left(p\right)}_{k}$ is disconnected).

We suggest to consider the following information bottleneck
\begin{align}
\mathcal{L}_{IB}=\underbrace{\mathcal{I}\left(\mathbf{X}^{\left(p\right)}_{k};\mathbf{X}^{\left(p\right)}_{1}\right)}_{\text{Preserved information}}-\psi\underbrace{\mathcal{I}\left(\mathbf{X}^{\left(p\right)}_{k};\mathbf{X}^{\left(p\right)}_{i}\right)}_{\text{Complexity}},\;\forall\psi>0\label{EQ31}
\end{align}
where $i=k-1$ if $k>1$ and $i=k$ if $k=1$. The parameter $\psi$ defines the regularization strength. We denote $\mathcal{I}\left(\cdot;\cdot\right)$ as the mutual information (see Appendix \ref{ASec1} for its definition), which can be estimated
by the mutual information estimator with local
non-uniformity corrections (we set the correction
strength as $10^{-3}$ such that corrections only happen when
necessary) \cite{gao2015efficient}. This approach has been included in the non-parametric entropy estimation toolbox \cite{gao2022toolbox}. Note that Eq. (\ref{EQ31}) is different from the information bottleneck considered in Ref. \cite{kline2022gaussian} due to the discrepancy between our concerned question and Ref. \cite{kline2022gaussian}.

\begin{figure*}[!t]
\includegraphics[width=1\columnwidth]{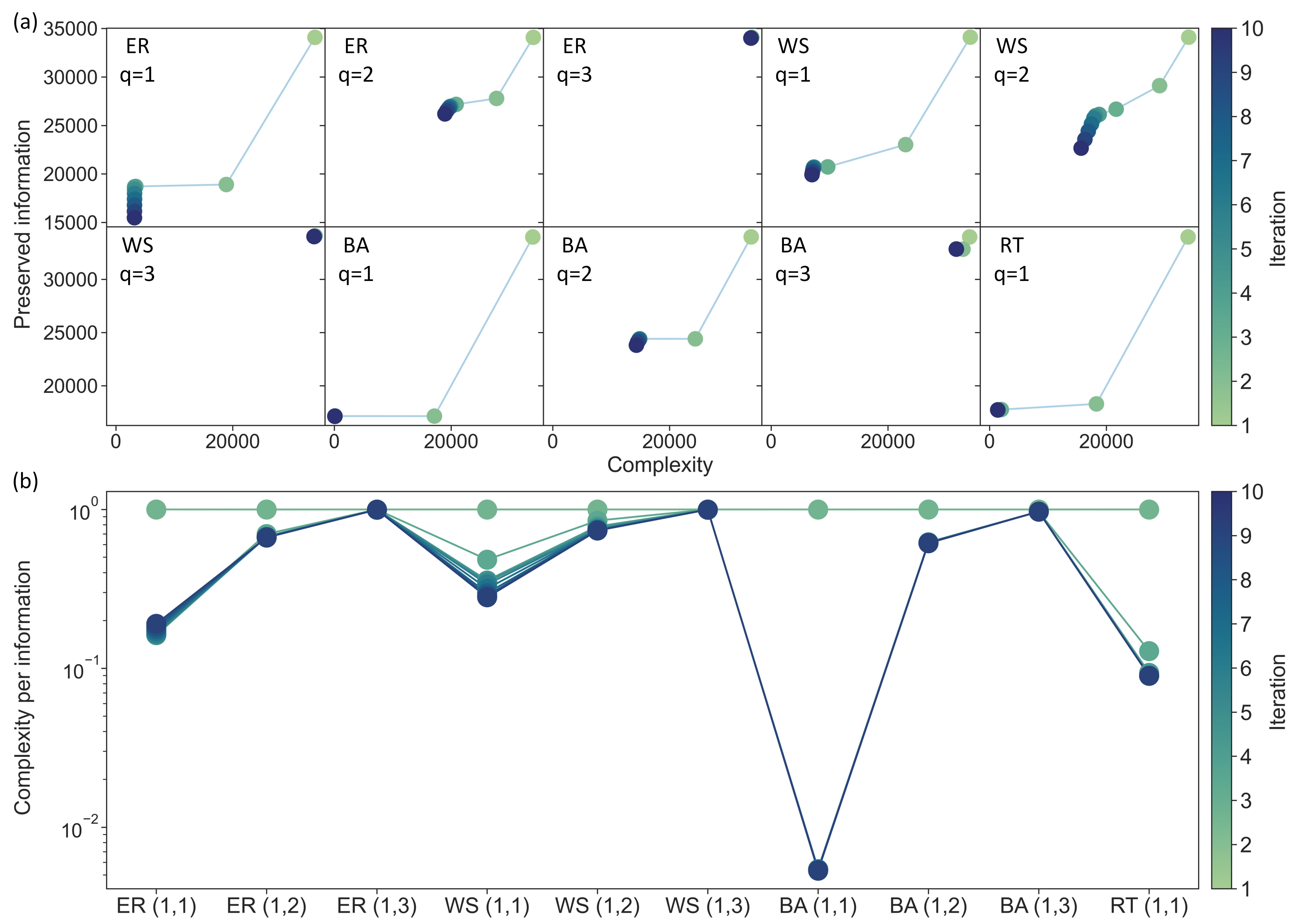}
\caption{\label{G9}The information curve of the SRG. (a) The SRG is applied on four synthetic interacting systems, whose pairwise interactions follow the Erdos-Renyi network (ER, each pair of units share an edge with a probability of $0.02$), the Watts-Strogatz network (WS, each unit initially has $5$ neighbors and edges are rewired according to a probability of $0.05$), the Barab{\'a}si-Albert network (BA, $c=4$), and the random tree (RT), respectively. Each kind of system consists of $500$ units and has $5$ replicas. The SRG is defined with the multi-order Laplacian and $\left(p,q\right)=\left(1,1\right)$ for RT while $\left(p,q\right)\in\{\left(1,1\right),\left(1,2\right),\left(1,3\right)\}$ is set for all other systems. The generated information flows are used to derive information curves. (b) The complexity per preserved information bit is shown after averaging across all replicas.} 
\end{figure*}

Intuitively, the objective function in Eq. (\ref{EQ31}) evaluates whether the renormalized system in the $k$-th iteration, $\mathbf{G}^{\left(p\right)}_{k}$, preserves the information of the original system, $\mathbf{G}^{\left(p\right)}_{1}$, by maximizing $\mathcal{I}\left(\mathbf{X}^{\left(p\right)}_{k};\mathbf{X}^{\left(p\right)}_{1}\right)$ and reduces the trivial dependence on its previous state, $\mathbf{G}^{\left(p\right)}_{i}$, by minimizing $\mathcal{I}\left(\mathbf{X}^{\left(p\right)}_{k};\mathbf{X}^{\left(p\right)}_{i}\right)$. For a valid renormalization group, being able to preserve some information of the original system is an essential capability. Because a trivial solution of maximizing $\mathcal{I}\left(\mathbf{X}^{\left(p\right)}_{k};\mathbf{X}^{\left(p\right)}_{1}\right)$ is to make $\mathbf{G}^{\left(p\right)}_{k}=\mathbf{G}^{\left(p\right)}_{k-1}$ for any $k>1$ (i.e., there is no renormalization at all), we need to include the complexity term in Eq. (\ref{EQ31}) to avoid this trivial result. 

Meanwhile, the possibility of maximizing $\mathcal{I}\left(\mathbf{X}^{\left(p\right)}_{k};\mathbf{X}^{\left(p\right)}_{1}\right)$ depends on the properties of the system as well. If the system is scale-invariant, the preserved information does not rapidly vanish when we reduce the complexity since the information of $\mathbf{G}^{\left(p\right)}_{1}$ holds across scales (note that $\mathcal{I}\left(\mathbf{X}^{\left(p\right)}_{k};\mathbf{X}^{\left(p\right)}_{1}\right)$ still reduces slightly because the decreasing dimensionality of $\mathbf{X}^{\left(p\right)}_{k}$ affects the numerical behaviours of the estimator). If the system is scale-dependent, the preserved information crucially relies on the complexity and vanishes unless the complexity is high. Therefore, we can divide $\mathcal{I}\left(\mathbf{X}^{\left(p\right)}_{k};\mathbf{X}^{\left(p\right)}_{1}\right)$ by $\mathcal{I}\left(\mathbf{X}^{\left(p\right)}_{k};\mathbf{X}^{\left(p\right)}_{i}\right)$ to measure the complexity required by a single bit of preserved information in the $k$-th iteration. A high value of complexity per information bit suggests that the system is scale-dependent because we can not find a low-complexity coarse-grained system to represent it efficiently.

In Fig. \ref{G9}(a), we show the information curves of the renormalization flows generated on four kinds of interacting systems. As suggested by our results, the renormalization flows of weakly scale-invariant (e.g., BA) and scale-invariant (e.g., RT) systems lead to sufficiently small complexity values and numerous preserved information bits hile the renormalization flows of scale-dependent systems (e.g., ER and WS) do not always achieve so (although WS realizes complexity reductions during renormalization, it is still not competitive with BA and RT). The information bits of scale-dependent systems are preserved at the cost of maintaining high complexity while scale-invariant and weakly scale-invariant systems progressively become more complexity-saving in preserving information during renormalization. These results can be quantitatively validated based on the complexity per information bit measured in Fig. \ref{G9}(b). These findings suggest the possibility of verifying scale-invariance from an informational perspective. Moreover, we observe that the SRG guided by pairwise interactions (i.e., $q=1$) frequently achieve more significant complexity reductions while preserving information. This phenomenon may arise from the fact that high-order interactions are distributed in systems in a sparser manner.

\section{Discussions}
Various real interacting systems share universal characteristics that govern their dynamics and phase transition behaviours \cite{cimini2019statistical}. These characteristics remain elusive because fundamental physics tools, such as path integrals and renormalization group theories, are not well-established when \emph{a priori} knowledge about system mechanisms is absent \cite{villegas2023laplacian}. This gap may be accountable for diverse controversies concerning the critical phenomena in biological or social systems (e.g., the controversies about brain criticality \cite{tian2022theoretical}) and has induced a booming field that is devoted to developing statistical physics approaches for the empirical data generated by unknown mechanisms \cite{villegas2023laplacian}. Notable progress is accomplished in the computational implementations of path integral and renormalization group theories on real systems with pairwise interactions \cite{meshulam2018coarse,garcia2018multiscale,villegas2023laplacian,bradde2017pca,lahoche2022generalized,zheng2020geometric}, supporting remarkable applications in analyzing neural dynamics \cite{meshulam2019coarse} and swarm behaviours \cite{cavagna2019dynamical}. Nevertheless, there exists no apparent equivalence between these pioneering works and an appropriate approach for analyzing un-decomposable high-order interactions because classic network or geometric representations are invalid in characterizing polyadic relations \cite{benson2016higher,lambiotte2019networks,majhi2022dynamics,lucas2020multiorder,battiston2021physics}. 

In this work, we contribute to the field by developing the natural generalizations of path integral and renormalization group on un-decomposable high-order interactions. Our main contributions are summarized below:

    \begin{itemize}
        \item[(1) ]Our theoretical derivations lead us to an intriguing perspective that the system evolution governed by high-order interactions can be fully formalized by path integrals on simplicial complexes (e.g., the return probability of an arbitrary system state is proportional to the path integral of all closed curves that start from and end with this state in high-order networks), which suggests the possibility for studying high-order interactions applying the tools of quantum field theory \cite{feynman2010quantum,kleinert2009path}. 
        \item[(2) ]We suggest that the contributions of microscopic fluctuations to macroscopic states in the moment space are precisely characterized by a function of the multi-order Laplacian \cite{lucas2020multiorder} or our propsed high-order path Laplacian. Consequently, a renormalization group, the SRG, can be directly developed based on the diffusion on simplicial complexes, which does not require any assumption on latent metric for mapping units into target spaces (e.g., the hyperbolic space \cite{garcia2018multiscale,zheng2020geometric}). This property ensures the general applicability of our theory on real complex systems, where \emph{a priori} knowledge is rare and any assumption can be unreliable. Different from classic renormalization groups, the SRG can renormalize $p$-order interactions in a system based on the structure and dynamics associated with $q$-order interactions ($p\leq q$), enabling us to analyze the effects of $q$-order interactions on the $p$-order ones and study the scaling properties across different orders.
        \item[(3) ]We propose a divide-and-conquer approach to deal with non-ergodic cases, where some systems states are never reachable by the evolution started from other systems states (i.e., without global connectivity). The developed framework improves the validity of our theory for high-order interactions, which usually feature sparse distributions that break system ergodicity on a high order. 
        
        \item[(4) ]We seek for a comprehensive analysis of the scaling relation in both ergodic and non-ergodic cases. Our theoretical derivations and computational experiments suggest that the scaling relation can help differentiate among scale-invariant, weakly scale-invariant, and scale-dependent systems on an arbitrary order (consistent with verification using the behaviours of the Laplacian eigenvalue spectrum). Meanwhile, the effects of high-order interactions on the scaling relation can be measured as well. 

        \item[(5) ]We propose a way to use the SRG to verify the existence of topological invariance, suggesting the possibility to relate the SRG with the persistent homology analysis \cite{wang2020persistent,edelsbrunner2008persistent,aktas2019persistence,horak2009persistent}. As empirically demonstrated in our results, the SRG can serve as a tool similar to the filtration process \cite{wang2020persistent,edelsbrunner2008persistent,aktas2019persistence,horak2009persistent} during studying the lifetimes of different topological properties. In our experiments, we observe a correlated relation between the multi-order scale-invariance and the topological invariance (i.e., multi-order scale-invariant systems are more likely to satisfy the topological invariance), which suggests the potential role of topology in shaping the statistical physics properties of complex systems.
        \item[(6) ]Apart from validating the essential difference between the SRG and clustering \cite{gfeller2007spectral}, our experiments also suggest the applicability of the SRG in identifying organizational structures (e.g., latent communities) of real complex systems and analyzing the crucial roles of the interactions of different orders in characterizing these structures.
        \item[(7) ]We suggest that the SRG can be analyzed from the perspective of information bottleneck, which quantifies how the renormalized system preserves the information of its original properties while reducing complexity. We discover that the complexity required for maintaining one information bit can distinguish between scale-invariant and scale-dependent systems in an informational aspect.
    \end{itemize}

In sum, by extending classic path integrals and renormalization groups to simplicial complexes, our research reveals a novel route to studying the universality classes of complex systems with intertwined high-order interactions. The proposed simplex path integral and simplex renormalization group can serve as precise tools for characterizing system dynamics, discovering intrinsic scales, and verifying potential scale invariance on different orders. The revealed information by our theory may shed light on the intricate effects of the
interplay among multi-order interactions on system dynamics properties (e.g., phase transitions or scale invariance). To ensure the capacity of our theory in analyzing the real data sets that are governed by unknown mechanisms or lack clear network structures, we suggest to consider \emph{a-priori}-knowledge-free framework in future studies. This framework begins with applying specific non-negative non-parametric metrics (e.g., distance correlation \cite{chaudhuri2019fast} or co-mutual information \cite{kraskov2004estimating}) to evaluate the coherence between units and define the adjacency matrix of pairwise interactions. Then, the high-order representation and the simplex renormalization group can be progressively calculated following our theory.

\section*{Data availability statement}
Any data that support the findings of this study are included within the article.

 \section*{Acknowledgements}
This work is a part of the Topophy program. Y.T develops the theory, designs computational experiments, writes the manuscript, and leads the Topophy program. A.H.C contributes to theoretical derivations, programmatic implementation, and computational experiment realization. Y.H.X contributes to theoretical derivations and proof reading. P.S supervises the research project, proofreads and revises the manuscript, contributes to conceptualization, and offers technical supports. This project is supported by the Artificial and General Intelligence Research Program of Guo Qiang Research Institute at Tsinghua University (2020GQG1017). Authors appreciate Dr. Hedong Hou, who studies at the Institut de Math{\'e}matiques d'Orsay, for his discussions. Authors thank Mr. Kangyu Weng, who studies at Tsinghua University, for his help in proof reading. Authors also appreciate anonymous reviewers for their inspiring suggestions, especially those about the topological invariance analysis, the SRG in a form of conventional renormalization group theories, and the nature of high-order interactions. 

\appendix
\begin{widetext}

\section{Auxiliary concept definitions}\label{ASec1}
There are numerous auxiliary concepts used to support our analysis. In this section, we elaborate their basic definitions and offer necessary references.

\paragraph{Simplex and simplicial complex. } For clarity, we first recall the elementary definitions of simplex and simplicial complex \cite{ribando2022graph,lim2020hodge}. A $q$-simplex, as visualized in Fig. \ref{G1}, is a convex hull of $q+1$ affinely independent points in $\mathbb{R}^{h}$ ($h\geq q$). A simplicial complex, $K$, is a collection of simplices in $\mathbb{R}^{h}$ (see Fig. \ref{G1}) that satisfies the following properties: 
\begin{itemize}
    \item[(1) ] Each face $\sigma^{\left(p\right)}$ of simplex $\sigma^{\left(q\right)}\in K$ also belongs to $K$.
    \item[(2) ] The non-empty intersection of each pair simplices, $\sigma^{\left(p\right)}$ and $\sigma^{\left(q\right)}$, in $K$ must be a face of both simplices $\sigma^{\left(p\right)}$ and $\sigma^{\left(q\right)}$ (here $p$ and $q$ are not necessarily the same). 
\end{itemize}
The simplicial complex considered in our work, as shown in Fig. \ref{G1}, is usually referred to as the clique complex in the persistent homology analysis \cite{wang2020persistent,ribando2022graph}. In general, a $q$-simplex is a clique of $q+1$ units (i.e., a $q$-order interaction requires the simultaneous participation of $q+1$ units).

\paragraph{Random network models. } In our computational experiments, we use the following random networks:
\begin{itemize}
    \item[(1) ] The Erdos-Renyi network (ER) \cite{erdHos1960evolution} is an elementary type of random network in which two units share an edge with a probability of $p$ or are disconnected with a probability of $1-p$. We have used the ER networks in Fig. \ref{G6}, Fig. \ref{G9}, Fig. \ref{AG1}, and Figs. \ref{AG3}-\ref{AG4}.
    \item[(2) ] The Watts-Strogatz network (WS) \cite{watts1998collective} is a typical model of small-world phenomenon. It is generated by an edge rewiring process, during which we first initialize a regular network (i.e., a network where all units have the same number of neighbors) and rewire each edge according to a probability of $p$ (i.e., the edge remains unchanged with a probability of $1-p$). Several WS networks are implemented in Figs. \ref{G5}-\ref{G6}, Fig. \ref{G9}, Fig. \ref{AG1}, and Figs. \ref{AG2}-\ref{AG4}.
    \item[(3) ] The Barab{\'a}si-Albert network (BA) \cite{albert2002statistical,barabasi2009scale} is a representative example of weakly scale-invariant system \cite{villegas2023laplacian}. It is generated by the network growth process (i.e., a process for progressively adding new units and edges into a null network), where the number of random edges to attach from a new unit to existing units is denoted as $c$. One can see the BA networks in Figs. \ref{G1}-\ref{G3}, Fig. \ref{G6}, Fig. \ref{G9}, Fig. \ref{AG1}, and Figs. \ref{AG3}-\ref{AG4}.
    \item[(4) ] The random tree (RT) \cite{drmota2009random} is an instance of scale-invariance, which is generated from the uniformly random Pr{\"u}fer sequence in computational implementation. The instances of RT are implemented in Figs. \ref{G5}-\ref{G6}, Fig. \ref{G9}, Fig. \ref{AG1}, and Figs. \ref{AG3}-\ref{AG4}.
    \item[(5) ] The triangular lattice (TL) \cite{bille2023random} is another instance of scale-invariance when system size approaches infinity. Under the periodic boundary condition, a finite triangular lattice system is also a regular network where each unit has six neighbors. When the periodic boundary condition is absent, only the units within boundary form a regular network while the units on boundary may exhibit different behaviours under finite size effects. One can see the instances of TL in Figs. \ref{G5}-\ref{G6}, Fig. \ref{AG1}, and Figs. \ref{AG3}-\ref{AG4}.
    \item[(6) ] The square lattice (SL) \cite{grimmett2018probability} is also an instance of scale-invariance when system size approaches infinity. It shares similar properties with the triangular lattice except that each unit in a finite square lattice system has four rather than six neighbors under the the periodic boundary condition. The SL has been analyzed in Fig. \ref{G5} and Fig. \ref{AG1}.
\end{itemize}

Among these random networks, the ER, WS, and BA networks may contain arbitrary orders of interactions, depending on the settings of network parameters. The RT only contains the interactions of the first order due to its tree structure. The TL only includes the interactions of the first two orders (i.e., we can not find any tetrahedron or any other simplicial complex of a higher order in the TL system). The SL is formed by the interactions of the first order (i.e., there is no triangle or any other simplicial complex of a higher order in the SL system).

\paragraph{Organizational structure identification, preserving, and evaluation. }
In our experiments, we tackle the identification and preserving tasks of organizational structures of interacting systems, where we compare our SRG framework with the Louvain community detection algorithm \cite{blondel2008fast,reichardt2006statistical,clauset2004finding}. The Louvain community detection algorithm searches for an ideal partition by maximizing the modularity of communities, which is essentially different from a renormalization group because it is purely optimization-based \cite{blondel2008fast,reichardt2006statistical,clauset2004finding}. For convenience, we refer to the system partition identified by the Louvain community detection algorithm \cite{blondel2008fast,reichardt2006statistical,clauset2004finding} as the Louvain communities.

When an organizational structure has been given (e.g., if the Louvain communities have been derived), we can explore whether the SRG can preserve this organizational structure during renormalization. The corresponding experiment is shown in Fig. \ref{AG5} and Appendix \ref{ASec12}. We do not repeatedly elaborate the experiment in this section.

If the organizational structure is unknown, we can explore if the SRG can identify an appropriate structure. To evaluate the effectiveness of organizational structure identification, we consider the following metrics:
\begin{itemize}
    \item[(1) ] The correctness of communities (i.e., or referred to as the performance) measures the proportion of correctly interpreted pairs of units among all pairs of units, i.e., two units within the same community are expected to share an edge or two units belonging to different communities are expected to be disconnected \cite{fortunato2010community}. An ideal organizational structure should make all pairs of units correctly interpreted, implying a correctness score of $1$.
    \item[(2) ] The disconnectivity between communities (i.e., or referred to as the coverage) measures the proportion of intra-community edges among all edges \cite{fortunato2010community}. An ideal organizational structure is expected to make all edges fall within communities (i.e., all communities are perfectly separable from each other), leading to a disconnectivity score of $1$.
\end{itemize}
One can see the corresponding experiment in Fig. \ref{G8}, where the Louvain community detection algorithm \cite{blondel2008fast,reichardt2006statistical,clauset2004finding} is implemented for comparison. Apart from the identification effectiveness mentioned above, the quantitative comparison between the SRG communities and the Louvain communities can be also realized by the following metrics:
\begin{itemize}
    \item[(1) ] The adjusted mutual information \cite{vinh2009information} is a mutual information score (i.e., similarity degree) between the SRG and the Louvain communities, which is regulated by chance and is independent of permutation \cite{vinh2009information}. The adjusted mutual information approaches $1$ when the compared communities are identical.
    \item[(2) ] The completeness score \cite{rosenberg2007v} measures the consistency between the SRG and the Louvain communities, which approaches $1$ if all units within each SRG community also belong to the same Louvain community.
    \item[(3) ] The homogeneity score \cite{rosenberg2007v}, similar to the completeness score, measures the consistency between the SRG and the Louvain communities, which approaches $1$ if all units within each Louvain community also belong to the same SRG community.
\end{itemize}

\paragraph{Information quantity measurement. } In our analysis, we use the mutual information to quantify the information of one random variable, $A$, successfully encoded by another random variable, $B$. Mathematically, the mutual information is defined as a Kullback-Leibler divergence $\mathcal{I}\left(A;B\right)=\mathcal{D}_{\text{KL}}\left[\text{Prob}\left(A,B\right)\Vert \text{Prob}\left(A\right)\text{Prob}\left(B\right)\right]$ \cite{gao2015efficient}. Computationally, the mutual information is estimated by the mutual information estimator with local non-uniformity corrections \cite{gao2015efficient}, which is an approach included in the non-parametric entropy estimation toolbox \cite{gao2022toolbox} and proven as efficient in dealing with high-dimensional variables. The corresponding results are used in the information bottleneck analysis shown in Fig. \ref{G9}.

\section{Theoretical backgrounds of high-order Laplacian operators}\label{ASec2}
Given a system with diverse orders of interactions, we can treat it as a simplicial complex, $K$, following Appendix \ref{ASec1}. In this simplicial complex, we can search a $n$-simplex to verify its potential existence (here $n+1$ is no more than the size of the system). By assigning the identified simplices with different physics meanings, we can use them to model various types of high-order interactions. In our work, we primarily focus on two kinds of basic high-order interactions. The first type of $n$-order interactions occur if and only if the associated $n+1$ units are simultaneously engaged in (i.e., the interaction becomes ineffective unless all the $n+1$ units participate in it at the same time). The second type of $n$-order interactions refer to the situations where $n+1$ units form an irreducible sequence of participation (i.e., the interaction becomes ineffective unless these $n+1$ units participate in it following a sequence). In this section, we elaborate the high-order Laplacian operators associated with them.

\paragraph{Multi-order Laplacian operator}
We begin with the first type of $n$-order interactions that manifest as the simultaneous participation of $n+1$ related units. We expect to see a $n$-simplex, $\sigma^{\left(n\right)}$, if there exists a certain $n$-order interaction. This $n$-simplex is observable if we analyze the system on $h$-order ($h\geq n$). The corresponding Laplacian operator is referred to as the multi-order Laplacian operator, $\mathbf{L}_{M}^{\left(n\right)}$, which is originally proposed by Ref. \cite{lucas2020multiorder}. To define such an operator, we actually need to measure the difference between the following numbers
\begin{align}
\left[\mathbf{L}_{M}^{\left(n\right)}\right]_{ij}=\delta_{ij}\cdot\text{the number of }\sigma^{\left(n\right)}\text{ containing unit }v_{i}-\text{the number of }\sigma^{\left(n\right)}\text{ containing units }v_{i}\text{ and }v_{j},\label{A2EQ1}
\end{align}
where $\delta_{ij}$ denotes the Kronecker delta function. The non-triviality is that there may exist diverse simplices in the system and $\sigma^{\left(n\right)}$ can be a face of multiple simplices. Therefore, a unit can participate in many $n$-order interactions. Two coupled units do not necessarily share only one $n$-order interaction. In the most general situation, we need to go through all the possibilities to count the numbers mentioned in Eq. (\ref{A2EQ1}). 

To search these possibilities, we let $\{i\}_{n+1}\subset V$ be an arbitrary set of $n+1$ units that contains unit $v_{i}$, where $V$ is the set of all units. Similarly, we denote $\{i,j\}_{n+1}\subset V$ as the set of $n+1$ units where units $v_{i}$ and $v_{j}$ are included. Given a set $\{i\}_{n+1}$ during the search process, we check whether the units contained by it form a $n$-simplex or not. For clarity, we denote $I_{\sigma^{\left(n\right)}}\left(\{i\}_{n+1}\right)=1$ if there is a $n$-simplex, $\sigma^{\left(n\right)}$, formed by these units and define $I_{\sigma^{\left(n\right)}}\left(\{i\}_{n+1}\right)=0$ if there is no such $n$-simplex. Based on these definitions, we can count the number of $n$-simplices containing units $v_{i}$ and $v_{j}$ as
\begin{align}
    \mathbf{A}_{ij}^{\left(n\right)}= (1-\delta_{ij})\sum_{\{i,j\}_{n+1}\subset V}I_{\sigma^{\left(n\right)}}\left(\{i,j\}_{n+1}\right).\label{A2EQ2}
\end{align}
A non-trivial property of Eq. (\ref{A2EQ2}) is that summing Eq. (\ref{A2EQ2}) over all possible units $v_{j}$ does not directly obtain the number of $n$-simplices containing unit $v_{i}$ unless $n=1$ (i.e., for pairwise interactions)
\begin{align}
    \sum_{j}\mathbf{A}_{ij}^{\left(n\right)}&=\sum_{j}\left(1-\delta_{ij}\right)\sum_{\{i,j\}_{n+1}\subset V}I_{\sigma^{\left(n\right)}}\left(\{i,j\}_{n+1}\right),\label{A2EQ3}\\
&=\sum_{j\neq i}\sum_{\{i,j\}_{n+1}\subset V}I_{\sigma^{\left(n\right)}}\left(\{i,j\}_{n+1}\right),\label{A2EQ4}\\
&=n\sum_{\{i\}_{n+1}\subset V}I_{\sigma^{\left(n\right)}}\left(\{i\}_{n+1}\right).\label{A2EQ5}
\end{align}
We obtain Eq. (\ref{A2EQ5}) from Eq. (\ref{A2EQ4}) because we repeatedly count each $n$-simplex $n$ times while summing $I_{\sigma^{\left(n\right)}}\left(\{i,j\}_{n+1}\right)$ over all units $v_{j}$ in Eq. (\ref{A2EQ4}). As a simple instance, let us consider a case with $n=3$. We assume that $v_{i}$, $v_{j}$, $v_{k}$, and $v_{l}$ form a $3$-simplex, $\sigma^{\left(3\right)}$. This simplex is considered when we analyze $I_{\sigma^{\left(3\right)}}\left(\{i,j\}_{4}\right)$, $I_{\sigma^{\left(3\right)}}\left(\{i,k\}_{4}\right)$, and $I_{\sigma^{\left(3\right)}}\left(\{i,l\}_{4}\right)$. Thus, it is repeatedly counted $3$ times. There is no repetition for pairwise interactions (i.e., $n=1$) because a pairwise interaction only consists of unit $v_{i}$ and another one unit, which is analyzed once.

In Eq. (\ref{A2EQ5}), the term scaled by factor $n$ equals the number of $n$-simplices containing unit $v_{i}$. Therefore, we have 
\begin{align}
    \sum_{j}\mathbf{A}_{ij}^{\left(n\right)}&=n\mathbf{D}_{i}^{\left(n\right)},\label{A2EQ6}
\end{align}
where $\mathbf{D}_{i}^{\left(n\right)}$ counts the $n$-simplices that include unit $v_{i}$. Eq. (\ref{A2EQ6}) directly leads to 
\begin{align}
   \left[\mathbf{L}_{M}^{\left(n\right)}\right]_{ij}=n \delta_{ij}\mathbf{D}_{i}^{\left(n\right)}-\mathbf{A}_{ij}^{\left(n\right)}.\label{A2EQ7}
\end{align}

A remaining question is how to determine the values of $I_{\sigma^{\left(n\right)}}\left(\{i\}_{n+1}\right)$ and $I_{\sigma^{\left(n\right)}}\left(\{i,j\}_{n+1}\right)$ in Eqs. (\ref{A2EQ2}-\ref{A2EQ5}). For convenience, we define $S^{\{i\}}_{n+1}$ as the set of all permutations on $\{i\}_{n+1}$. Because the $n$-simplex studied in our work is a clique of $n+1$ units, we can calculate the indicator function as
\begin{align}
   I_{\sigma^{\left(n\right)}}\left(\{i\}_{n+1}\right)=\prod_{\left(k_{1}\ldots k_{n}\right)\in 
S^{\{i\}}_{n+1}}\mathbf{A}_{k_{1}k_{2}}\cdots\mathbf{A}_{k_{n-1}k_{n}},\label{A2EQ8}
\end{align}
where $\left(k_{1}\ldots k_{n}\right)$ denotes a permutation $\mathbf{A}=\mathbf{A}^{\left(1\right)}$ is the classic adjacency matrix of units (i.e., this matrix can be derived on pairwise interactions). According to the recursive attribute of a simplex mentioned in Appendix \ref{ASec1} (i.e., each face of a simplex is also a simplex), the units in $\{i\}_{n+1}$ can form a $n$-simplex if and only if each sub-set of $\{i\}_{n+1}$ that contains $m+1$ units also forms a $m$-simplex ($m\leq n$). By going through all permutations in $S^{\{i\}}_{n+1}$, we can check the existence of all possible $m$-simplices. Therefore, starting from pairwise interactions, Eq. (\ref{A2EQ8}) enables us to verify the existence of $2$-order, $3$-order, and eventually $n$-order interactions. For simplicity, we mark $\omega=\left(k_{1}\ldots k_{n}\right)$ and define 
\begin{align}
\mathbf{M}_{\omega}=\mathbf{A}_{k_{1}k_{2}}\cdots\mathbf{A}_{k_{n-1}k_{n}},\label{A2EQ9}
\end{align}
which supports us to write Eq. (\ref{A2EQ9}) as $I_{\sigma^{\left(n\right)}}\left(\{i\}_{n+1}\right)=\prod_{\omega\in S^{\{i\}}_{n+1}}\mathbf{M}_{\omega}$.

\paragraph{High-order path Laplacian operator}
Then, we formulate the second type of $n$-order interactions, which manifest as the sequential actions of $n+1$ related units after they participate in the interplay progressively. The second type of $n$-order interactions differ from the first type because the participation of units happen sequentially rather than simultaneously. In both types of high-order interactions, none of the involved units is dispensable. Therefore, we need to analyze both the faces of simplices (i.e., to ensure that all related units are involved) and the paths defined on these faces (i.e., to describe the sequential participation) to model the second type of $n$-order interactions.

The Laplacian operator corresponding to the second type of $n$-order interactions is referred to as the high-order path Laplacian operator, whose general form is defined as
\begin{align}
\left[\mathbf{L}_{H}^{\left(n\right)}\right]_{ij}=&\frac{1}{n}\big(\delta_{ij}\cdot\text{the number of paths that traverse all units on }\sigma^{\left(n\right)}\text{ without repetition and terminate at }v_{i}\notag\\&-\text{the number of paths that traverse all units on }\sigma^{\left(n\right)}\text{ without repetition, initiate at }v_{j}\text{, and terminate at }v_{i}\big).\label{A2EQ10}
\end{align}
Here a path can be also treated as a sequence of $1$-simplices. The factor $\frac{1}{n}$ is used to normalize the propagation speed of interactions along paths (i.e., there are $n$ end-to-end $1$-simplices in each path).

In Eq. (\ref{A2EQ10}), the number of paths that traverse all units on a $n$-simplex without repetition, initiate at unit $v_{j}$, and terminate at unit $v_{i}$ can be calculated as
\begin{align}
\mathbf{B}_{ij}^{\left(n\right)}= \left(1-\delta_{ij}\right)\sum_{\{i,j\}_{n+1}\subset V}\left(n-1\right)!I_{\sigma^{\left(n\right)}}\left(\{i,j\}_{n+1}\right),\label{A2EQ11}
\end{align}
where the coefficient $\left(n-1\right)!$ arises from the fact that the rest part of units in $\{i,j\}_{n+1}$ can form $\left(n-1\right)!$ kinds of unique sequences after we fix units $v_{i}$ and $v_{j}$ as the endpoints of these sequences. Based on simple derivations, we have
\begin{align}
\mathbf{B}_{ij}^{\left(n\right)} = \left(n-1\right)! \mathbf{A}_{ij}^{\left(n\right)} ,\label{A2EQ12}
\end{align}
which indicates the difference in freedom degrees between the first and the second type of $n$-order interactions. Specifically, if units $v_{i}$ and $v_{j}$ share a single $n$-order interaction of the first type, there are $\left(n-1\right)!$ possibilities for the second type of $n$-order interactions to occur and include units $v_{i}$ and $v_{j}$.

Another difference between the first and the second type of $n$-order interactions is that summing Eq. (\ref{A2EQ11}) over all possible units $v_{j}$ directly leads to the number of paths that traverse all units on a $n$-simplex without repetition and terminate at unit $v_{i}$, irrespective of whether these interactions are pairwise ($n=1$) or high-order ($n>1$)
\begin{align}
    \sum_{j}\mathbf{B}_{ij}^{\left(n\right)}&=\sum_{j}\left(1-\delta_{ij}\right)\sum_{\{i,j\}_{n+1}\subset V}\left(n-1\right)!I_{\sigma^{\left(n\right)}}\left(\{i,j\}_{n+1}\right),\label{A2EQ13}\\
&=\left(n-1\right)!\sum_{j\neq i}\sum_{\{i,j\}_{n+1}\subset V}I_{\sigma^{\left(n\right)}}\left(\{i,j\}_{n+1}\right),\label{A2EQ14}\\
&=\left(n-1\right)!\left[n\sum_{\{i\}_{n+1}\subset V}I_{\sigma^{\left(n\right)}}\left(\{i\}_{n+1}\right)\right],\label{A2EQ15}\\
&=n!\sum_{\{i\}_{n+1}\subset V}I_{\sigma^{\left(n\right)}}\left(\{i\}_{n+1}\right).\label{A2EQ16}
\end{align}
Eq. (\ref{A2EQ15}) is derived from Eq. (\ref{A2EQ14}) using Eqs. (\ref{A2EQ4}-\ref{A2EQ5}). We can discover that Eq. (\ref{A2EQ16}) equals to the number of paths that traverse all units on a $n$-simplex without repetition and terminate at unit $v_{i}$. This is because the rest part of units in $\{i\}_{n+1}$ can form $n!$ unique sequences if we fix unit $v_{i}$ as the endpoint. For convenience, we define $\mathbf{P}_{i}^{\left(n\right)}=n!\sum_{\{i\}_{n+1}\subset V}I_{\sigma^{\left(n\right)}}\left(\{i\}_{n+1}\right)$ as the number these paths and write Eq. (\ref{A2EQ17}) as
\begin{align}
   \left[\mathbf{L}_{H}^{\left(n\right)}\right]_{ij}=\frac{1}{n}\left( \delta_{ij}\mathbf{P}_{i}^{\left(n\right)}-\mathbf{B}_{ij}^{\left(n\right)}\right).\label{A2EQ17}
\end{align}
According to Eqs. (\ref{A2EQ5}-\ref{A2EQ6}) and Eq. (\ref{A2EQ16}), we can see that 
\begin{align}
\mathbf{P}_{i}^{\left(n\right)}=n!\mathbf{D}_{i}^{\left(n\right)}.\label{A2EQ18}
\end{align}

\paragraph{Differences between two kinds of high-order interactions}

Based on Eq. (\ref{A2EQ12}) and Eq. (\ref{A2EQ18}), the multi-order Laplacian operator and the high-order path Laplacian can be related by
\begin{align}
   \left[\mathbf{L}_{H}^{\left(n\right)}\right]_{ij}&=\frac{1}{n}\left( \delta_{ij}n!\mathbf{D}_{i}^{\left(n\right)}-\left(n-1\right)! \mathbf{A}_{ij}^{\left(n\right)} \right),\label{A2EQ19}\\
   &=\frac{\left(n-1\right)!}{n}\left[\mathbf{L}_{M}^{\left(n\right)}\right]_{ij}.\label{A2EQ20}
\end{align}
Please note that the factorial satisfies $0!=1$. Based on Eq. (\ref{A2EQ20}), we can see that the joint effects of the first type of $n$-order interactions (i.e., simultaneous participation) are no less than the second type of $n$-order interactions (i.e., sequential participation) only when $n\leq 2$. Specifically, we can consider the evolution of system dynamics mentioned in Sec. \ref{Sec2}. We assume that systems units follow the sequential participation
\begin{align}
  \mathbf{x}=&\exp\left(-\tau\mathbf{L}_{H}^{\left(n\right)}\right)\mathbf{x}_{0}.\label{A2EQ21}
\end{align}
We calculate $\mathbf{L}_{M}^{\left(n\right)}=\sum_{\lambda_{M}^{\left(n\right)}}\lambda_{M}^{\left(n\right)}\big\vert\lambda_{M}^{\left(n\right)}\big\rangle\big\langle\lambda_{M}^{\left(n\right)}\big\vert$, where each $\lambda_{M}^{\left(n\right)}$ is an eigenvalue of $\mathbf{L}_{M}^{\left(n\right)}$. Then, we can reformulate Eq. (\ref{A2EQ21}) as
\begin{align}
  \mathbf{x}&=\exp\left(-\tau\frac{\left(n-1\right)!}{n}\mathbf{L}_{M}^{\left(n\right)}\right)\mathbf{x}_{0},\label{A2EQ22}\\&=\left(\sum_{\lambda^{\left(n\right)}}\exp\left(-\tau\frac{\left(n-1\right)!}{n}\lambda_{M}^{\left(n\right)}\right)\big\vert\lambda_{M}^{\left(n\right)}\big\rangle\big\langle\lambda_{M}^{\left(n\right)}\big\vert\right)\mathbf{x}_{0}\label{A2EQ23}
\end{align}
using the property of matrix exponential. Based on Eq. (\ref{A2EQ23}), we can see that the decay rate of dynamic evolution, which is measured by $\tau\frac{\left(n-1\right)!}{n}\lambda_{M}^{\left(n\right)}$, depends on the type of $n$-order interactions and the actual value of $n$. When $n=1$ (i.e., pairwise interactions), the decay rate maintains the same no matter units follow the simultaneous or the sequential participation since $\frac{\left(n-1\right)!}{n}=1$. When $n=2$, the sequential participation leads to a smaller decay rate compared with the simultaneous participation since $\frac{\left(n-1\right)!}{n}<1$. When $n>2$, the sequential participation causes a larger decay rate since $\frac{\left(n-1\right)!}{n}>1$.

\section{Theoretical relations with the persistent homology theory and related concepts}\label{ASec3}
There are two closely related mathematical questions concerning the SRG: 
\begin{itemize}
    \item[(1) ] What are the relations among the multi-order Laplacian, the high-order path Laplacian, the combinatorial Laplacian \cite{ribando2022graph}, and the Hodge Laplacian \cite{lim2020hodge,schaub2020random}? 
    \item[(2) ] How is the SRG related to the persistent homology theory \cite{wang2020persistent}?
\end{itemize}
 Although these questions are not the essential parts of our theory, discussing them may help understand the SRG from other perspectives. Below, we sketch our main ideas and offer necessary references.

\paragraph{Similarities and differences among four kinds of Laplacian operators. } Before we summarize key similarities and differences among four kinds of Laplacian operators, we first present the basic definitions of the combinatorial Laplacian \cite{ribando2022graph} and the Hodge Laplacian \cite{lim2020hodge,schaub2020random}.

Generalized from the graph theory, a $q$-simplex also has its own adjacency relations within a simplicial complex, whose number is referred to as the degree of this $q$-simplex. The non-trivial part lies in that a $q$-simplex can be adjacent to both $\left(q-1\right)$-simplices (i.e., lower adjacency) and $\left(q+1\right)$-simplices (i.e., upper adjacency). Therefore, the degree of a $q$-simplex is the sum of its lower and upper degrees (i.e., the number of adjacent simplices of lower and higher orders) \cite{ribando2022graph,lim2020hodge}
\begin{align}
\operatorname{deg}\left(\sigma^{\left(q\right)}\right)&=\operatorname{deg}_{L}\left(\sigma^{\left(q\right)}\right)+\operatorname{deg}_{U}\left(\sigma^{\left(q\right)}\right),\label{A3EQ1}\\
&=q+1+\operatorname{deg}_{U}\left(\sigma^{\left(q\right)}\right).\label{A3EQ2}
\end{align}
Here we can derive Eq. (\ref{A3EQ2}) because $\operatorname{deg}_{L}\left(\sigma^{\left(q\right)}\right)$ measures the number of non-empty $\left(q-1\right)$-simplices in $K$ that are the faces of $\sigma^{\left(q\right)}$, which always equals $q+1$. 

To construct the combinatorial Laplacian \cite{ribando2022graph} and the Hodge Laplacian \cite{lim2020hodge,schaub2020random}, we need to define orderings on the units of simplex and simplicial complex \cite{ribando2022graph,lim2020hodge,schaub2020random}. Let us consider a $q$-simplex $\sigma^{\left(q\right)}=\left[v_{1},\ldots,v_{q+1}\right]$, an ordering of units $v_{1},\ldots,v_{q+1}$ defines an orientation of $\sigma^{\left(q\right)}$. We treat two orderings on $\left[v_{1},\ldots,v_{q+1}\right]$ are equivalent if and only if they differ from each other by an even permutation. Simplex $\sigma^{\left(q\right)}$ is oriented if an orientation is given. Apart from the orientation, another concept to define is the boundary operator, which is closely related to the chain complex in topology \cite{ribando2022graph,lim2020hodge,schaub2020random}. Let us consider the $q$-chain, a formal weighted sum of $q$-simplices in $K$ whose weight coefficients are selected from field $\mathbb{Z}_{2}$. A $q$-chain with a zero boundary is referred to as the $q$-cycle in topology. Given the addition operation in field $\mathbb{Z}_{2}$, the set of all $q$-chains form a chain group, $C^{\left(q\right)}\left(K\right)$. The boundary operator is a mapping between different chain groups, $\partial^{\left(q\right)}:C^{\left(q\right)}\left(K\right)\rightarrow C^{\left(q-1\right)}\left(K\right)$. Specifically, it maps a $q$-chain, a weighted sum of $q$-simplices, to a new sum of $\left(q-1\right)$-simplices with the same weights
\begin{align}
\partial^{\left(q\right)}\sigma^{\left(q\right)}=\sum_{i=1}^{q+1}\left(-1\right)^{i}f\left(\sigma^{\left(q\right)},i\right),\label{A3EQ3}
\end{align}
where $f\left(\cdot,\cdot\right)$ denotes a removing operation that creates a $\left(q-1\right)$-simplex from $\sigma^{\left(q\right)}$. Specifically, we let $f\left(\sigma^{\left(q\right)},i\right)=\left[v_{1},\ldots,v_{i-1}, v_{i+1},\ldots,v_{q+1}\right]$ be a $\left(q-1\right)$-simplex generated from $\sigma^{\left(q\right)}$ by omitting unit $v_{i}$. The chain complex mentioned above is a series of chain groups transmitted by boundary operators \cite{ribando2022graph,lim2020hodge,schaub2020random}
\begin{align}
\cdots\xrightarrow{\partial^{\left(q+2\right)}}C^{\left(q+1\right)}\left(K\right)\xrightarrow{\partial^{\left(q+1\right)}}C^{\left(q\right)}\left(K\right)\xrightarrow{\partial^{\left(q\right)}}C^{\left(q-1\right)}\left(K\right)\xrightarrow{\partial^{\left(q-1\right)}}\cdots\label{A3EQ4}
\end{align}
which plays an important role in the persistent homology theory \cite{wang2020persistent}. Given a $q$-boundary operator $\partial^{\left(q\right)}$, we can also define the corresponding $q$-adjoint boundary operator, $\partial^{\left(q,*\right)}:C^{\left(q-1\right)}\left(K\right)\rightarrow C^{\left(q\right)}\left(K\right)$. The above definitions can be directly applied on simplex, simplicial complex, and their oriented versions. Based on these definitions, the combinatorial Laplacian \cite{ribando2022graph} is given as
\begin{align}
\Delta^{\left(q\right)}_{C}=\partial^{\left(q+1\right)}\partial^{\left(q+1,*\right)}+\partial^{\left(q,*\right)}\partial^{\left(q\right)}.\label{A3EQ5}
\end{align}
The matrix representation of the combinatorial Laplacian, $\mathbf{L}^{\left(q\right)}_{C}$, when $q>0$ is \cite{ribando2022graph}
\begin{align}
\left[\mathbf{L}^{\left(q\right)}_{C}\right]_{ij}=\begin{cases}
\operatorname{deg}\left(\sigma^{\left(q\right)}_{i}\right), & i=j\\
1, & i\neq j\;\;\text{and}\;\;\sigma^{\left(q\right)}_{i}\overset{U}{\nsim}\sigma^{\left(q\right)}_{j}\;\;\text{and}\;\;\sigma^{\left(q\right)}_{i}\overset{L}{\sim}\sigma^{\left(q\right)}_{j}\;\;\text{with similar orientation}\\
-1, & i\neq j\;\;\text{and}\;\;\sigma^{\left(q\right)}_{i}\overset{U}{\nsim}\sigma^{\left(q\right)}_{j}\;\;\text{and}\;\;\sigma^{\left(q\right)}_{i}\overset{L}{\sim}\sigma^{\left(q\right)}_{j}\;\;\text{with dissimilar orientations}\\
0, & i\neq j\;\;\text{either}\;\;\sigma^{\left(q\right)}_{i}\overset{U}{\sim}\sigma^{\left(q\right)}_{j}\;\;\text{or}\;\;\sigma^{\left(q\right)}_{i}\overset{L}{\nsim}\sigma^{\left(q\right)}_{j}
\end{cases}.\label{A3EQ6}
\end{align}
When $q=0$, the combinatorial Laplacian matrix reduces to the classic Laplacian matrix \cite{gross2018graph}
\begin{align}
\left[\mathbf{L}^{\left(q\right)}_{C}\right]_{ij}=\begin{cases}
\operatorname{deg}\left(\sigma^{\left(q\right)}_{i}\right), & i=j\\
1, & \sigma^{\left(q\right)}_{i}\overset{U}{\sim}\sigma^{\left(q\right)}_{j}\\
-1, & \text{otherwise}
\end{cases}.\label{A3EQ7}
\end{align}
In Eqs. (\ref{A3EQ6}-\ref{A3EQ7}), we use $\overset{U}{\sim}$ to denote the upper adjacent relation, i.e.,  two $q$-simplices are incident to the same $\left(q+1\right)$-simplex, and $\overset{L}{\sim}$ stands for the lower adjacent relation, i.e., two $q$-simplices are incident to the same $\left(q-1\right)$-simplex. If two $q$-simplices are upper adjacent (e.g., incident to simplex $\sigma^{\left(q+1\right)}$), we say they have similar orientations if they share the same sign in $\partial^{\left(q+1\right)}\sigma^{\left(q+1\right)}$ and they have dissimilar orientations if their signs are opposite.

The derivation of the Hodge Laplacian requires more knowledge about the Rham-Hodge theory \cite{lim2020hodge,schaub2020random}, which will not be elaborated here due to its theoretical complexity. We refer to Ref. \cite{lim2020hodge} for more details. In physics and applied mathematics, the discretized Hodge Laplacian is more frequent to see, which is defined as \cite{ribando2022graph}
\begin{align}
\mathbf{L}^{\left(q\right)}_{D}=\left[\mathbf{Z}^{\left(q\right)}\right]^{\top}\mathbf{S}^{\left(q+1\right)}\mathbf{Z}^{\left(q\right)}+\mathbf{S}^{\left(q\right)}\mathbf{Z}^{\left(q-1\right)}\left[\mathbf{S}^{\left(q-1\right)}\right]^{-1}\left[\mathbf{Z}^{\left(q-1\right)}\right]^{\top}\mathbf{S}^{\left(q\right)}.\label{A3EQ8}
\end{align}
In Eq. (\ref{A3EQ8}), matrix $\mathbf{Z}^{\left(q\right)}$ is defined as $\mathbf{Z}^{\left(q\right)}=\left[\mathbf{F}^{\left(q+1\right)}\right]^{\top}$, where boundary matrix, $\mathbf{F}^{\left(q+1\right)}$, and its transpose, $\left[\mathbf{F}^{\left(q+1\right)}\right]^{\top}$, are the discrete counterparts of $\partial^{\left(q+1\right)}$ and $\partial^{\left(q+1,*\right)}$ such that the combinatorial Laplacian matrix in Eqs. (\ref{A3EQ6}-\ref{A3EQ7}) can be expressed as $\mathbf{L}^{\left(q\right)}_{C}=\mathbf{F}^{\left(q+1\right)}\left[\mathbf{F}^{\left(q+1\right)}\right]^{\top}+\left[\mathbf{F}^{\left(q\right)}\right]^{\top}\mathbf{F}^{\left(q\right)}$. Matrix $\mathbf{S}^{\left(q\right)}$ is a diagonal matrix that discretizes the Hodge star \cite{ribando2022graph} on simplicial complexes. The Hodge star is an operator that transforms $q$-forms to $\left(h-q\right)$-forms in differential geometry theories \cite{ribando2022graph}. Meanwhile, the original Hodge Laplacian, $\Delta^{\left(q\right)}_{D}$, can be approximated by a discrete counterpart, $\left[\mathbf{S}^{\left(q\right)}\right]^{-1}\mathbf{L}^{\left(q\right)}_{C}$ \cite{ribando2022graph}.

Although the (discretized) Hodge Laplacian and the combinatorial Laplacian seem to share multiple similarities, they are not necessarily equivalent in mathematics. On most types of simplicial complexes, the spectra of these two Laplacian operators significantly differ from each other, which may arise from the fact that the combinatorial Laplacian is completely determined by network connectivity and has no geometrical constraint \cite{ribando2022graph}. Here we emphasize a common property shared by the (discretized) Hodge Laplacian and the combinatorial Laplacian, which is also one of the intrinsic differences between these two Laplacian operators and our proposed Laplacian operators (i.e., the multi-order Laplacian the high-order path Laplacian in the SRG). This difference lies in the following aspects:
\begin{itemize}
    \item[(1) ]The $\left(i,j\right)$-th element of the combinatorial Laplacian, $\mathbf{L}^{\left(q\right)}_{C}$, and the (discretized) Hodge Laplacian, $\mathbf{L}^{\left(q\right)}_{D}$, reflects the relations between simplex $\sigma^{\left(q\right)}_{i}$ and simplex $\sigma^{\left(q\right)}_{j}$ \cite{ribando2022graph,lim2020hodge,schaub2020random}, which can involve with multiple system units. The dimensions of these two operators are determined by the number of simplices, which can differ from the number of system units. Because a $q$-simplex denotes the un-decomposable high-order interaction among $q+1$ units, each element of the (discretized) Hodge Laplacian and the combinatorial Laplacian actually describes the relations between two high-order interactions. Matrix $\mathbf{L}^{\left(q\right)}_{C}$ reduces to the classic Laplacian in graph theory when $q=0$ because each system unit is also a $0$-simplex. On most kinds of simplicial complexes, matrix $\mathbf{L}^{\left(q\right)}_{D}$ is not same as the classic Laplacian even when $q=0$.
    \item[(2) ]The $\left(i,j\right)$-th element of the multi-order Laplacian, $\mathbf{L}^{\left(q\right)}_{M}$, and the high-order path Laplacian, $\mathbf{L}^{\left(q\right)}_{H}$, describes the $q$-order interactions between unit $i$ and unit $j$. In other words, these operators characterize the un-decomposable interactions among system units on different orders. Therefore, the dimensions of these two operators are determined by the number of system units. When $q=1$, matrices $\mathbf{L}^{\left(q\right)}_{M}$ and $\mathbf{L}^{\left(q\right)}_{H}$ reduce to the classic graph Laplacian because $1$-simplex is same as edge in networks.
\end{itemize}
Consequently, if we establish a renormalization group (RG) framework on the (discretized) Hodge Laplacian or the combinatorial Laplacian, the targets to renormalize are high-order interactions and the criteria used to guide coarse graining processes are the relations among high-order interactions. When we define a RG framework on the multi-order Laplacian or the high-order path Laplacian, as described in the main text, the objects to renormalize are system units and the criteria of renormalization are determined by high-order correlations among units. Please note that the difference between these two perspectives is neither superior nor inferior to each other. We follow the second perspective in the SRG theory only because it is normal to choose system units as renormalization objects in practice.

The similarity shared by the combinatorial Laplacian and the (discretized) Hodge Laplacian is that the multiplicity of their zero eigenvalues equals the Betti number, i.e., the number of $q$-dimensional holes \cite{ribando2022graph,lim2020hodge,schaub2020random}. For the multi-order Laplacian and the high-order path Laplacian, the multiplicity of their zero eigenvalues equals the number of connected components in the high-order network sketches of the system. Here a $q$-order network sketch of the system is a network where two units are connected if they share at least one $q$-order interaction, which is useful in visualizing high-order systems of large sizes.

\paragraph{Relations with the persistent homology theory. } A common objective of the persistent homology theory is to identify specific topological properties of simplicial complexes that have long lifetime when we add or remove simplices from the system \cite{wang2020persistent,edelsbrunner2008persistent,aktas2019persistence,horak2009persistent}. Although the multi-order Laplacian and the high-order path Laplacian have not been used in the persistent homology theory \cite{wang2020persistent,edelsbrunner2008persistent,aktas2019persistence,horak2009persistent} yet, the combinatorial Laplacian and the (discretized) Hodge Laplacian have been extensively included in persistent homology analysis \cite{xia2015multiscale,bramer2018multiscale,wang2020persistent,davies2023persistent,memoli2022persistent,wang2023persistent}. Below, we suggest a possible idea to relate the SRG with the persistent homology theory.

Our idea arises from a simple fact. No matter we implement a SRG framework with the multi-order Laplacian or the high-order path Laplacian, when this SRG renormalizes the system on a $p$-order according to the properties of $q$-order interactions ($p\leq q$), we can always record the set of units in $\mathbf{G}^{\left(p\right)}_{k}$ that are aggregated into each macro-unit in $\mathbf{G}^{\left(p\right)}_{k+1}$. By repeating this recording in every iteration, we can recursively find the set of initial units in $\mathbf{G}^{\left(p\right)}_{1}$ aggregated into each macro-unit in $\mathbf{G}^{\left(p\right)}_{k+1}$ for each $k\geq 1$. In other words, we can know how the initial structure formed by system units is changed by the SRG after $k$ iterations. If we set $p=1$ and search simplices in the initial system, $\mathbf{G}^{\left(1\right)}_{1}$, we can use the recursively tracked information to identify how many simplices in $\mathbf{G}^{\left(1\right)}_{1}$ are added, removed, or modified by the SRG after $k$ iterations. Therefore, similar to the filtration process in the persistent homology theory \cite{wang2020persistent,horak2009persistent}, the SRG also offers an opportunity to study the lifetime of different topological properties (e.g., the Betti number). Given a simplicial complex $K$, the SRG also creates a filtration sequence of sub-complexes $\left(K_{k}\right)_{k=0}^{m}$ (here $m$ denotes the number of iterations in the SRG). Certainly, the SRG is different from the common filtration processes that persistently add or remove simplices following deterministic rules \cite{wang2020persistent,horak2009persistent}. It has more intricate effects on the topological properties of the system.

\section{The SRG in a form of conventional renormalization group theories}\label{ASec4}
In this section, we contextualize the SRG in the background of conventional renormalization group (RG) theories.

Before presenting a more in-depth analysis, we first review the Gaussian model used in complex system analysis (e.g., see Refs. \cite{tuncer2015spectral,bradde2017pca} for instances). Let us consider a system state, $\mathbf{x}$, whose component associated with each system unit $i$ is denoted as $\mathbf{x}\left(i\right)$. For convenience, we assume that the system size is infinitely large and satisfies continuity. For an interacting system, the conventional Gaussian model can be established by omitting the fourth and higher coupling terms of the Landau model, whose functional Hamiltonian can be expressed as \cite{tuncer2015spectral,bradde2017pca}
\begin{align}
\mathcal{H}=\int\left[\frac{1}{2}\left(\vert \nabla \mathbf{x}\left(i\right)\vert^{2}+r\mathbf{x}\left(i\right)^{2}\right)-J\mathbf{x}\left(i\right)\right]\mathsf{d}i\label{A4EQ1},
\end{align}
where $r$ and $J$ denote some parameters concerning the temperature and the external field. We assume that the system described in Eq. (\ref{A4EQ1}) is distributed on a network whose adjacency matrix is $\mathbf{A}$. Given this assumption, we can rewrite Eq. (\ref{A4EQ1}) as 
\begin{align}
\mathcal{H}&\simeq \frac{1}{2}\sum_{i,j}\left[\frac{1}{2}\mathbf{A}_{ij}\left(\mathbf{x}\left(i\right)-\mathbf{x}\left(j\right)\right)^{2}+r\mathbf{x}\left(i\right)^{2}\delta_{ij}\right]-J\sum_{i}\mathbf{x}\left(i\right)\label{A4EQ2}.
\end{align}
We notice that Eq. (\ref{A4EQ2}) shares similar terms with the Laplacian dynamics analyzed in one of our earlier works \cite{tian2023laplacian}, which inspires us to reformulate Eq. (\ref{A4EQ2}) into
\begin{align}
    \mathcal{H}&\simeq \frac{1}{2}\sum_{i,j}\Bigg[\mathbf{x}\left(i\right)\underbrace{\Bigg(\delta_{i,j}\sum_{j}\mathbf{A}_{ij}-\mathbf{A}_{ij}\Bigg)}_{\text{A Laplacian operator}}\mathbf{x}\left(j\right)+r\mathbf{x}\left(i\right)^{2}\delta_{ij}\Bigg]-J\sum_{i}\mathbf{x}\left(i\right)\label{A4EQ3}.
\end{align}
According to Eq. (\ref{A4EQ2}), we can naturally obtain a Gaussian model governed by the Laplacian operator if we place the system on a network. In fact, this same idea has been proposed and validated in existing studies extensively (e.g., see Refs. \cite{tuncer2015spectral,burioni2005random} for instances).

Now, let us relate the above analysis with the SRG framework. We can consider high-order interactions and insert the multi-order Laplacian, $\mathbf{L}^{\left(q\right)}_{M}$, or the high-order path Laplacian, $\mathbf{L}^{\left(q\right)}_{H}$, into Eq. (\ref{A4EQ3}) to derive 
\begin{align}
    \mathcal{H}^{\left(q\right)}&\simeq \frac{1}{2}\sum_{i,j}\Bigg[\mathbf{x}\left(i\right)\left[\mathbf{L}^{\left(q\right)}\right]_{ij}\mathbf{x}\left(j\right)+r\mathbf{x}\left(i\right)^{2}\delta_{ij}\Bigg]-J\sum_{i}\mathbf{x}\left(i\right)\label{A4EQ4},\\
    &\simeq \frac{1}{2}\sum_{i,j}\Bigg[\mathbf{x}\left(i\right)\left(\left[\mathbf{L}^{\left(q\right)}\right]_{ij}+r\delta_{ij}\right)\mathbf{x}\left(j\right)\Bigg]-J\sum_{i}\mathbf{x}\left(i\right)\label{A4EQ5},
\end{align}
where $\left[\mathbf{L}^{\left(q\right)}\right]_{ij}$ can be selected as $\left[\mathbf{L}^{\left(q\right)}_{M}\right]_{ij}$ or $\left[\mathbf{L}^{\left(q\right)}_{H}\right]_{ij}$. We refer to $\mathcal{H}^{\left(q\right)}$ as the $q$-order functional Hamiltonian, which characterizes the system behaviours on the $q$-order. A more strict form of $\mathcal{H}^{\left(q\right)}$ is shown in Eq. (\ref{A4EQ6}), where we indicate that the linear sum over all possible pairs of $\left(i,j\right)$ in the first term of Eq. (\ref{A4EQ5}) is not really a linear combination of independent items.

If we write $\mathbf{K}_{ij}^{\left(q\right)}=\left[\mathbf{L}^{\left(q\right)}\right]_{ij}+r\delta_{ij}$, we notice that $\mathbf{K}_{ij}^{\left(q\right)}$ is essentially equivalent to the term of $\phi^{4}$ theory analyzed in Ref. \cite{bradde2017pca}. The main difference lies in that the PCA-like RG proposed by Ref. \cite{bradde2017pca} does not study high-order interactions (the key difference created by introducing high-order interactions is highlighted by Eq. (\ref{A4EQ6}) in our subsequent analysis). Given this close relation, we can have following ideas after learning Ref. \cite{bradde2017pca}:
\begin{itemize}
    \item[(1) ]The operator $\mathbf{L}^{\left(q\right)}$ used in the SRG mainly serves as a foundation for determining relevant scales based on its spectrum, whose role is similar to the precision matrix (i.e., the inverse of covariance matrix) studied in the PCA-like RG under the condition of translation invariance \cite{bradde2017pca}. To certain extents, integrating out interactions associated with large eigenvalues of $\mathbf{L}^{\left(q\right)}$ in the SRG shares the same idea with eliminating the modes with small variances in the PCA-like RG \cite{bradde2017pca}. This validates our idea of designing the SRG based on the spectrum of $\mathbf{L}^{\left(q\right)}$. Apart from the relation with the PCA-like RG, the usage of the spectrum of $\mathbf{L}^{\left(q\right)}$ is also similar with the ideas of functional RG theories, where a regulator is used to govern the renormalization \cite{kopietz2010introduction,polonyi2003lectures,dupuis2021nonperturbative}. Then main difference between the SRG and functional RGs lies in that there is no explicit functional or related formalism given in the SRG because it is hard to explicitly derive these concepts for complex systems.
    \item[(2) ]The non-trivial point is how to define an appropriate cutoff of the spectrum \cite{bradde2017pca}. Different from a flexible and empirically selected cutoff in the PCA-like RG \cite{bradde2017pca}, we generalize the idea of Ref. \cite{villegas2023laplacian} to define the specific heat in both ergodic and non-ergodic cases and search for a cutoff based on the maxima of specific heat. We choose the specific heat because it is demonstrated as an indicator of second-order phase transitions in diffusion processes and is informative about characteristic diffusion scales \cite{villegas2022laplacian,villegas2023laplacian}. Meanwhile, as suggested in Ref. \cite{villegas2022laplacian}, the specific heat is proportional to the amplitudes of entropy fluctuations if we consider the thermal fluctuation-dissipation theorem \cite{christensen2005complexity}. Moreover, the specific heat can be directly calculated using $\mathbf{L}^{\left(q\right)}$ since diffusion on networks is fully characterized by the Laplacian operator. In sum, the specific heat is a valid and convenient criterion to determine a cutoff. Certainly, one can consider other criteria in practice (e.g., the critical slowing-down of order parameter fluctuations and the ensuing time scale separation \cite{christensen2005complexity}).
    \item[(3) ]By considering a Gaussian model, we omit the fourth and higher coupling terms among units \cite{tuncer2015spectral,bradde2017pca}. Why can we still obtain non-trivial results under this condition? Below, we present possible explanations:
    \begin{itemize}
        \item[(a) ]As shown by the perturbation analysis of the PCA-like RG \cite{bradde2017pca}, the behaviours of an interacting system during renormalization are not necessarily governed by the central limit theorem even though the coarse graining procedure only considers Gaussian fluctuations. Instead, more non-trivial behaviours are possible to occur during renormalization. This property makes the PCA-like RG as well as our framework be applicable to more complex systems and identify non-Gaussian behaviours \cite{bradde2017pca}. For instance, the original probability distribution to renormalize in diverse complex systems can be non-Gaussian (although the explicit theoretical expression of this initial probability distribution may be unknown to us). Meanwhile, the departures from a Gaussian distribution can become either more or less important under the scale transformation by the RG. These conditions are all possible to create non-trivial phenomena.
        \item[(b) ]Most importantly, perhaps a frequently neglected point is that $\mathbf{K}^{\left(q\right)}$ can be defined with $q>1$ (in fact, this is the most common case in the SRG). When $q>1$, the term $\mathbf{x}\left(i\right)\mathbf{K}^{\left(q\right)}_{ij}\mathbf{x}\left(j\right)$ can not be treated as the conventional pairwise relation between units $v_{i}$ and $v_{j}$ that only depends on these two units due to the following two reasons:
        \begin{itemize}
            \item[(b1) ]First, extra units are involved in the relation between units $v_{i}$ and $v_{j}$ when $q>1$. Because the system is analyzed on the $q$-order, this relation between units $v_{i}$ and $v_{j}$ implicitly require the participation of $q+1$ units in total (i.e., there are $q-1$ extra units required by this relation). If there is no enough unit satisfying this requirement, it is impossible for units $v_{i}$ and $v_{j}$ to share a $q$-order interaction since the associated $q$-simplex can not be formed. Therefore, although $\mathbf{x}\left(i\right)\mathbf{K}^{\left(q\right)}_{ij}\mathbf{x}\left(j\right)$ seems to have a dyadic form, it is not equivalent to the classic pairwise relation that only involves with units $i$ and $j$.
            \item[(b2) ]Second, the off-diagonal elements of $\mathbf{K}^{\left(q\right)}$ can be non-linearly coupled with each other when $q>1$. 
            \begin{itemize}
                \item[(b2-1) ]Before further analysis, we recall that a conventional linear combination of multiple items, such as $w=x+y+z$, usually requires the combined items to be independent from each other. For instance, if we reduce $x$ by $k$, we are expected to see that $w$ is reduced by $k$ as a direct consequence of our action and other items (e.g., $y$ and $z$) remain unaffected. In an opposite case where some items are coupled with each other, we can not treat $x+y+z$ as a conventional linear combination. For instance, we assume that reducing $x$ by $k$ will also reduce $y$ by $\frac{x-k}{x+z/k}$ since $x$, $y$, and $z$ are coupled by certain non-linear mechanisms. In this situation, the action of reducing $x$ by $k$ does not directly leads us to $w-k=\left(x-k\right)+y+z$. Instead, it triggers a series of subsequent changes and we need to consider all these changes to ensure that the equality still holds. 
                \item[(b2-2) ]We recall this basic property because Eq. (\ref{A4EQ5}) includes a sum over non-linearly coupled items. Let us consider a simple instance with $q=3$. We assume that there exists a $3$-simplex formed by units $a$, $b$, $c$, $d$ in the system (i.e., these units share $3$-order interactions). If we change $\mathbf{K}^{\left(3\right)}_{ab}$ by deleting the interaction between units $v_{a}$ and $v_{b}$ on $1$-order, we also affect the values of $\mathbf{K}^{\left(3\right)}_{ac}$, $\mathbf{K}^{\left(3\right)}_{ad}$, $\mathbf{K}^{\left(3\right)}_{bc}$, $\mathbf{K}^{\left(3\right)}_{bd}$, and $\mathbf{K}^{\left(3\right)}_{cd}$ because the original $3$-simplex inevitably collapses after removing the interaction between units $v_{a}$ and $v_{b}$ (i.e., the definition of clique simplex shown in Appendix \ref{ASec1} is no longer satisfied). In fact, these changes will propagate over units to modify other $3$-simplices and higher-order simplices until all the simplices involving units $v_{a}$ and $v_{b}$ are changed. These cascade changes after a single modification create non-linear dependencies among the summed items in Eq. (\ref{A4EQ5}). Meanwhile, they make the analyzed freedom degrees discontinuous (i.e., since some freedom degrees appear and disappear together). To highlight these dependencies, we introduce a more formal definition of $\mathcal{H}^{\left(q\right)}$ 

            \begin{align}
    \mathcal{H}^{\left(q\right)}&\simeq \frac{1}{2}\sum_{\sigma^{\left(q\right)}\in K}\sum_{i,j\in\sigma^{\left(q\right)}}h\left(\sigma^{\left(q\right)}/\{i,j\},q\right)\Bigg[\mathbf{x}\left(i\right)\mathbf{K}^{\left(q\right)}_{ij}\mathbf{x}\left(j\right)\Bigg]-J\sum_{i}\mathbf{x}\left(i\right),\label{A4EQ6}
\end{align}
where we treat the system as a simplicial complex, $K$. The first sum goes through every $q$-simplex, $\sigma^{\left(q\right)}$, in the system and the second sum goes through each pair of units, $v_{i}$ and $v_{j}$, in this $q$-simplex. We denote $\sigma^{\left(q\right)}/\{i,j\}$ as the set of all other units in this $q$-simplex apart from units $v_{i}$ and $v_{j}$. The non-linear map $h\left(\cdot,q\right)$ is referred to as the response function on the $q$-order, which checks through each pair of units in $\sigma^{\left(q\right)}/\{i,j\}$ to identify potential changes. We define $h\left(\sigma^{\left(q\right)}/\{i,j\},q\right)=1$ if and only if there is no change on $\sigma^{\left(q\right)}/\{i,j\}$. If there is any change on any pair of units $v_{k}$ and $v_{l}$ in $\sigma^{\left(q\right)}/\{i,j\}$ (e.g., their interactions are removed on any $p$-order and $p\leq q$), we use $h\left(\sigma^{\left(q\right)}/\{i,j\},q\right)\in\left[0,\infty\right)/\{1\}$ to reflect the cascade effects of this change on $\sigma^{\left(q\right)}$. In a special case where units $v_{i}$ and $v_{j}$ only participate in a single $q$-order interaction (i.e., the one described by $\sigma^{\left(q\right)}$) and at least one pair of units in $\sigma^{\left(q\right)}/\{i,j\}$ become decoupled on the $p$-order ($p\leq q$), we have $h\left(\sigma^{\left(q\right)}/\{i,j\},q\right)=0$ because the only $q$-simplex containing $v_{i}$ and $v_{j}$ has collapsed. Unfortunately, we find it difficult to derive an explicit and universal expression of $h\left(\cdot,q\right)$. This function is highly structure-dependent (i.e., determined by the actual properties of $K$). Defining the response function $h\left(\cdot,q\right)$ on representative systems remains as a meaningful question for future studies, which may suggest a new perspective to study the cascade effects of perturbations using topology.
\item[(b2-3) ] In brief, there are numerous non-linear coupling relations between $\mathbf{x}\left(i\right)\mathbf{K}^{\left(q\right)}_{ij}\mathbf{x}\left(j\right)$ for different pairs of $\left(i,j\right)$. In a sense, these coupling relations are the reflections of the un-decomposable properties of high-order interactions. Although we can write $\mathcal{H}^{\left(q\right)}$ in the form of Eq. (\ref{A4EQ5}) to relate it with conventional renormalization group theories, we can not neglect these coupling relations that enable cascade effects to happen over a large region and distinguish the SRG from classic frameworks. In Eq. (\ref{A4EQ6}), we have shown that these un-decomposable properties are implicitly reflected by $\mathcal{H}^{\left(q\right)}$, where analyzing the relation between units $i$ and $j$ in a $q$-simplex, $\mathbf{x}\left(i\right)\mathbf{K}^{\left(q\right)}_{ij}\mathbf{x}\left(j\right)$, also requires us to check the statuses of other units in $\sigma^{\left(q\right)}/\{i,j\}$. In other words, the $q+1$ units contained in $\sigma^{\left(q\right)}$ can not be studied independently. This property is consistent with conventional definitions of high-order interactions based on simplicial complexes \cite{battiston2021physics}. 
            \end{itemize} 
        \end{itemize}
        \item[(c) ]Another frequently neglected point is that $\mathcal{H}^{\left(q\right)}$ is not the actual Hamiltonian of the interacting system under study. Instead, $\mathcal{H}^{\left(q\right)}$ is a representational Hamiltonian obtained by placing the interacting system on $q$-order alone. In real cases, diverse orders of interactions can coexist in an interacting system to shape its dynamics, which can not be completely described by $\mathcal{H}^{\left(q\right)}$. The actual Hamiltonian of system dynamics may explicitly consist of different polynomial terms to describe high-order interactions, which can be seen in existing frameworks of high-order interactions in statistical and quantum field theories (e.g., $q$-spin interactions \cite{pachos2004three,muller2011simulating,bermudez2009competing}) and synchronization models (e.g., $q$-oscillator couplings \cite{skardal2020higher,kundu2022higher,kovalenko2021contrarians}). In these frameworks, a $q$-order polynomial term is explicitly proposed to describe $q$-unit interactions (e.g., $3$-unit interactions are characterized by $g\left(i,j,k\right)\mathbf{x}\left(i\right)\mathbf{x}\left(j\right)\mathbf{x}\left(k\right)$, where $g\left(i,j,k\right)$ reflects $3$-unit correlations). In the SRG, we use $\mathcal{H}^{\left(q\right)}$ to guide the renormalization of $\mathcal{H}^{\left(p\right)}$ ($p\leq q$). During the renormalization process, we only deal with the properties of the interacting system on two orders, yet there can exist other orders that affect the renormalization (e.g., as the most simple instance, if $p<w<q$, the renormalization of $p$-order interactions guided by $q$-order interactions is affected by $w$-order interactions as well). Therefore, there exist numerous latent factors to provide possibilities for non-trivial behaviours to occur during renormalization.
    \end{itemize}
\end{itemize}

\section{The behaviour of the specific heat}\label{ASec5}
As mentioned in our main text and explained in Appendix \ref{ASec4}, the specific heat is used to define the cutoff of the spectrum of operator $\mathbf{L}^{\left(q\right)}$, which helps determine relevant scales in the system. Specifically, the $q$-order specific heat in Eq. (\ref{EQ25}), derived using the first derivative of the $q$-order spectral entropy, is used to define the optimal time scale, $\widehat{\tau}^{\left(q\right)}$, of the SRG in Eq. (\ref{EQ26}). Then, the inverse of time scale $\widehat{\tau}^{\left(q\right)}$ is utilized to help define the number of characteristic modes for describing the system near criticality (see step (4) in the SRG).

In this section, we mainly show how to derive the specific heat, $X^{\left(q\right)}_{1}\left(\tau^{\left(q\right)}\right)$, in Eq. (\ref{EQ25}) from Eq. (\ref{EQ24}) when $\tau^{\left(q\right)}$ is sufficiently large. Formally, we can solve the differential equation, $X^{\left(q\right)}_{1}\left(\tau^{\left(q\right)}\right)=-\frac{\mathsf{d}}{\mathsf{d}\log\left(\tau^{\left(q\right)}\right)}S^{\left(q\right)}_{1}\left(\tau^{\left(q\right)}\right)$, in Eq. (\ref{EQ24}) as
\begin{align}
X^{\left(q\right)}_{1}\left(\tau^{\left(q\right)}\right)=
-\tau^{\left(q\right)}\big\langle\lambda^{\left(q\right)}_{1}\big\rangle_{\rho}-\left(\tau^{\left(q\right)}\right)^{2}\frac{\mathsf{d}}{\mathsf{d}\tau^{\left(q\right)}}\big\langle\lambda^{\left(q\right)}_{1}\big\rangle_{\rho}-\tau^{\left(q\right)}\frac{\mathsf{d}}{\mathsf{d}\tau^{\left(q\right)}}\log \left(\sum_{\lambda^{\left(q\right)}_{1}}\exp\left(-\tau^{\left(q\right)}\lambda^{\left(q\right)}_{1}\right)\right),\label{A5EQ1}
\end{align}
where the third term can be reformulated as
\begin{align}
-\tau^{\left(q\right)}\frac{\mathsf{d}}{\mathsf{d}\tau^{\left(q\right)}}\log \left(\sum_{\lambda^{\left(q\right)}_{1}}\exp\left(-\tau^{\left(q\right)}\lambda^{\left(q\right)}_{1}\right)\right)=
&-\tau^{\left(q\right)}\frac{1}{\sum_{\lambda^{\left(q\right)}_{1}}\exp\left(-\tau^{\left(q\right)}\lambda^{\left(q\right)}_{1}\right)}\frac{\mathsf{d}}{\mathsf{d}\tau^{\left(q\right)}}\sum_{\lambda^{\left(q\right)}_{1}}\exp\left(-\tau^{\left(q\right)}\lambda^{\left(q\right)}_{1}\right),\label{A5EQ2}\\
=&\tau^{\left(q\right)}\frac{\sum_{\lambda^{\left(q\right)}_{1}}\lambda^{\left(q\right)}_{1}\exp\left(-\tau^{\left(q\right)}\lambda^{\left(q\right)}_{1}\right)}{\sum_{\lambda^{\left(q\right)}_{1}}\exp\left(-\tau^{\left(q\right)}\lambda^{\left(q\right)}_{1}\right)},\label{A5EQ3}\\
=& \tau^{\left(q\right)}\big\langle\lambda^{\left(q\right)}_{1}\big\rangle_{\rho}.\label{A5EQ4}
\end{align}

Inserting Eq. (\ref{A5EQ4}) into Eq. (\ref{A5EQ1}), we can readily derive
\begin{align}
X^{\left(q\right)}_{1}\left(\tau^{\left(q\right)}\right)&=-\left(\tau^{\left(q\right)}\right)^{2}\frac{\mathsf{d}}{\mathsf{d}\tau^{\left(q\right)}}\big\langle\lambda^{\left(q\right)}_{1}\big\rangle_{\rho}
,\label{A5EQ5}
\end{align}
which is the specific heat shown in our main text.

\section{Bias correction in time scale calculation}\label{ASec6}
To understand the bias of time scale calculation, we need to recall the derivation process of $\widetilde{\tau}^{\left(q\right)}$, a time scale under the infinite deceleration condition
\begin{align}
\frac{\mathsf{d}}{\mathsf{d}\tau^{\left(q\right)}}X^{\left(q\right)}_{1}\left(\tau^{\left(q\right)}\right)\Big\vert_{\widetilde{\tau}^{\left(q\right)}}&=0,\label{A6EQ1}\\
\frac{\mathsf{d}}{\mathsf{d}\tau^{\left(q\right)}}\left[-\left(\tau^{\left(q\right)}\right)^{2}\frac{\mathsf{d}}{\mathsf{d}\tau^{\left(q\right)}}\big\langle\lambda^{\left(q\right)}_{1}\big\rangle_{\rho}\right]\bigg\vert_{\widetilde{\tau}^{\left(q\right)}}&=0,\label{A6EQ2}\\
\frac{\mathsf{d}}{\mathsf{d}\tau^{\left(q\right)}}\left[-\left(\tau^{\left(q\right)}\right)^{2}\frac{\mathsf{d}}{\mathsf{d}\tau^{\left(q\right)}}\operatorname{tr}\left(\mathbf{L}^{\left(q\right)}_{1}\rho^{\left(q\right)}_{1}\right)\right]\bigg\vert_{\widetilde{\tau}^{\left(q\right)}}&=0,\label{A6EQ3}
\end{align}
where each $\lambda^{\left(q\right)}_{1}$ is an eigenvalue of $\mathbf{L}^{\left(q\right)}_{1}$. Eq. (\ref{A6EQ2}) is directly obtained from Eq. (\ref{EQ25}). As shown by Eqs. (\ref{A6EQ2}-\ref{A6EQ3}), the selection of $\widetilde{\tau}^{\left(q\right)}$ is fully determined by the Laplacian operator $\mathbf{L}^{\left(q\right)}_{1}$, its eigenvalue spectrum, and its associated density operator $\rho^{\left(q\right)}_{1}$. 

Similar to what we have discussed in Sec. \ref{Sec4} (i.e., see step (3) in the SRG), the effectiveness of diffusion may not necessarily be reflected by Eqs. (\ref{A6EQ2}-\ref{A6EQ3}). Specifically, in the density operator $\rho^{\left(q\right)}_{1}$, there exist numerous possibilities for its $\left(i,j\right)$-th element to be smaller than both the $\left(i,i\right)$-th and the $\left(j,j\right)$-th elements. In these cases, the $q$-order interactions between $v_{i}$ and $v_{j}$ are less stronger than the $q$-order self-interactions of these two units. Given a time scale $\tau^{\left(q\right)}$, the accumulations of the $q$-order interactions between $v_{i}$ and $v_{j}$ are less important than the accumulations of the $q$-order self-interactions in determining the behaviours of $v_{i}$ and $v_{j}$. Therefore, we should treat the diffusion process from $v_{i}$ to $v_{j}$ (or vice versa) as less effective. Eqs. (\ref{A6EQ2}-\ref{A6EQ3}) may neglect these possibilities because they use $\rho^{\left(q\right)}_{1}$ without further modification. This property leads to a bias in time scale calculation.

\begin{figure*}[!t]
\includegraphics[width=1\columnwidth]{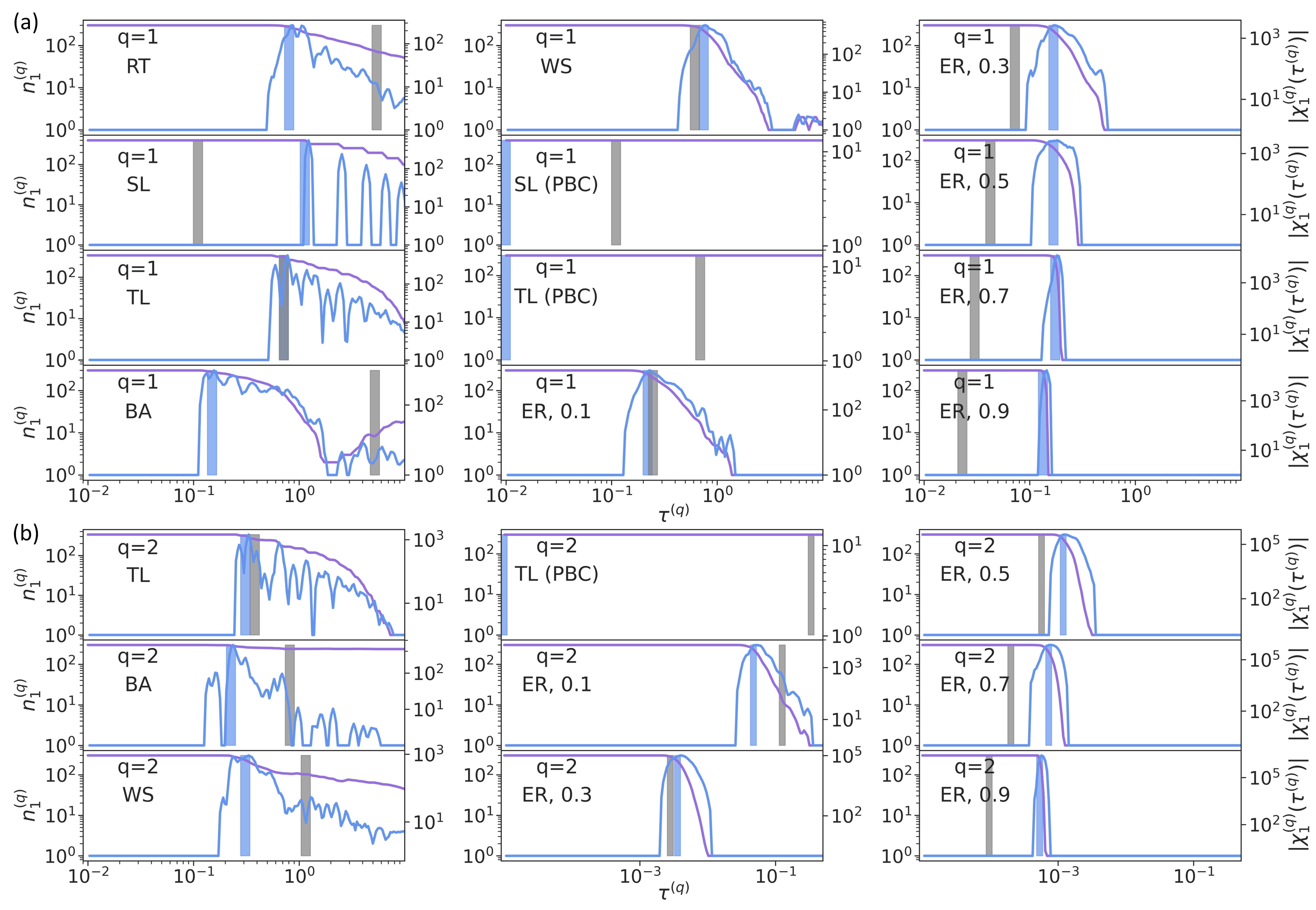}
\caption{\label{AG1}Bias in time scale calculation. (a) shows $n_{1}^{\left(q\right)}\left(\tau^{\left(q\right)}\right)$ and $\big\vert\chi_{1}^{\left(q\right)}\left(\tau^{\left(q\right)}\right)\big\vert$ as the functions of the time scale $\tau^{\left(q\right)}$ based on the results obtained on the Erdos-Renyi network (ER), the Watts-Strogatz network (WS), the Barab{\'a}si-Albert network (BA), the random tree (RT), the triangular lattice (TL), and the square lattice (SL), respectively. Each network is initialized with $300$ units. Among these networks, TL and SL can be defined with or without the periodic boundary condition (PBC). The SRG is defined with $\left(p,q\right)=\left(1,1\right)$. (b) shows the same analysis on these networks, where the SRG is defined with $\left(p,q\right)=\left(1,2\right)$. Note that RT is no longer included in our analysis since it contains no $2$-order interaction. In (a-b), ER networks are defined with different connection probabilities. For instance, ``ER, 0.1" denotes the case where a ER network is defined with a connection probability of $0.1$. Under each condition, we represent $n_{1}^{\left(q\right)}\left(\tau^{\left(q\right)}\right)$ by a purple line and represent $\big\vert\chi_{1}^{\left(q\right)}\left(\tau^{\left(q\right)}\right)\big\vert$ by a blue line. Meanwhile, we mark $\widetilde{\tau}^{\left(q\right)}$ and $\tau^{\left(q\right)}_{*}$ by a gray bar and a blue bar, respectively. Therefore, we have $\tau^{\left(q\right)}_{*}>\widetilde{\tau}^{\left(q\right)}$ if the blue bar occurs on the right side of the gray bar, which indicates a bias in time scale calculation.} 
\end{figure*}

To explore when this bias occurs, we use $n_{1}^{\left(q\right)}\left(\tau^{\left(q\right)}\right)$, the number of Kadanoff blocks in $\mathbf{M}^{\left(q\right)}_{1}\left(\tau^{\left(q\right)}\right)$, as an observable. Here $\mathbf{M}^{\left(q\right)}_{1}\left(\tau^{\left(q\right)}\right)$ is a reference network derived in a way similar to the network $\mathbf{H}^{\left(q\right)}_{k}$ shown in the step (2) of the SRG in Sec. \ref{Sec4}. Specifically, we initialize $\mathbf{M}^{\left(q\right)}_{1}\left(\tau^{\left(q\right)}\right)$ as a null network and calculate 
\begin{align}
\rho^{\left(q\right)}_{1}\left(\tau^{\left(q\right)}\right)=\frac{\sum_{\lambda_{1}^{\left(q\right)}}\exp\left(-\tau^{\left(q\right)}\lambda_{1}^{\left(q\right)}\right)\big\vert\lambda_{1}^{\left(q\right)}\big\rangle\big\langle\lambda_{1}^{\left(q\right)}\big\vert}{\sum_{\lambda_{1}^{\left(q\right)}}\exp\left(-\tau^{\left(q\right)}\lambda_{1}^{\left(q\right)}\right)}.\label{A6EQ4}
\end{align}
We go through every element in $\rho^{\left(q\right)}_{1}\left(\tau^{\left(q\right)}\right)$. If the $\left(i,j\right)$-th element is greater than the $\left(i,i\right)$-th or the $\left(j,j\right)$-th element, we add an edge between units $v_{i}$ and $v_{j}$ in $\mathbf{M}^{\left(q\right)}_{1}\left(\tau^{\left(q\right)}\right)$. We count the number of clusters in $\mathbf{M}^{\left(q\right)}_{1}\left(\tau^{\left(q\right)}\right)$ as $n_{1}^{\left(q\right)}\left(\tau^{\left(q\right)}\right)$. To reflect how $n_{1}^{\left(q\right)}\left(\tau^{\left(q\right)}\right)$ changes with $\tau^{\left(q\right)}$, we measure the absolute value of a susceptibility-like quantity
\begin{align}
\Big\vert\chi_{1}^{\left(q\right)}\left(\tau^{\left(q\right)}\right)\Big\vert=\bigg\vert\frac{\partial }{\partial\tau}n_{1}^{\left(q\right)}\left(\tau^{\left(q\right)}\right)\bigg\vert_{\tau^{\left(q\right)}}\bigg\vert.\label{A6EQ5}
\end{align}
As $\tau^{\left(q\right)}$ increases, there is a higher possibility for $n_{1}^{\left(q\right)}\left(\tau^{\left(q\right)}\right)$ to decrease (note that $n_{1}^{\left(q\right)}\left(\tau^{\left(q\right)}\right)$ is not strictly decreasing) since we progressively add more edges into $\mathbf{M}^{\left(q\right)}_{1}\left(\tau^{\left(q\right)}\right)$.

In Fig. \ref{AG1}, we use the results obtained on several representative random network models to show $n_{1}^{\left(q\right)}\left(\tau^{\left(q\right)}\right)$ and $\big\vert\chi_{1}^{\left(q\right)}\left(\tau^{\left(q\right)}\right)\big\vert$ as the functions of the time scale $\tau^{\left(q\right)}$. Meanwhile, we show the time scale $\widetilde{\tau}^{\left(q\right)}$ selected based on Eqs. (\ref{A6EQ2}-\ref{A6EQ3}) as a reference. In our analysis, we focus on analyzing the relation between $\widetilde{\tau}^{\left(q\right)}$ and $\tau^{\left(q\right)}_{*}$, where $\tau^{\left(q\right)}_{*}$ denotes the time scale that maximizes $\big\vert\chi_{1}^{\left(q\right)}\left(\tau^{\left(q\right)}\right)\big\vert$. As shown in Fig. \ref{AG1}, a time scale $\tau^{\left(q\right)}$ below $\tau^{\left(q\right)}_{*}$ generally corresponds to a null network $\mathbf{M}^{\left(q\right)}_{1}\left(\tau^{\left(q\right)}\right)$ where there exists no effective diffusion. Once the time scale $\tau^{\left(q\right)}$ exceeds $\tau^{\left(q\right)}_{*}$, the value of $n_{1}^{\left(q\right)}\left(\tau^{\left(q\right)}\right)$ exhibits significant reductions on all random network models. Therefore, the time scale $\tau^{\left(q\right)}_{*}$ can serve as an indicator of the emergence of effective diffusion. When we compare between $\widetilde{\tau}^{\left(q\right)}$ and $\tau^{\left(q\right)}_{*}$, we can see that $\widetilde{\tau}^{\left(q\right)}$ is larger than $\tau^{\left(q\right)}_{*}$ in most random network models except for the Erdos-Renyi network (ER) under some conditions. This phenomenon occurs on both pairwise and $2$-order interactions. Consequently, in most random network models, the time scale $\widetilde{\tau}^{\left(q\right)}$ selected according to the infinite deceleration condition shown in Eqs. (\ref{A6EQ2}-\ref{A6EQ3}) is sufficient to create effective diffusion. As for the ER network, we define it with different connection probability (i.e., two units share an edge with a given probability). When this probability enlarges, the ER network becomes increasingly dense. It can be seen that $\widetilde{\tau}^{\left(q\right)}$ gradually becomes increasingly smaller than $\tau^{\left(q\right)}_{*}$ as the system becomes more dense. A value of $\widetilde{\tau}^{\left(q\right)}$ that is smaller than $\tau^{\left(q\right)}_{*}$ inevitably leads to a bias because there is no effective diffusion included while measuring the specific heat in Eq. (\ref{A6EQ1}). To avoid this bias, we suggest a practical way to re-scale $\widetilde{\tau}^{\left(q\right)}$ when it is smaller than $\tau^{\left(q\right)}_{*}$. Specifically, we define a simple coefficient
\begin{align}
\varepsilon=\frac{\tau^{\left(q\right)}_{*}}{\min\{\widetilde{\tau}^{\left(q\right)},\tau^{\left(q\right)}_{*}\}}\label{A6EQ6}
\end{align}
and always re-scale $\widetilde{\tau}^{\left(q\right)}$ as $\varepsilon\widetilde{\tau}^{\left(q\right)}$ (i.e., if $\widetilde{\tau}^{\left(q\right)}$ is smaller than $\tau^{\left(q\right)}_{*}$, we re-scale $\widetilde{\tau}^{\left(q\right)}$ to $\tau^{\left(q\right)}_{*}$ directly. If $\widetilde{\tau}^{\left(q\right)}$ is no less than $\tau^{\left(q\right)}_{*}$, there is no further adjustment). This re-scaling procedure leads to Eq. (\ref{EQ26}) in Sec. \ref{Sec4}.

Apart from the bias in time scale calculation, Fig. \ref{AG1} also conveys an important message about the effects of the periodic boundary condition. It can be seen that $n_{1}^{\left(q\right)}\left(\tau^{\left(q\right)}\right)$ generally maintains constant under the periodic boundary condition (there are only small perturbations caused by numerical errors). This is because all units and interactions become identical and the effective diffusion no longer exists. These results provide an explanation for the exactly invariant property of the lattice systems defined with the periodic boundary condition during renormalization. Note that the selection of $\tau^{\left(q\right)}_{*}$ under the periodic boundary condition, as shown by Fig. \ref{AG1}, is mainly determined by the occasional and small numerical errors while measuring $n_{1}^{\left(q\right)}\left(\tau^{\left(q\right)}\right)$.

\section{Comparison of the SRG across different settings of $\left(p,q\right)$}\label{ASec7}

In this section, we compare the SRG across different settings of $\left(p,q\right)$ to validate the meaning of distinguishing between $p$ and $q$ during renormalization. In Fig. \ref{AG2}, we use the SRG to process a system whose pairwise interactions follow the Watts-Strogatz network (WS), a random network model exhibiting small-world properties \cite{watts1998collective}.

In our experiment, we renormalize the system on the $p$-order under the guidance of $q$-order interactions. We first consider a case with $p=q$ (i.e., we renormalize the system on an order according to its properties on this order). When $p=q=1$, our analysis becomes similar with other conventional renormalization group frameworks that focus on pairwise interactions. When $p=q>1$, the SRG is implemented on high-order interactions but our analysis involves no cross-order information since we do not distinguish between $p$ and $q$ during renormalization. In these cases, the SRG is only equipped with the information of a single order.

\begin{figure*}
\includegraphics[width=1\columnwidth]{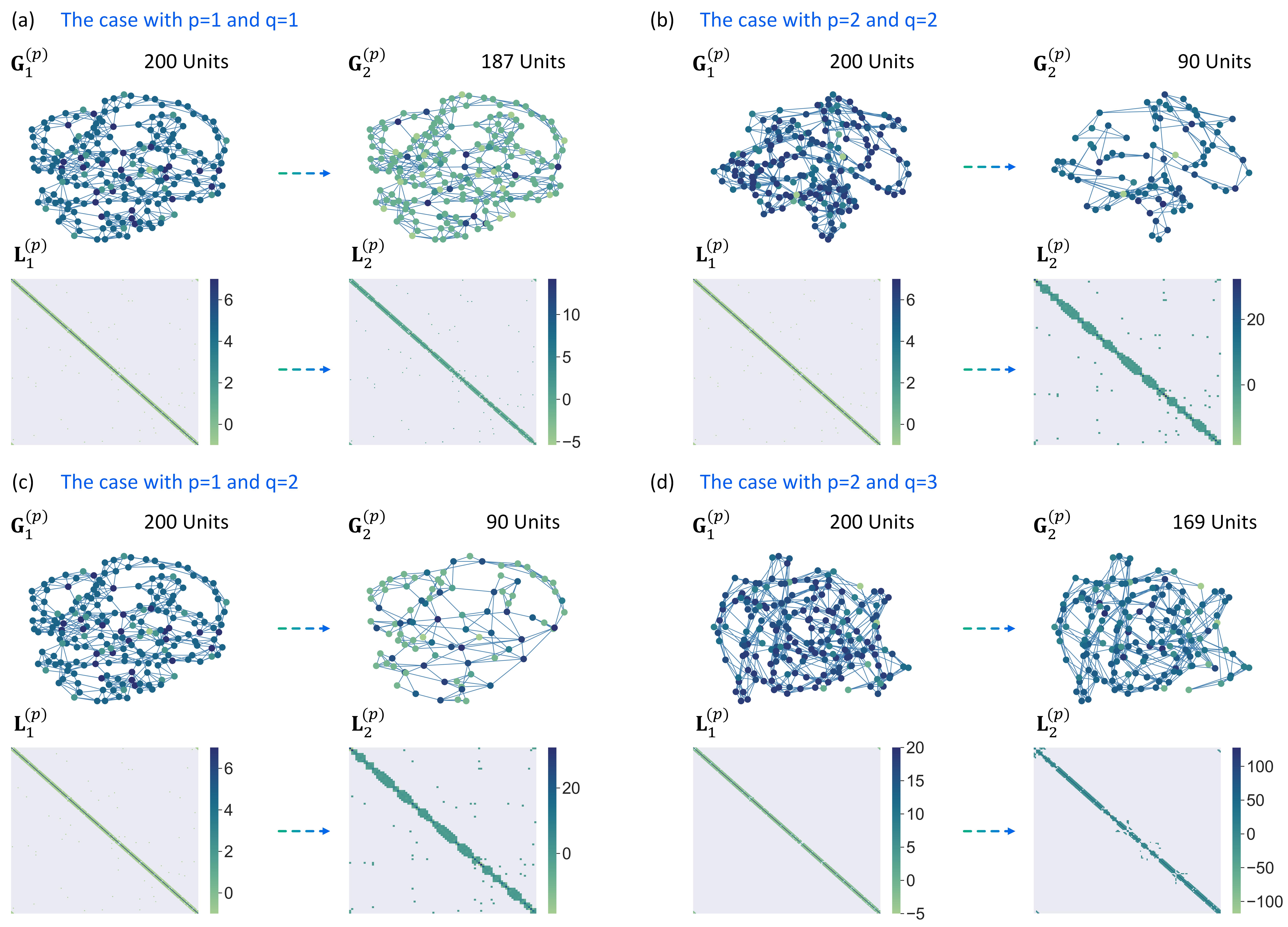}
\caption{\label{AG2}The simplex renormalization group (SRG) with different $\left(p,q\right)$. We generate a system with $200$ units whose pairwise interactions follow a Watts-Strogatz network (each unit initially has $6$ neighbors and edges are rewired according to a probability of $0.05$) \cite{watts1998collective}. Then, the SRG is applied to renormalize the system on the $p$-order according to the properties of $q$-order interactions. (a-d) show the renormalizations flows in both moment and real spaces, which are generated by the SRG defined with $\left(p,q\right)=\left(1,1\right)$, $\left(p,q\right)=\left(2,2\right)$, $\left(p,q\right)=\left(1,2\right)$, and $\left(p,q\right)=\left(2,3\right)$, respectively.} 
\end{figure*}

For comparison, we distinguish between $p$ and $q$ and ensure that $p\neq q$. Thus, we create a possibility for analyzing the system based on cross-order information. Specifically, we consider $\left(p,q\right)=\left(1,2\right)$ and $\left(p,q\right)=\left(2,3\right)$ as two instances. These two cases have not been studied by existing renormalization group frameworks, allowing us to see the nontrivial effects of the interactions on a higher order on the interactions on a lower order.

Our results are presented in Fig. \ref{AG2}, where we illustrate the renormalization flows in both real (i.e., $\mathbf{G}^{\left(p\right)}_{k}$) and moment (i.e., $\mathbf{L}^{\left(p\right)}_{k}$) spaces. By comparing between Figs. \ref{AG2}(a-b) and Figs. \ref{AG2}(c-d), we can directly observe systematic differences between the case with $p=q$ and the case with $p\neq q$. Distinguishing between $p$ and $q$ during renormalization is meaningful and effective, enabling us to explore more general cases uncovered by classic frameworks. Moreover, although high-order interactions are usually distributed in a sparser manner, renormalizing them frequently leads to more drastic system reductions. This is because high-order interactions involve with more units than low-order ones. 

\section{Scaling relation of the Laplacian eigenvalue spectrum in the ergodic case}\label{ASec8}

Given the general form of the Laplacian eigenvalue spectrum, i.e., $\operatorname{Prob}\left(\lambda^{\left(q\right)}_{1}\right)\sim \left(\lambda^{\left(q\right)}_{1}\right)^{\beta^{\left(q\right)}}=\alpha\left(\lambda^{\left(q\right)}_{1}\right)^{\beta^{\left(q\right)}}$, we can first reformulate Eq. (\ref{EQ27}) in the main text as
\begin{align}
\frac{\widetilde{X}^{\left(q\right)}}{\widetilde{\tau}^{\left(q\right)}}&=\big\langle\lambda^{\left(q\right)}_{1}\big\rangle_{\widetilde{\rho}}+\mu,\label{A8EQ1}\\
&=\frac{\frac{1}{N^{\left(q\right)}_{k}}\sum_{\lambda^{\left(q\right)}_{1}}\lambda^{\left(q\right)}_{1}\exp\left(-\widetilde{\tau}^{\left(q\right)}\lambda^{\left(q\right)}_{1}\right)}{\frac{1}{N^{\left(q\right)}_{k}}\sum_{\lambda^{\left(q\right)}_{1}}\exp\left(-\widetilde{\tau}^{\left(q\right)}\lambda^{\left(q\right)}_{1}\right)}+\mu
.\label{A8EQ2}
\end{align}
When the system is near criticality, the predominant domain of each eigenvalue tends to be similar, making it reasonable to assume that the eigenvalue spectrum is quasi-continuous
\begin{align}
    \frac{\widetilde{X}^{\left(q\right)}}{\widetilde{\tau}^{\left(q\right)}}&=\frac{\int_{0}^{\infty}\alpha\left(\lambda^{\left(q\right)}_{1}\right)^{\beta^{\left(q\right)}+1}\exp\left(-\widetilde{\tau}^{\left(q\right)}\lambda^{\left(q\right)}_{1}\right)\mathsf{d}\lambda^{\left(q\right)}_{1}}{\int_{0}^{\infty}\alpha\left(\lambda^{\left(q\right)}_{1}\right)^{\beta^{\left(q\right)}}\exp\left(-\widetilde{\tau}^{\left(q\right)}\lambda^{\left(q\right)}_{1}\right)\mathsf{d}\lambda^{\left(q\right)}_{1}}
+\mu,\;\exists\mu\in\mathbb{R}.\label{A8EQ3}
\end{align}
In Eq. (\ref{A8EQ3}), we can also understand $\mathsf{d}\lambda^{\left(q\right)}_{1}$ as the continuous limit of the weight associated with each eigenvalue. Using the Euler's gamma function, i.e., $\Gamma\left(z\right)=\int_{0}^{\infty}x^{z-1}\exp\left(-x\right)\mathsf{d}x$, we can transform Eq. (\ref{A8EQ3}) as
\begin{align}
\frac{\widetilde{X}^{\left(q\right)}}{\widetilde{\tau}^{\left(q\right)}}&=\frac{\frac{\alpha}{\left(\widetilde{\tau}^{\left(q\right)}\right)^{\beta^{\left(q\right)}+2}}\Gamma\left(\beta^{\left(q\right)}+2\right)}{\frac{\alpha}{\left(\widetilde{\tau}^{\left(q\right)}\right)^{\beta^{\left(q\right)}+1}}\Gamma\left(\beta^{\left(q\right)}+1\right)}+\mu,\label{A8EQ4}
\end{align}
which directly leads to
\begin{align}
\widetilde{X}^{\left(q\right)}&=\frac{\Gamma\left(\beta^{\left(q\right)}+2\right)}{\Gamma\left(\beta^{\left(q\right)}+1\right)}+\mu \widetilde{\tau}^{\left(q\right)},\label{A8EQ5}\\
&=\beta^{\left(q\right)}+1+\nu,\;\exists\nu\in\left[0,\infty\right).\label{A8EQ6}
\end{align}
Note that we have reduce $\mu \widetilde{\tau}^{\left(q\right)}$ in Eq. (\ref{A8EQ5}) to $\nu$ in Eq. (\ref{A8EQ6}) for convenience, where $\nu$ denotes an arbitrary non-negative number. The scope of $\nu\in\left[0,\infty\right)$ is derived from the fact that $\beta^{\left(q\right)}\leq-1$ (i.e., as required by the power-law distribution) and $\widetilde{X}^{\left(q\right)}\geq 0$ (i.e., the specific heat is non-negative).

 \begin{figure*}
\includegraphics[width=1\columnwidth]{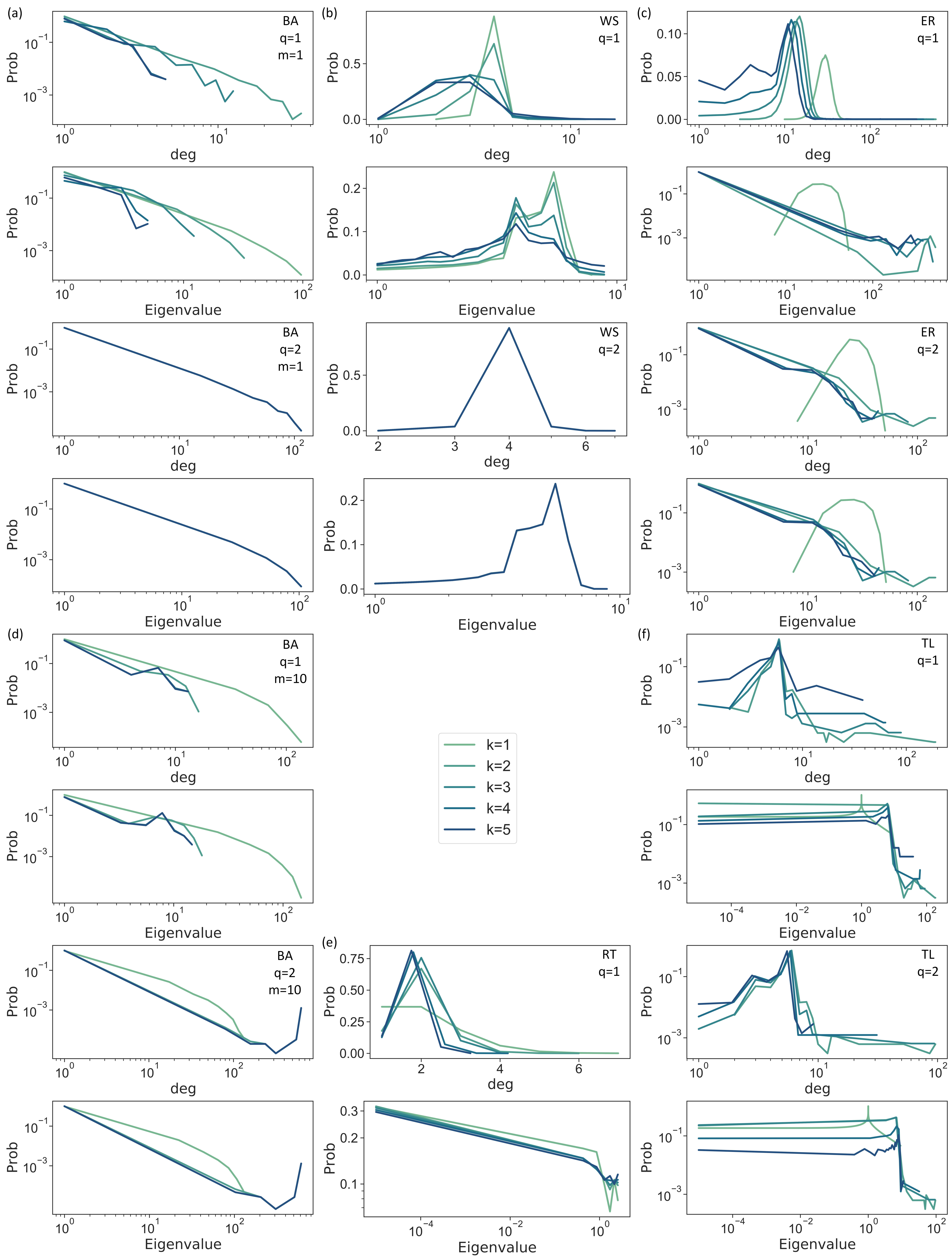}
\end{figure*}
\begin{figure*}
\begin{adjustwidth}{-0cm}{}
    \caption[]{\label{AG3}The existence of the multi-order scale-invariance (Part I). (a-f) The degree distributions and the Laplacian eigenvalue spectra of different synthetic interacting systems under the transformation of the SRG are illustrated after being averaged across all replicas. The multi-order Laplacian operator in Eq. (\ref{EQ1}) is applied to implement the SRG. Note that the original Laplacian eigenvalue spectra of all triangular lattices (TL) are calculated using the analytic formula developed by Ref. \cite{bille2023random}.}
     \end{adjustwidth}
\end{figure*}

 \begin{figure*}
\includegraphics[width=1\columnwidth]{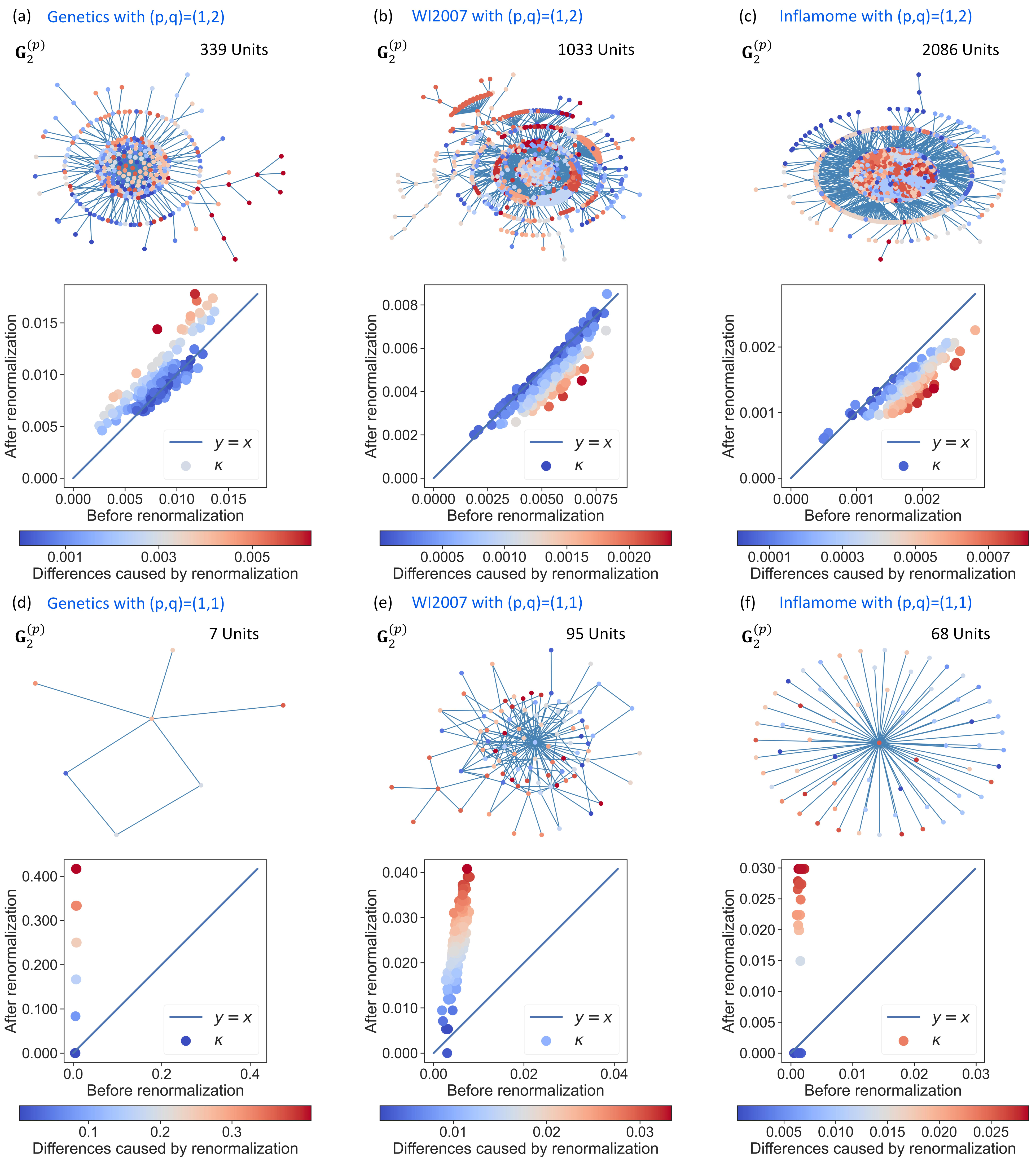}
\end{figure*}
\begin{figure*}
\begin{adjustwidth}{-0cm}{}
    \caption[]{\label{AG4}The existence of the multi-order scale-invariance (Part II). (a-f) The degree distributions and the Laplacian eigenvalue spectra of different synthetic interacting systems under the transformation of the SRG are illustrated after being averaged across all replicas. The high-order path Laplacian operator in Eq. (\ref{EQ5}) is applied to implement the SRG. Note that the original Laplacian eigenvalue spectra of all triangular lattices (TL) are calculated using the analytic formula developed by Ref. \cite{bille2023random}.}
     \end{adjustwidth}
\end{figure*}

\section{Scaling relation of the Laplacian eigenvalue spectrum in the non-ergodic case}\label{ASec9}

Here we present the derivations of the scaling relation in the non-ergodic case. For convenience, we denote $\mathbf{L}^{\left(q\right)}_{k}\left(i\right)$ as the Laplacian of cluster $C_{i}$ of $\mathbf{G}^{\left(q\right)}_{k}$. It is clear that the Laplacian of $\mathbf{G}^{\left(q\right)}_{k}$ can be formulated in a block diagonal form
\begin{align}
\mathbf{L}^{\left(q\right)}_{k}=\operatorname{diag}\left(\left[\mathbf{L}^{\left(q\right)}_{k}\left(1\right),\ldots,\mathbf{L}^{\left(q\right)}_{k}\left(r\right)\right]\right)
\label{A9EQ1}
\end{align}
because there exists no interaction between each pair of clusters. Eq. (\ref{A9EQ1}) leads to an important property of the determinant
\begin{align}
\operatorname{det}\left(\mathbf{L}^{\left(q\right)}_{k}\right)&=\prod_{i=1}^{r}\operatorname{det}\left(\mathbf{L}^{\left(q\right)}_{k}\left(i\right)\right)
,\label{A9EQ2}\\
\prod_{\lambda^{\left(q\right)}_{k}\in\Lambda^{\left(q\right)}_{k}}\lambda^{\left(q\right)}_{k}&=\prod_{i=1}^{r}\prod_{\lambda^{\left(q\right)}_{k}\in\Lambda^{\left(q\right)}_{k}\left(i\right)}\lambda^{\left(q\right)}_{k},\label{A9EQ3}
\end{align}
where $\Lambda^{\left(q\right)}_{k}$ and $\Lambda^{\left(q\right)}_{k}\left(i\right)$ denote the eigenvalue sets of $\mathbf{L}^{\left(q\right)}_{k}$ and $\mathbf{L}^{\left(q\right)}_{k}\left(i\right)$, respectively. According to Eqs. (\ref{A9EQ2}-\ref{A9EQ3}), the characteristic polynomial roots of $\mathbf{L}^{\left(q\right)}_{k}$ can be constructed using the combination of the characteristic polynomial roots of $\{\mathbf{L}^{\left(q\right)}_{k}\left(1\right),\ldots,\mathbf{L}^{\left(q\right)}_{k}\left(r\right)\}$
\begin{align}
\Lambda^{\left(q\right)}_{k}=\bigcup_{i=1}^{r}\Lambda^{\left(q\right)}_{k}\left(i\right).
\label{A9EQ4}
\end{align}

Because each eigenvalue set $\Lambda^{\left(q\right)}_{k}\left(i\right)$ follows an independent distribution $\operatorname{Prob}_{i}\left(\lambda^{\left(q\right)}_{k}\right)\sim \left(\lambda^{\left(q\right)}_{k}\right)^{\beta^{\left(q\right)}\left(i\right)}=\alpha\left(i\right)\left(\lambda^{\left(q\right)}_{k}\right)^{\beta^{\left(q\right)}\left(i\right)}$, the eigenvalue set $\Lambda^{\left(q\right)}_{k}$ defined by the weighted mixture (i.e., weighted according to cluster size) of $\{\Lambda^{\left(q\right)}_{k}\left(1\right),\ldots,\Lambda^{\left(q\right)}_{k}\left(r\right)\}$ in Eq. (\ref{A9EQ4}) is expected to follow
\begin{align}
\operatorname{Prob}\left(\lambda^{\left(q\right)}_{k}\right)=\sum_{i=1}^{r}\frac{\vert\Lambda^{\left(q\right)}_{k}\left(i\right)\vert}{\sum_{j=1}^{r}\vert\Lambda^{\left(q\right)}_{k}\left(j\right)\vert}\operatorname{Prob}_{i}\left(\lambda^{\left(q\right)}_{k}\right).
\label{A9EQ5}
\end{align}
This property enable us to relate the expectation of $\Lambda^{\left(q\right)}_{k}$ with the expectations of $\{\Lambda^{\left(q\right)}_{k}\left(1\right),\ldots,\Lambda^{\left(q\right)}_{k}\left(r\right)\}$
\begin{align}
\big\langle\lambda^{\left(q\right)}_{k}\big\rangle_{\widetilde{\rho}}
=\sum_{i=1}^{r}\frac{\vert\Lambda^{\left(q\right)}_{k}\left(i\right)\vert}{\sum_{j=1}^{r}\vert\Lambda^{\left(q\right)}_{k}\left(j\right)\vert}\big\langle\lambda^{\left(q\right)}_{k}\big\rangle_{\widetilde{\rho}\left(i\right)},
\label{A9EQ6}
\end{align}
where $\widetilde{\rho}^{\left(q\right)}\left(i\right)$ is the density operator associated with $\mathbf{L}^{\left(q\right)}_{k}\left(i\right)$. Given that each eigenvalue set $\Lambda^{\left(q\right)}_{k}\left(i\right)$ satisfies 
\begin{align}
\big\langle\lambda^{\left(q\right)}_{k}\big\rangle_{\widetilde{\rho}\left(i\right)}=\frac{\beta^{\left(q\right)}\left(i\right)+1}{\widetilde{\tau}^{\left(q\right)}\left(i\right)}+\nu,\;\exists \nu\in\left[0,\infty\right)\label{A9EQ7}
\end{align}
according to Eqs. (\ref{EQ27}-\ref{EQ28}) in the main text, we can derive 
\begin{align}
\big\langle\lambda^{\left(q\right)}_{k}\big\rangle_{\widetilde{\rho}}
&=\sum_{i=1}^{r}\frac{\vert\Lambda^{\left(q\right)}_{k}\left(i\right)\vert}{\sum_{j=1}^{r}\vert\Lambda^{\left(q\right)}_{k}\left(j\right)\vert}\frac{\beta^{\left(q\right)}\left(i\right)+1}{\widetilde{\tau}^{\left(q\right)}\left(i\right)}+\sum_{i=1}^{r}\frac{\vert\Lambda^{\left(q\right)}_{k}\left(i\right)\vert}{\sum_{j=1}^{r}\vert\Lambda^{\left(q\right)}_{k}\left(j\right)\vert}\nu,\label{A9EQ8}\\&=\Big\langle\frac{\beta^{\left(q\right)}\left(i\right)+1}{\widetilde{\tau}^{\left(q\right)}\left(i\right)}\Big\rangle_{i}+\nu,
\label{A9EQ9}
\end{align}
where $\langle\cdot\rangle_{i}$ denotes the weighted average across all clusters. Note that we have reduced the second term of Eq. (\ref{A9EQ8}) into $\nu$ given the arbitrariness of $\nu$. Eq. (\ref{A9EQ9}) further leads to
\begin{align}
\widetilde{X}^{\left(q\right)}
=\widetilde{\tau}^{\left(q\right)}\Big\langle\frac{\beta^{\left(q\right)}\left(i\right)+1}{\widetilde{\tau}^{\left(q\right)}\left(i\right)}\Big\rangle_{i}+\nu,\;\exists \nu\in\left[0,\infty\right),
\label{A9EQ10}
\end{align}
where $\widetilde{X}^{\left(q\right)}$ is the global specific heat measured on $\mathbf{L}^{\left(q\right)}_{k}$ under the infinite deceleration condition and $\widetilde{\tau}^{\left(q\right)}$ denotes the corresponding time scale.

\section{Extended results of the multi-order scale-invariance}\label{ASec10}

In this section, we present the extended results of multi-order scale-invariance verification. Our verification is implemented using the behaviours of Laplacian eigenvalue spectra and degree distributions. 

We implement our experiments on multiple types of synthetic interacting systems, whose pairwise interactions follow the Barab{\'a}si-Albert network (BA, $c=1$ or $c=2$), the Watts-Strogatz network (WS, each unit initially has $10$ neighbors and edges are rewired according to a probability of $0.1$), the Erdos-Renyi network (ER, each pair of units share an edge with a probability of $0.01$), the triangular lattice (TL), and the random tree (RT), respectively. Each kind of system consists of $1000$ units and has $100$ replicas. We use the SRG defined with different settings of $\left(p,q\right)$ to process these systems on different orders. Based on the obtained renormalization flows, we calculate eigenvalue spectra and degree distributions using $\mathbf{L}^{\left(p\right)}_{k}$ and $\mathbf{G}^{\left(p\right)}_{k}$. Please note that the degree distribution of $\mathbf{G}^{\left(p\right)}_{k}$ gradually becomes weighted during renormalization.

In Fig. \ref{AG3}, we show the results generated by the SRG defined with the multi-order Laplacian operator (MOL). As shown in our results, the interacting systems whose pairwise interactions follow weakly scale-invariant (e.g., BA) and scale-invariant (e.g., TL and RT) structures exhibit multi-order scale invariance because their Laplacian eigenvalue spectra generally follow fixed power-law distributions during renormalization (note that noises in Laplacian eigenvalue spectra are unavoidable due to the numerical instability of eigen-decomposition in sparse matrices). For comparisons, the absence of multi-order scale invariance can be observed in the interacting systems with scale-dependent pairwise interactions (e.g., ER and WS). These results are consistent with our findings in Fig. \ref{G6}, where the degrees to which the scaling relation holds are higher in scale-invariant and weakly scale-invariant systems (i.e., with smaller deviations) and are lower in other systems. In fact, the departures from the scaling relation of scale-dependent systems arise from the absence of power-law behaviours (e.g., see Fig. \ref{AG3}(b-c)).

Apart from the verification of multi-order scale invariance, the results shown in Fig. \ref{AG3} also support the study of the effects of high-order interactions on power-law behaviours. When the SRG is guided by $2$-order interactions, the Laplacian eigenvalue spectra of the interacting systems whose pairwise interactions follow the ER model seem to exhibit power-law behaviours after one iteration of renormalization (i.e., $k>1$). We speculate that specific scale-invariant structures may emerge in the ER model after some high-order interactions are reduced.

The above findings can also be validated by the observations in Fig. \ref{AG4}, where all settings in the computational experiments are same as those in Fig. \ref{AG3} except that the high-order path Laplacian operator (HOPL) is applied to design the SRG.

\begin{figure*}
\includegraphics[width=1\columnwidth]{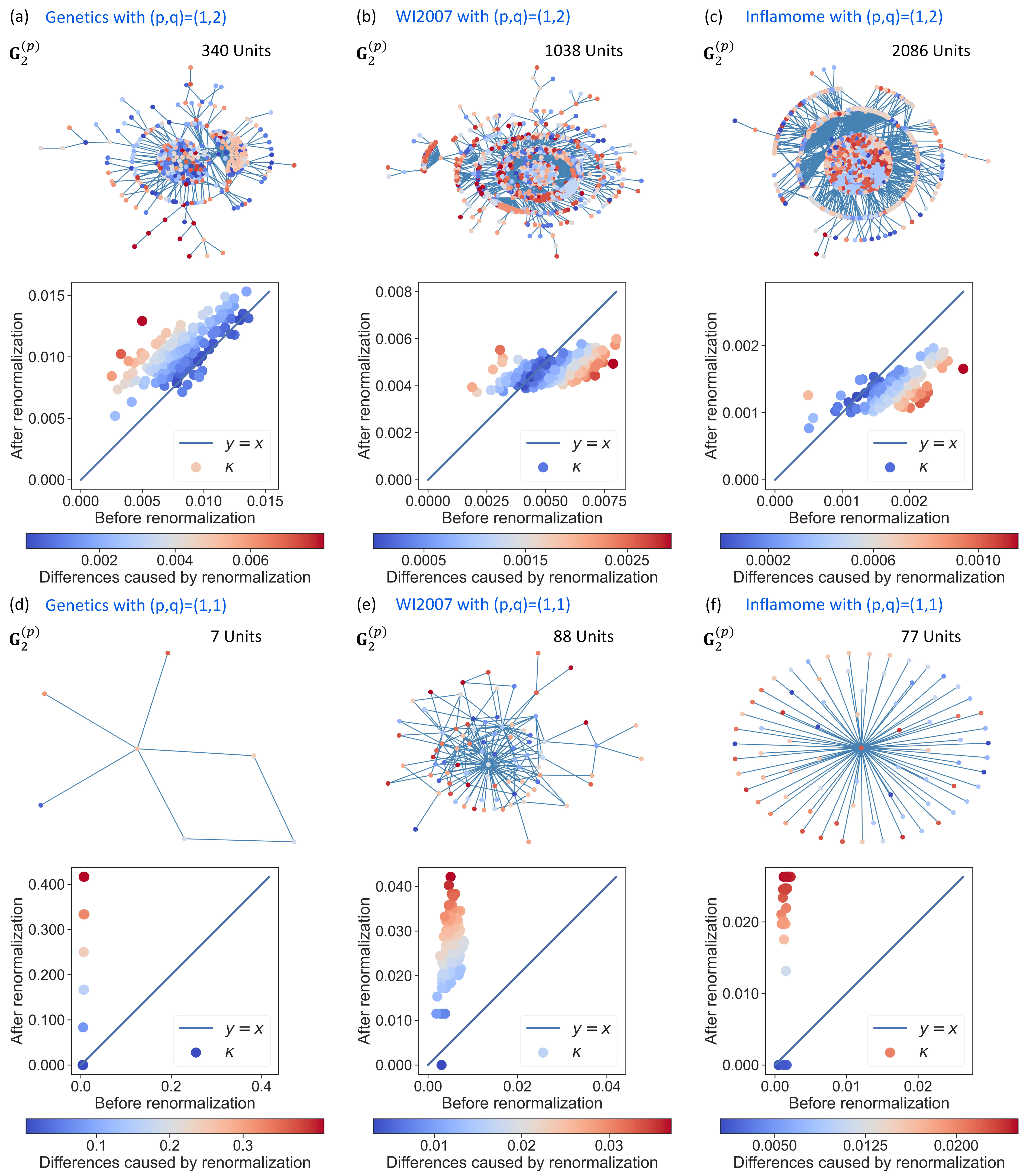}
\caption{\label{AG5}The organizational structure transformed by renormalization flows. The Louvain communities of all data sets are adopted from Fig. \ref{G8} directly. (a-c) The SRG is applied to renormalize these systems with $\left(p,q\right)=\left(2,1\right)$. The community label of each renormalized macro-unit is determined according to the most frequent community label of the units aggregated into it. The values of average path distance between communities, $\kappa$, are measured before and after renormalization, whose absolute differences (i.e., see color-bars) quantify the deviation degrees from the initial organizational structures caused by renormalization. (d-f) The analysis same as (a-c) is implemented with $\left(p,q\right)=\left(2,1\right)$.} 
\end{figure*}

\section{On the scaling relation and high-order effects}\label{ASec11}

Here we elaborate the methodology for illustrating the scaling relation defined in Eq. (\ref{EQ28}) and the high-order effects on it when $k\in\{1,2,3\}$. In computational experiments, different types of interacting systems are generated according to their pairwise interaction properties, where each type of system has $500$ replicas (i.e., realizations) and each replica consists of $1000$ units.

For each $k$-th iteration, we first derive $\beta^{\left(q\right)}\left(i\right)$ on $\mathbf{G}^{\left(q\right)}_{k}$ applying a maximum likelihood estimation approach proposed by Refs. \cite{alstott2014powerlaw,clauset2009power} and measure $\nu$ using the least square method. Given each replica, we can derive the eigenvalue set $\Lambda^{\left(q\right)}_{k}\left(i\right)$ for its $i$-th cluster and apply the maximum likelihood estimation approach \cite{alstott2014powerlaw,clauset2009power} to measure exponent $\beta^{\left(q\right)}\left(i\right)$ for the eigenvalue spectrum of this cluster such that $\beta^{\left(q\right)}\left(i\right)$ defines a power-law distribution. This estimation is implemented using the toolbox provided by Ref. \cite{alstott2014powerlaw}. After deriving $\beta^{\left(q\right)}\left(i\right)$ and $\widetilde{\tau}^{\left(q\right)}\left(i\right)$ for each cluster $C_{i}$, we can calculate the value of $\widetilde{\tau}^{\left(q\right)}\Big\langle\frac{\beta^{\left(q\right)}\left(i\right)+1}{\widetilde{\tau}^{\left(q\right)}\left(i\right)}\Big\rangle_{i}$ for the selected replica. Parallel to this process, we can measure $\widetilde{X}^{\left(q\right)}$ and $\widetilde{\tau}^{\left(q\right)}$ for this replica as well. In sum, these calculations enable us to obtain a pair of values, $\left(\widetilde{X}^{\left(q\right)}
,\widetilde{\tau}^{\left(q\right)}\Big\langle\frac{\beta^{\left(q\right)}\left(i\right)+1}{\widetilde{\tau}^{\left(q\right)}\left(i\right)}\Big\rangle_{i}\right)$, for the replica. By repeating the above calculations on all replicas, we can generate the sample sets of predictors, $\widetilde{\tau}^{\left(q\right)}\Big\langle\frac{\beta^{\left(q\right)}\left(i\right)+1}{\widetilde{\tau}^{\left(q\right)}\left(i\right)}\Big\rangle_{i}$, and responses, $\widetilde{X}^{\left(q\right)}$, respectively. These samples can be used to implement a vanilla linear fitting $\widetilde{X}^{\left(q\right)}
=\widetilde{\tau}^{\left(q\right)}\Big\langle\frac{\beta^{\left(q\right)}\left(i\right)+1}{\widetilde{\tau}^{\left(q\right)}\left(i\right)}\Big\rangle_{i}+\nu$, where $\nu$ can be solved by the least square method. After deriving $\nu$, we obtain a set of predictions, $\widetilde{\tau}^{\left(q\right)}\Big\langle\frac{\beta^{\left(q\right)}\left(i\right)+1}{\widetilde{\tau}^{\left(q\right)}\left(i\right)}\Big\rangle_{i}+\nu$, and their associated labels, $\widetilde{X}^{\left(q\right)}$. By respectively averaging these predictions and labels across all replicas, we can derive the mean observations presented in Figs. \ref{G6}(a-e). Meanwhile, we can quantify the departures from the scaling relation by the standard deviations of $\widetilde{X}^{\left(q\right)}$ and $\widetilde{\tau}^{\left(q\right)}\Big\langle\frac{\beta^{\left(q\right)}\left(i\right)+1}{\widetilde{\tau}^{\left(q\right)}\left(i\right)}\Big\rangle_{i}+\nu$.

\section{On organizational structure preservation}\label{ASec12}

Here we elaborate on the experiments of verifying whether given organizational structures can be preserved by the SRG framework.

Before renormalization, the communities of units are automatically detected by the Louvain community detection algorithm \cite{blondel2008fast}. After every iteration of renormalization, units are aggregated into macro-units according to the SRG. Here we do not define latent community structures by applying the algorithm \cite{blondel2008fast} on the renormalized system again. Instead, we determine latent community structures according to unit aggregation during renormalization, where the community label of each macro-unit is defined as the most frequent community label of the units aggregated into it. For instance, if six units are aggregated into a macro-unit during renormalization and four of them belong to the first community, then the macro-unit is defined as a member of the first community.

To characterize the organizational structure of a system with multiple communities, we can analyze the connectivity between communities in terms of normalized average path distance. Specifically, the average path distance between two communities is averaged across all path distances between the units that belong to these communities. Here the path distance between two units refers to the weighted length of the shortest paths between them (i.e., being weighted according to edge weights). A smaller average path distance suggests that two corresponding communities are closer to each other. Then, we normalize these average path distances by dividing them by the largest possible path distance (i.e., the number of units minus one). The above calculation can be implemented before and after renormalization to derive two sets of results for comparison. For instance, if the normalized average path distance between a pair of communities before renormalization is numerically similar to its counterpart after renormalization, the connectivity properties between communities are generally preserved under the transformation of the SRG. 

Consistent with Ref. \cite{gfeller2007spectral}, our results in Fig. \ref{AG5} suggest that the renormalized systems are not equivalent to the networks of communities (i.e., the networks where each unit is a community). Moreover, we observe qualitative similarities between apparent and latent community structures when the SRG is guided by high-order interactions (i.e., $q=2$). These similarities can be quantitatively validated by showing how the connectivity properties (e.g., the normalized average path distances) between communities are preserved under the transformation of the SRG. For comparison, the results with $q=1$ show that the renormalization guided by pairwise interactions, even though being applied on the same data, does not preserve community connectivity properties. On the contrary, the renormalization makes the latent community structures deviate from the apparent ones because the whole system is reduced significantly.

\section{Programmatic implementation of the SRG}\label{ASec13}

The SRG is programmed in Python, whose open-source code can be seen in \cite{aohua2023toolbox}. The SRG depends on several external libraries listed below. Users should prepare these libraries before using the SRG.

\begin{lstlisting}[language=Python]
## Dependency libraries used for the SRG:
import numpy as np
import networkx as nx
import scipy as spy
import itertools
import copy
from scipy.signal import argrelextrema
from scipy.linalg import expm
\end{lstlisting}

\paragraph{Main function and usage of the SRG framework}

In application, we have a system, \textbf{G}, to process. For the SRG, we need to ensure that \textbf{G} is a graph object in the $\mathbf{networkx}$ library. The SRG renormalizes the system on the $p$-order based on the $q$-order interactions ($p\leq q$), where $p$ and $q$ can be defined by specifying \textbf{p} and \textbf{q}. Meanwhile, to select the type of the Laplacian representation (i.e., the Multi-order Laplacian operator or the high-order path Laplacian in Sec.\ref{Sec2}), we need to specify \textbf{L\_Type} in the program. Given these settings, the SRG runs on the system for \textbf{IterNum} iterations to generate a renormalization flow. 

\begin{lstlisting}[language=Python]
def SRG_Flow(G,q,p,L_Type,IterNum):
    A=nx.adjacency_matrix(G).toarray()       
    Lq=HighOrderLaplician(A, L_Type, Order=q)
    Lp=HighOrderLaplician(A, L_Type, Order=p)
    Lq_List,Lp_List,Gq_List,Gp_List,C_List,Tracked_Alignment=SRG_Function(Lq,Lp,q,IterNum)
    return  Lq_List,Lp_List,Gq_List,Gp_List,C_List,Tracked_Alignment
\end{lstlisting}

The main function of the SRG generates six outputs after computation. The first two outputs, \textbf{Lq\_List} and \textbf{Lp\_List}, are the lists of operator $\mathbf{L}^{\left(q\right)}_{k}$ and operator $\mathbf{L}^{\left(p\right)}_{k}$, respectively. For instance, the first element of \textbf{Lq\_List} is $\mathbf{L}^{\left(q\right)}_{1}$, the second element is $\mathbf{L}^{\left(q\right)}_{2}$, and so on. The number of elements in \textbf{Lq\_List} and \textbf{Lp\_List} is determined by \textbf{IterNum}.

The third and forth outputs, \textbf{Gq\_List} and \textbf{Gp\_List}, are the lists of the associated network sketches $\mathbf{G}^{\left(q\right)}_{k}$ and $\mathbf{G}^{\left(p\right)}_{k}$ of the above Laplacian operators, respectively. For example, the first element of \textbf{Gq\_List} is $\mathbf{G}^{\left(q\right)}_{1}$, the second element is $\mathbf{G}^{\left(q\right)}_{2}$, and so on. The number of elements in \textbf{Gq\_List} and \textbf{Gp\_List} equals to \textbf{IterNum}.

The fifth output, \textbf{C\_List}, is the list of specific heat $X_{1}^{\left(q\right)}$ vector calculated by the initial $q$-order Laplacian, \textbf{Lq\_List[0]}. The number of specific heat vector in \textbf{C\_List} is determined by the number of connected clusters in \textbf{Lq\_List[0]}. For instance, in the ergodicity case, \textbf{C\_List} contains only one element, which is a vector of specific heat derived on the only one connected cluster. 

The last output of the main function is \textbf{Tracked\_Alignment}, which indicates the indexes of the initial units aggregated into each macro-unit in every connected cluster after the $k$-th iteration of the SRG. Below, we present a simple instance where the system \textbf{G} contains only six units and satisfies the ergodicity. 
\begin{lstlisting}[language=Python]
Tracked_Alignment[0]=[[[0, [0, 1, 3]], [2, [2]], [4, [4]], [5, [5]]]]
Tracked_Alignment[1]=[[[0, [0, 5]], [2, [2]], [4, [4]]]]
\end{lstlisting} 
Before renormalization, each macro-unit only contains itself (i.e., the initial unit), which can be represented by a structured list [[[0, [0]], [1, [1]], [2, [2]], [3, [3]], [4, [4]], [5, [5]]]] in the form of [macro-unit, list of the units aggregated into this macro-unit]. Note that this trivial list is not included in \textbf{Tracked\_Alignment} for convenience. After two iterations of renormalization, \textbf{Tracked\_Alignment} is a list of two elements. As shown in the instance presented above, the first element of \textbf{Tracked\_Alignment} is [[0, [0, 1, 3]], [2, [2]], [4, [4]], [5, [5]]]. This list means that there remain four macro-units after the first iteration. The first macro-unit, 0, is formed by three initial units, 0, 1, and 3. Initial units 2, 4, and 5 are not coarse grained with other units during the first iteration so they only contain themselves. The second element of is [[[0, [0, 5]], [2, [2]], [4, [4]]]], suggesting that there exist three macro-units after the second iteration. The first macro-unit is derived by grouping 0 and 5, two macro-units generated in the first iteration, together. Consequently, this macro-unit contains four initial units, whose indexes are 0, 1, 3, and 5. Other elements of $\mathbf{Tracked\_Alignment}$ can be understood in a similar way. In the non-ergodic case, $\mathbf{Tracked\_Alignment}[k][i][j]$ contains the indexes of the units aggregated into the $j$-th macro-unit in the $i$-th connected cluster after $\left(k+1\right)$-th iteration.

To run the SRG, one can consider the following instances:
\begin{lstlisting}[language=Python]
G=nx.random_graphs.barabasi_albert_graph(1000,3) # Generate a random BA network with 1000 units

# Multiorder Laplacian operator
Lq_List,Lp_List,Gq_List,Gp_List,C_List,Tracked_Alignment=SRG_Flow(G,q=1,p=1,L_Type='MOL',IterNum=5) # Run a SRG for 5 iterations, which renormalize the system on the 1-order based on the 1-order interactions

Lq_List,Lp_List,Gq_List,Gp_List,C_List,Tracked_Alignment=SRG_Flow(G,q=2,p=1,L_Type='MOL',IterNum=5) # Run a SRG for 5 iterations, which renormalize the system on the 1-order based on the 2-order interactions

Lq_List,Lp_List,Gq_List,Gp_List,C_List,Tracked_Alignment=SRG_Flow(G,q=3,p=1,L_Type='MOL',IterNum=5) # Run a SRG for 5 iterations, which renormalize the system on the 1-order based on the 3-order interactions

# High-order path Laplacian
Lq_List,Lp_List,Gq_List,Gp_List,C_List,Tracked_Alignment=SRG_Flow(G,q=1,p=1,L_Type='HOPL',IterNum=5) # Run a SRG for 5 iterations, which renormalize the system on the 1-order based on the 1-order interactions

Lq_List,Lp_List,Gq_List,Gp_List,C_List,Tracked_Alignment=SRG_Flow(G,q=2,p=1,L_Type='HOPL',IterNum=5) # Run a SRG for 5 iterations, which renormalize the system on the 1-order based on the 2-order interactions

Lq_List,Lp_List,Gq_List,Gp_List,C_List,Tracked_Alignment=SRG_Flow(G,q=3,p=1,L_Type='HOPL',IterNum=5) # Run a SRG for 5 iterations, which renormalize the system on the 1-order based on the 3-order interactions
\end{lstlisting}

\paragraph{Full code implementation}
For convenience, we attach the full code implementation below. One can also see \cite{aohua2023toolbox} for the official release of our framework, where we provide instances in the Jupyter notebook.

\begin{lstlisting}[language=Python]
def HighOrderLaplician(A, Type, Order):
    ## Input: 
    # A is the 1st-order adjacency matrix of the graph

    # Type is a str that determines which type of high order Laplician representation to use
    # Type='MOL': the function generates the multi-order Laplacian operator, which is 
    # Type='HOPL': the function generates the high-order path Laplacian operator proposed in our work

    # Order is a number that determines which order of interactions to analyze

    ## Output:
    # L is the high order Laplician representation, which is the Multiorder Laplacian operator or the high-order path Laplacian
    Num = A.shape[0]
    # Stores the vertices
    store = [0]* (Order+1)

    # Degree of the vertices
    # d = [deg for (_, deg) in G.degree()]
    d = list(np.sum(A,axis=0))

    Cliques = []
    # Function to check if the given set of vertices
    # in store array is a clique or not
    def is_clique(b) :
        # Run a loop for all the set of edges
        # for the select vertex
        for i in range(b) :
            for j in range(i + 1, b) :
                # If any edge is missing
                if (A[store[i]][store[j]] == 0) :
                    return False
        return True

    # Function to find all the cliques of size s
    def findCliques(i, l, s) :    
        # Check if any vertex from i+1 can be inserted as the l-th node in the simplex of size s
        for j in range( i + 1, Num) :
            # If the degree of the graph is sufficient
            if (d[j] >= s - 1) :
                # Add the vertex to store
                store[l] = j
                # If the graph is not a clique of size k
                # then it cannot be a clique
                # by adding another edge
                if (is_clique(l + 1)) :
                    # If the length of the clique is
                    # still less than the desired size
                    if (l < s-1) :
                        # Recursion to add vertices
                        findCliques(j, l + 1, s)
                    # Size is met
                    else :
                        Cliques.append(store[:s])
    findCliques(-1, 0, Order+1)

    if len(Cliques)>0:
        if Type=='MOL': # the multi-order Laplacian operator
            HO_A = np.zeros((Num, Num))
            for Simplex in Cliques:
                for [i, j] in itertools.combinations(Simplex,2):
                    HO_A[i,j] += 1
            HO_A += HO_A.T
            HO_D = np.zeros(Num)
            for Simplex in Cliques:
                for [i] in itertools.combinations(Simplex,1):
                    HO_D[i] += 1
            L = Order * np.diag(HO_D) - HO_A
        elif Type=='HOPL': # the high-order path Laplacian operator
            HO_B = np.zeros((Num, Num))
            for Simplex in Cliques:
                for [i, j] in itertools.combinations(Simplex,2):
                    HO_B[i,j] += np.math.factorial(Order-1)
            HO_B += HO_B.T
            HO_P = sum(HO_B)
            L = (np.diag(HO_P) - HO_B)/Order
    else:
        L=np.zeros((Num,Num))
    return L

def NetworkInfo(L):
    ## Input: 
    # L is the high order Laplician representation, which can be the Multiorder Laplacian operator or the high-order path Laplacian

    ## Output:
    # Sub_Ls the list of the sub-Laplician-matrices associated with all connected components from L
    # Sub_NodeIDs is the list of sub-node-indice associated with all connected components from L
    G=nx.from_numpy_array(np.abs(L)-np.diag(np.diag(L)))
    Sub_Ls=[L[list(SG),:][:,list(SG)] for SG in nx.connected_components(G)] 
    Sub_NodeIDs=[list(SG) for SG in nx.connected_components(G)]
    return Sub_Ls, Sub_NodeIDs

def NetworkUnion(Sub_Ls):
    L=np.ones((1,1))
    for SL in Sub_Ls:
        L=spy.linalg.block_diag(L,SL)
    L=L[1:,1:]
    return L

def SRG_Flow(G,q,p,L_Type,IterNum):
    A=nx.adjacency_matrix(G).toarray()       
    Lq=HighOrderLaplician(A, L_Type, Order=q)
    Lp=HighOrderLaplician(A, L_Type, Order=p)
    Lq_List,Lp_List,Gq_List,Gp_List,C_List,Tracked_Alignment=SRG_Function(Lq,Lp,q,IterNum)
    return  Lq_List,Lp_List,Gq_List,Gp_List,C_List,Tracked_Alignment

    
def SRG_Function(Lq,Lp,q,IterNum):
    ## Input: 
    # Lq is the initial guiding high order Laplician representation, which can be the Multiorder Laplacian operator or the high-order path Laplacian
    # Lp is the initial guided high order Laplician representation, which can be the Multiorder Laplacian operator or the high-order path Laplacian

    ## Output:
    # C_List is the list of specific heat vector calculated by the initial q - order Laplacian L_List[0] and a range of time scale
    # Lq_List is the list of the guiding q-order Laplacian over renormalization steps
    # Lp_List is the list of the guided p-order Laplacian over renormalization steps
    # Gq_List is the list of the guiding q-order network sketches over renormalization steps
    # Gp_List is the list of the guided p-order network sketches over renormalization steps
    # Tracked_Alignment is the indexes of the initial units aggregated into each macro-unit of all connected clusters after every iteration of the SRG
    
    Iter=1
    Gq=nx.from_numpy_array(np.abs(Lq)-np.diag(np.diag(Lq)))
    Gp=nx.from_numpy_array(np.abs(Lp)-np.diag(np.diag(Lp)))
    Lq_List=[Lq]
    Lp_List=[Lp]
    Gq_List=[Gq]
    Gp_List=[Gp]
    C_List=[]
    Init_Sub_Lqs,_=NetworkInfo(Lq)
    tau_Vec=np.zeros(len(Init_Sub_Lqs))
    for ID in range(len(Init_Sub_Lqs)):
        if np.size(Init_Sub_Lqs[ID],0)>1:
            tau,CVector=TauSelection(Init_Sub_Lqs[ID])
            tau=tau/((q+1)/2)
            tau_Vec[ID]=tau
            C_List.append(CVector)
    All_Alignment=[]
    while Iter<IterNum:
        ## step 1: initiate Laplacian operators and associated high-order network sketches in the k-th iteration
        Origin_Lq,Origin_Lp=Lq_List[Iter-1],Lp_List[Iter-1]
        Origin_Gq,Origin_Gp=Gq_List[Iter-1],Gp_List[Iter-1]
        Origin_Ap=nx.adjacency_matrix(Origin_Gp).toarray()
        New_Ap=copy.deepcopy(Origin_Ap)
        New_Lp=copy.deepcopy(Origin_Lp)
        Sub_Lqs, Sub_NodeIDs=NetworkInfo(Origin_Lq)
        Sub_Gqs=[Origin_Gq.subgraph(SNodeID).copy() for SNodeID in Sub_NodeIDs]
        NewSub_Lqs=[]
        NewSub_Gqs=[]
        New_Snodes_list=[]
        ClusterNodeAlignment=[]
        for SID in range(len(Sub_Lqs)):
            SLq=Sub_Lqs[SID]
            SGq=Sub_Gqs[SID]
            SNodeID = Sub_NodeIDs[SID]
            if np.size(SLq,0)==1:
                NodeAlignment=[]
                NewSub_Lqs.append(SLq)
                NewSub_Gqs.append(SGq)
                New_Snodes_list.extend(SNodeID)
                NodeAlignment.append([SNodeID[0],SNodeID])
            else:
                NodeAlignment=[]
                ## step 2: search targets to coarse grain in the real space
                Rho = expm(-tau*SLq)/np.trace(expm(-tau*SLq))
                diagonal_min = np.minimum(Rho.diagonal().reshape(-1, 1), Rho.diagonal())
                Rho_prime=Rho/diagonal_min
                Rho_prime=(Rho_prime+Rho_prime.T)/2
                Ref_adj=(Rho_prime>1).astype('int')
                # delete false edges not in subgraph of SLq
                Origin_SAq=nx.adjacency_matrix(SGq).toarray()
                Ref_mask=(Origin_SAq!=0).astype(int)
                Ref_adj=Ref_adj*Ref_mask
                RefG = nx.from_numpy_array(Ref_adj) # reference graph
                Clusters=[list(c) for c in list(nx.connected_components(RefG))]
                n_k=len(Clusters)
                ## step 3: implement a renormalization procedure for q-order and p-order network sketches in the real space
                # contracting SGq nodes
                New_SAq=copy.deepcopy(Origin_SAq)
                New_nodes=[Nodes[0] for Nodes in Clusters]
                for i in range(n_k):
                    iNodes=Clusters[i]
                    for j in range(i+1,n_k):
                        jNodes=Clusters[j]
                        new_weight=sum(Origin_SAq[iNodes,:][:,jNodes].reshape(-1))
                        New_SAq[iNodes[0],jNodes[0]]=new_weight
                        New_SAq[jNodes[0],iNodes[0]]=new_weight
                New_SAq=New_SAq[New_nodes,:][:,New_nodes]
                New_SGq=nx.from_numpy_array(New_SAq)
                NewSub_Gqs.append(New_SGq)
                # contracting Gp nodes
                New_Snodes=[SNodeID[Nodes[0]] for Nodes in Clusters]
                New_Snodes_list.extend(New_Snodes)
                Non_Nodes=[x for x in list(range(New_Ap.shape[0])) if x not in SNodeID]
                for i in range(n_k):
                    iSNodes = [SNodeID[node] for node in Clusters[i]]
                    New_Ap[iSNodes[0],Non_Nodes]=np.sum(Origin_Ap[iSNodes,:][:,Non_Nodes],axis=0) # merge edges from different connected components
                    for j in range(i+1,n_k):
                        jSNodes = [SNodeID[node] for node in Clusters[j]]
                        new_weight=sum(Origin_Ap[iSNodes,:][:,jSNodes].reshape(-1))  
                        New_Ap[iSNodes[0],jSNodes[0]]=new_weight
                        New_Ap[jSNodes[0],iSNodes[0]]=new_weight
                ## step 4: search modes to reduce in the moment space
                Evalqs, Evecqs=np.linalg.eig(SLq)
                tau=tau_Vec[SID]
                Evalq_idx=np.where(np.real(Evalqs)<1/tau)[0]
                m_k=len(Evalq_idx) if len(Evalq_idx)>=n_k else n_k
                ## step 5: calculate the re-scaled contributions of long-range eigenvectors
                SQq = np.zeros((SLq.shape[0],SLq.shape[1]))
                Evalq_idx=Evalqs.argsort()[:m_k]
                for ID in Evalq_idx:
                    Evecq = Evecqs[:,ID].reshape(-1,1)
                    SQq += np.real(Evalqs[ID]*np.matmul(Evecq,Evecq.T))

                SLp=Origin_Lp[SNodeID,:][:,SNodeID]
                Evalps, Evecps = np.linalg.eig(SLp)
                Evalp_idx=Evalps.argsort()[:m_k] 
                SQp = np.zeros((SLp.shape[0],SLp.shape[1]))
                for ID in Evalp_idx:
                    Evecp = Evecps[:,ID].reshape(-1,1)
                    SQp += np.real(Evalps[ID]*np.matmul(Evecp,Evecp.T))
                ## step 6: coarse grain q-order and p-order Laplacian in the momentum space 
                # calculate new SLq and SLp
                New_SLq = np.zeros((n_k,n_k))
                New_SLp = np.zeros((n_k,n_k))
                for i in range(n_k):
                    iNodes=Clusters[i]
                    iSNodes = [SNodeID[node] for node in Clusters[i]]
                    New_Lp[iSNodes[0],Non_Nodes]=np.sum(Origin_Lp[iSNodes,:][:,Non_Nodes],axis=0) # merge edges from different connected components
                    for j in range(i,n_k):
                        jNodes=Clusters[j]
                        new_weight=sum(SQq[iNodes,:][:,jNodes].reshape(-1))
                        new_weight0=sum(SQp[iNodes,:][:,jNodes].reshape(-1))
                        New_SLq[i,j]=New_SLq[j,i]=new_weight
                        New_SLp[i,j]=New_SLp[j,i]=new_weight0
                # remove false edges in SLq and SLp
                maskq=(New_SAq!=0).astype(int)
                New_SLq=New_SLq*maskq
                maskp=(New_Ap[New_Snodes,:][:,New_Snodes]).astype(int)
                New_SLp=New_SLp*maskp
                # ensure the sum of each row of q-order Laplacian operator is zero
                for i in range(n_k):
                    tmp=list(range(n_k))
                    tmp.remove(i)
                    New_SLq[i,i]=-np.sum(New_SLq[i,tmp])
                NewSub_Lqs.append(New_SLq)

                for i in range(n_k):
                    iNode1=New_Snodes[i]
                    for j in range(i+1,n_k):
                        jNode1=New_Snodes[j]
                        New_Lp[iNode1,jNode1]=New_SLp[i,j]
                        New_Lp[jNode1,iNode1]=New_SLp[j,i]
                # track node alignment
                for Nodes in Clusters:
                    SNodes = [SNodeID[node] for node in Nodes]
                    Node1 = SNodes[0]
                    NodeAlignment.append([Node1, SNodes])
            ClusterNodeAlignment.append(NodeAlignment)
        All_Alignment.append(ClusterNodeAlignment)

        New_Gq=nx.disjoint_union_all(NewSub_Gqs)
        New_Ap=New_Ap[New_Snodes_list,:][:,New_Snodes_list]
        New_Ap=np.maximum(New_Ap,New_Ap.T) # gaurantee the symmetry
        New_Gp=nx.from_numpy_array(New_Ap)
        New_Lp = New_Lp[New_Snodes_list,:][:,New_Snodes_list]
        New_Lp=np.minimum(New_Lp,New_Lp.T) # gaurantee the symmetry
        # ensure the sum of each row of p-order Laplacian operator is zero
        for i in range(New_Lp.shape[0]):
            tmp=list(range(New_Lp.shape[0]))
            tmp.remove(i)
            New_Lp[i,i]=-np.sum(New_Lp[i,tmp])
        New_Lq=NetworkUnion(NewSub_Lqs)
        print(New_Lq.shape)

        Lq_List.append(New_Lq)
        Lp_List.append(New_Lp)
        Gq_List.append(New_Gq)
        Gp_List.append(New_Gp)
        Iter=Iter+1
    Tracked_Alignment=TrackingUnitID(All_Alignment,Lq.shape[0])
    return C_List,Lq_List,Lp_List,Gq_List,Gp_List,All_Alignment

def TauSelection(L):
    ## Input: 
    # L is the initial high order Laplician representation, which can be the Multiorder Laplacian operator or the high-order path Laplacian

    ## Output:
    # tau is the ideal constant that determines the time scale
    TauVec=np.logspace(-2, np.log10(50), 200)
    A=np.abs(L)-np.diag(np.diag(L))
    n_ks=[]
    MLambdaVector=np.zeros_like(TauVec)
    for ID in range(len(TauVec)):
        Poss_tau=TauVec[ID]
        MatrixExp=expm(-Poss_tau*L)
        Rho=MatrixExp/np.trace(MatrixExp)
        MLambdaVector[ID]=np.trace(L @ Rho)

        diagonal_min = np.minimum(Rho.diagonal().reshape(-1, 1), Rho.diagonal())
        Rho_prime=Rho/diagonal_min
        Rho_prime=(Rho_prime+Rho_prime.T)/2
        Ref_adj=(Rho_prime>1).astype('int')
        # delete false edges not in subgraph of SL
        Ref_mask=(A!=0).astype(int)
        Ref_adj=Ref_adj*Ref_mask
        RefG = nx.from_numpy_array(Ref_adj) # reference graph
        n_ks.append(nx.number_connected_components(RefG))
    CVector=-np.power(TauVec[1:],2)*np.diff(MLambdaVector)/np.diff(TauVec)
    CVector = CVector[np.where(~np.isnan(CVector))[0]]
    TauVec = TauVec[np.where(~np.isnan(CVector))[0]]

    potential_localmax_id = argrelextrema(CVector,np.greater)[0]
    true_localmax_id = potential_localmax_id[np.where(CVector[potential_localmax_id]>0.5*np.max(CVector))[0]]
    tau=TauVec[true_localmax_id[0]]

    dCVector=np.diff(CVector)/np.diff(TauVec)
    plateau_idx = np.where((np.abs(dCVector)<5e-2)&(TauVec[1:]>=tau))[0]
    plateau_idx = [i for i in plateau_idx if i+1 in plateau_idx] # find consecutive index as true plateau
    if len(plateau_idx)>50 and (CVector[plateau_idx]>0.1*np.max(CVector)).all():
        tau = TauVec[plateau_idx[0]]

    n_ks=np.array(n_ks)
    n_ks=n_ks[np.where(~np.isnan(CVector))[0]]
    dn_ks=np.abs(np.diff(n_ks)/np.diff(TauVec))
    bi_tau=TauVec[np.argmax(dn_ks)]
    tau=tau if tau>bi_tau else bi_tau
    print(['our method: tau is ',tau,'!!'])
    return tau,CVector

def TrackingUnitID(All_Alignment,UnitNum):
    # convert nodeID within an iteration (All_Alignment) to global nodeID (Tracked_Alignment)
    Tracked_Alignment=copy.deepcopy(All_Alignment)
    UnitIDVec=list(range(UnitNum))
    for IterID in range(len(All_Alignment)):
        if IterID>0:
            for ClusterID in range(len(All_Alignment[IterID])):
                for CoarseID in range(len(All_Alignment[IterID][ClusterID])):
                    NodesToTrack=All_Alignment[IterID][ClusterID][CoarseID][1]
                    for IDT in range(len(NodesToTrack)):
                        TrackedID=UnitIDVec[NodesToTrack[IDT]]
                        Tracked_Alignment[IterID][ClusterID][CoarseID][1][IDT]=TrackedID
                    Tracked_Alignment[IterID][ClusterID][CoarseID][0]=Tracked_Alignment[IterID][ClusterID][CoarseID][1][0]
        # delete Coarsed node
        NodetoDelete=[]
        for ClusterID in range(len(Tracked_Alignment[IterID])):
            for CoarseID in range(len(Tracked_Alignment[IterID][ClusterID])):
                Nodes=Tracked_Alignment[IterID][ClusterID][CoarseID][1][1:]
                NodetoDelete.extend(Nodes)

        for IDD in NodetoDelete:
            UnitIDVec.remove(IDD)

    return Tracked_Alignment
\end{lstlisting}

\end{widetext}

\bibliography{apssamp}
\end{document}